\documentclass[11pt,a4paper]{article}
\pdfoutput=1
\usepackage{jheppub}
\usepackage{graphicx}
\usepackage{epsfig}
\usepackage{amsmath, amssymb}
\usepackage{dcolumn}
\usepackage{bm}
\usepackage{amsfonts}

\usepackage{subfigure}
\usepackage{amssymb}
\usepackage{epstopdf}

\newcommand{\bea}[1]{\begin{eqnarray} \mbox{$\label{#1}$}}
\newcommand{\eea}{\end{eqnarray}}
\newcommand{\be}[1]{\begin{equation} \mbox{$\label{#1}$}}
\newcommand{\ee}{\vspace{0.1cm}\end{equation}}

\newcommand{\bi}{\begin{itemize}}
\newcommand{\ei}{\end{itemize}}

\newcommand{\mn}{{\mu\nu}}

\newcommand{\hI}{\hspace{1cm}}

\newcommand{\hV}{\hspace{.5cm}}

\newcommand{\nn}{\nonumber}
\newcommand{\TT}{{\rm TT}}
\newcommand{\la}{\left\langle}
\newcommand{\ra}{\right\rangle}
\newcommand{\GW}{{_{\rm GW}}}

\newcommand{\bk}{{\mathbf k}}

\newcommand{\bp}{{\mathbf p}}

\newcommand{\bx}{{\mathbf x}}

\title{Stochastic Background of Gravitational Waves from Fermions, -- Theory and Applications --}

\author[a,b]{Daniel G. Figueroa}
\affiliation[a]{D\'epartement de Physique Th\'eorique and Center for Astroparticle Physics, Universit\'e de Gen\`eve, 24 quai Ernest Ansermet, CH--1211 Gen\`eve 4, Switzerland}
\affiliation[b]{Physics Department, University of Helsinki and Helsinki Institute of Physics, P.O. Box 64, FI-00014, Helsinki, Finland}
\emailAdd{daniel.figueroa@unige.ch}
\author[b]{Tuukka Meriniemi}
\emailAdd{tuukka.meriniemi@helsinki.fi}

\date{\today}

\abstract{Out-of-equilibrium fermions can be created in the early Universe by non-perturbative parametric effects, both at preheating or during the thermal era. An anisotro-pic stress is developed in the fermion distribution, acting as a source of a stochastic background of gravitational waves (GW). We derive a general formalism to calculate the spectrum of GW produced by an ensemble of fermions, which we apply to a variety of scenarios after inflation. We discuss in detail the regularization of the source, i.e.~of the unequal-time-correlator of the fermions' transverse-traceless anisotropic stress. We discuss how the GW spectrum builds up in time and present a simple parametrization of its final amplitude and peak frequency. We find that fermions may generate a GW background with a significant amplitude at very high frequencies, similarly to the case of preheating with scalar fields. A detection of this GW background would shed light on the physics of the very early Universe, but new technology at high 
frequencies is required, beyond the range accessible to currently planned detectors. 
}

\begin{document}
\maketitle

\section{Introduction}
\label{sec:Intro}

The most recent observations of the cosmic microwave background (CMB) by the Planck satellite~\cite{Planck} strongly support the inflation paradigm, i.e.~that the Universe underwent an early phase of accelerated expansion. Following inflation, there was -- necessarily -- a period of "reheating", during which (almost) all the particles in the Universe were created out of the energy responsible for inflation. The produced particle species eventually thermalized, and  after that the history of the Universe followed the hot Big Bang theory. Reheating, however, is far from being understood. The initial stages are typically expected to be driven by non-perturbative quantum effects, generally coined as $preheating$. Moreover, the physics since the end of Reheating till the Big Bang Nucleosynthesis (BBN) is uncertain, and there is plenty of room for speculation. Several backgrounds of gravitational waves (GW) are indeed expected to have been generated during these early stages. If detected, these backgrounds would 
bring us invaluable information about many of the unknowns in the early Universe physics. 

The existence of GW constitute a basic prediction of general relativity. To date, unfortunately, GW have yet not been directly detected.  The measured decay of the orbital period of the first binary pulsar ever discovered, PSR 1913+16, has nevertheless provided indirect evidence for their existence~\cite{HulseTaylor}. Actually, based on theoretical considerations, it is expected that the Universe should be permeated by a variety of GW backgrounds of diverse origin. For instance, from astrophysical phenomena, we expect GW from the collapse of supernovas or the coalescence of compact binaries. From the early Universe, we expect GW from non-equilibrium phenomena. The presence today of a GW background can be characterized by its spectrum, defined as the energy density in GW per logarithmic interval of frequency, $\Omega_\GW(f) \equiv {1\over \rho_c}{d\rho_\GW\over d\log f}$, normalized to the actual critical energy density $\rho_c$. Due to the uncertainty in the actual expansion rate of the universe $H_o$, and 
since $\rho_c \propto H_o^2$, constraints are often set on the combination $h^2\Omega_\GW(f)$, where $h$ is the little-Hubble parameter defined as $H_o = h \times 100$ Km s$^{-1}$ Mpc$^{-1}$.

A number of constraints on GW have already been derived from a variety of observations. From BBN, there is an upper bound on the total amount of GW as $\int h^2\Omega_\GW(f)\,d\log f$  $< 5.6\cdot10^{-6}(N_{\rm eff}-3)$, with $N_{\rm eff}$ the effective number of relativistic species. Ignoring the unlikely possibility that the GW spectrum may have a very narrow peak, one can take as a rule of thumb that for frequencies $f \gtrsim 10^{-10}$ Hz, $h^2\Omega_\GW(f) \lesssim 10^{-5}$~\cite{Allen}. From CMB observations there is the so called COBE bound, $h^2\Omega_\GW(f) \lesssim 7\cdot10^{-11}$, valid for the range $3\cdot10^{-18}$ Hz $< f < 10^{-16}$ Hz. This bound becomes stronger, as $h^2\Omega_\GW(f) \lesssim 10^{-14}$, at $f \simeq 10^{-16}$ Hz~\cite{Maggiore}. For other CMB constraints see also~\cite{Elena}. From individual millisecond pulsars and the European Pulsar Timing Array (EPTA), it have been found respectively $h^2\Omega_\GW(f) \lesssim 2\cdot10^{-8}$~\cite{mscPulsar} and $h^2\Omega_\GW(f) \lesssim 6\cdot10^{-9}$~\cite{Sanidas}, both at $f \simeq 4\cdot10^{-9}$ Hz. 

There are also many plans for direct detection experiments, such as the Laser Interferometer Gravitational Wave Observatory (LIGO)~\cite{LIGO}, the European Laser Interferometer Space Antenna (eLISA)~\cite{eLISA}, the Big Bang Observer (BBO)~\cite{BBO}, the Decihertz Interferometer Gravitational Wave Observatory (DECIGO)~\cite{DECIGO}, or the Einstein Telescope (ET)~\cite{ET}. All these observatories operate at some typical frequencies ranging from $\sim 10^{-5}$ Hz to $\sim 10^3$ Hz. From the lack of a direct detection, the LIGO collaboration has already provided an upper limit as $h^2\Omega_\GW(f) \lesssim 3.6\cdot10^{-6}$, valid at $f \simeq 10^{2}$ Hz. From the polarization of the CMB one could also infer upper bounds on the amplitudes of potential GW backgrounds. These arise because metric perturbations generated by inflation (or cosmic defects) include also tensor modes, which are GW. If the scale of inflation (or the vacuum expectation value -- VEV -- of the defects) is sufficiently high, one might 
even hope to detect directly~\cite{Kamionkowski,JuanRuthEliDaniMartin} the tensor perturbations with the Planck Surveyor (expected to release its polarization results in 2014), or with polarization optimized ground-based CMB experiments such as PolarBear or QUIET~\cite{PolarBearQUIET}.

During the post-inflationary period, but previous to BBN, several GW backgrounds are expected to be generated by causal mechanisms. These include preheating~\cite{GWpreheating1}-\cite{GWpreheating6}, phase transitions~\cite{PT1}-\cite{PT5}, turbulent motions~\cite{Turbulence1}-\cite{Turbulence3} and cosmic defects~\cite{GWdefects1}-\cite{GWdefects3}, which are all non-equilibrium phenomena. The resulting GW backgrounds are very different in nature from the inflationary one, which was originated from quantum fluctuations~\cite{StarobinskyGW}. In particular, just after the end of inflation, the (almost) homogeneous inflaton field begins to oscillate around the minimum of its potential. If it is coupled to other fields -- and it should be, since it must decay --, the oscillations provide an effective time dependent mass to these fields. Due to this there can be, under certain circumstances, a significant burst of particle production from the time dependent inflaton field. This is precisely what we were 
referring to before as preheating~\cite{Preheating1},\cite{Preheating2}. As mentioned, the production of particles during preheating takes place out-of-equilibrium. Besides, it corresponds to a non-perturbative effect, which cannot be described by standard quantum field theory perturbation techniques, so it has to be studied by other means. With the help, for instance, of lattice simulations, the non-thermal distribution of the excited fields at preheating has been studied in detail in several scenarios.

Energetically speaking, the excitations of the fields during preheating are very violent, what gives rise to the production of a significant background of GW~\cite{GWpreheating1}-\cite{GWpreheating6}. In the case of chaotic inflation models the fields coupled to the inflaton are excited through a phenomenon known as parametric resonance~\cite{Preheating1},\cite{Preheating2}. In other models like Hybrid inflation, there is a tachyonic instability~\cite{Preheating3}. Either way, the field distributions develop a non-trivial anisotropic stress due to the non-equilibrium nature of the process, and that acts precisely as the source of GW. Something similar occurs in the case of a phase transition. There the collisions of the nucleated bubbles in the true vacuum, also produce to a non-equilibrium distribution of the order-parameter field (the "Higgs" of the model), which develops an anisotropic stress sourcing correspondingly GW~\cite{PT1}-\cite{PT5}.

Each source of GW produce a spectrum with a distinctive shape and amplitude. It is then important to characterize all the potential sources. However, most works studying post-inflationary early universe phenomena responsible for the generation of GW, have focused only on bosonics sources, often scalar fields. Thus, we want to complete the picture here, by considering fermionic fields as the source of GW. Not only it is conceivable that fermions were produced in the early stages of the Universe, it is also necessary that they were created somewhen between the end of inflation and BBN. All matter fields of which we have direct evidence, the quarks and leptons of the Standard Model, are indeed fermions. Dark Matter candidates include heavy fermions, and the realization of Baryo/Lepto-genesis can be attained through mechanisms involving fermions. Fermionic preheating~\cite{Fermions1}-\cite{Fermions3} after inflation is indeed no less natural than the usual bosonic preheating~\cite{Preheating1},\cite{
Preheating2}, but simply more difficult to treat.

It is then expected that a distribution of fermions was created by some mechanism(s) somewhen during the evolution of the early Universe, before BBN. Usual effects by which fermions are created correspond to out-of-equilibrium phenomena, as in the case of bosonic species. The spectrum of such fermions is then non-thermal, and their energy-momentum tensor acts as an anisotropic-stress over the background. It is then natural to expect that such non-thermally produced fermions might generate a GW background. The question then is: how big is the amplitude, and what is the frequency? The aim of the present paper is to answer in detail these questions. We have developed a general formalism to compute the GW spectrum generated by an ensemble of fermions, which then we have applied to several post-inflationary/pre-BBN scenarios. There are two distinct possibilities: first, fermions generated in preheating due to their interactions with the inflaton, and second, fermions excited because of their interactions with a scalar field (other 
than the inflaton), spectator in the post-reheating thermal era. We treat both possibilities.

In a recent letter~\cite{KariDaniTuukka} we presented the GW spectrum generated by fermions created at some post-inflationary scenarios. In this paper we will go through the physics in detail, describing carefully the technicalities of our formalism and considering its application within several scenarios, including a systematic exploration of the parameters involved.
With our present work, we want to demonstrate that, contrary to naive expectations based on Pauli-blocking arguments, fermions can indeed act as very effective generators of GW. The natural frequencies of the backgrounds we predict lie, unfortunately, in a high-frequency window inaccessible to currently planned GW detectors. This is, of course, certainly a concern. However, we would like to see this fact as a reason to stimulate our experimental colleagues to start an active seek for the possibility of developing high-frequency GW detectors. A detection of a GW background like the one we predict here, would provide a direct observational access to the physics of the very early Universe. Therefore, with no doubt, the endeavor is certainly worth to be pursued.

The paper is divided as follows. In section~\ref{sec:GW} we review the formalism of stochastic backgrounds of GW. In section~\ref{sec:Theory}, we first present a description of fermions in an expanding universe (section~\ref{subsec:FermionsFRWdynamics}), and then develop a general formalism for computing the spectrum of GW that they generate (section~\ref{subsec:GWspectrumFromFermions}). We discuss the problem of ultraviolet divergences which arise naturally in the problem, and propose a method to regularize the fermionic source of GW (section~\ref{subsec:Regularization}). In section~\ref{sec:Applications} we first discuss general aspects of the dynamics of fermions and scalar fields interacting in an expanding universe (section~\ref{subsec:YukawaAndScalarDynamics}), followed by a parametric estimation of the frequency and amplitude of the spectrum of GW created by fermions (section~\ref{subsec:ParametricAntsazs}). We then analyze in detail the scenarios where fermions are excited during preheating from a 
massless (section~\ref{subsec:FermionicMasslessPreheating}) and a massive (section~\ref{subsec:FermionicMassivePreheating}) inflaton, as well as the case when the fermion production takes place from a massive scalar field in the thermal era after the completion of reheating (section~\ref{subsec:ThermalEraScenarios}). Finally in section~\ref{sec:Conclusions} we summarize all our results and draw our final conclusions. In Appendices~\ref{app:Traceology} and~\ref{app:AuxFermField} we present the technical details of our calculations, in Appendix~\ref{app:FermionsMass} we discuss the case of fermions with a bare mass, and finally in Appendix~\ref{app:GWbosons} we review -- for completeness -- the formalism of stochastic GW backgrounds from scalar fields.

All through the paper we will work in $\hbar = c = 1$ units, with $M_p \simeq 1.2\times10^{19}\,{\rm GeV}$ the Planck mass, related to the Newton constant as $G = 1/M_p^2$. Summation will be assumed over repeated indices. We will use the Fourier transform convention
\begin{equation}
f(\mathbf{x},t)=\int\frac{d\mathbf{k}}{\left(2\pi\right)^{3}}~e^{-i\mathbf{k\cdot x}}~f(\mathbf{k},t) \hV\leftrightarrow \hV f(\bk,t) = \int d\bx ~e^{+i\mathbf{k\cdot x}}~f(\bx,t).
\end{equation}
Note also that the following acronyms will be often used: BBN for big bang nucleosynthesis, GW for gravitational waves, CMB for cosmic microwave background, IR for infrared, UV for ultraviolet, $dof$ for degrees of freedom, TT for transverse-traceless, FRW for Friedman-Robertson-Walker, RD for radiation-dominated, MD for matter-dominated, VEV for vacuum expectation value, tNO for time-dependent normal-ordering, and UTC for unequal time correlator.

\section{Gravitational waves}
\label{sec:GW}

Gravitational Waves (GW) are tensor perturbations of the space-time metric. More specifically, they are the transverse-traceless (TT) degrees of freedom ($dof$) of the metric perturbations. After inflation the Universe is well described by a spatially flat Friedman-Robertson-Walker (FRW) background. The perturbed FRW line element with GW as the only perturbation, can then be written as
\begin{equation}
ds^{2}=a^{2}(t)\left[-dt^{2}+\left(\delta_{ij}+h_{ij}\right)dx^{i}dx^{j}\right],\label{eq:metric}
\end{equation}
with $a(t)$ the scale factor and $t$ the conformal time. The perturbations $h_{ij}$ verify the conditions $\partial_{i}h_{ij} = 0$ (transversality) and $h_{i}^{i} = 0$ (tracelessness), required to identify them with GW.

Splitting the Einstein equations into background and linearized equations, one finds that the GW $eom$~\cite{Maggiore} in a FRW background are given by
\begin{eqnarray}
\ddot{h}_{ij}(\bx,t)+2\mathcal{H}\dot{h}_{ij}(\mathbf{x},t)-\nabla h_{ij}(\mathbf{x},t)=16\pi G\,\Pi_{ij}^{\rm TT}(\mathbf{x},t),
\end{eqnarray}
with $\mathcal{H} \equiv \dot a/a$ the (comoving) Hubble rate and dots denoting derivatives with respect conformal time. The source $\Pi_{ij}^{\rm TT}$ is the TT-part of the anisotropic stress $\Pi_{ij}$, which we define below. The conditions $\partial_i \Pi_{ij}^{\rm TT} = \Pi_{ii}^{\rm TT} = 0$ then hold $\forall\,\bx, \forall\,t$. 

Obtaining the TT-part of a tensor in configuration space amounts to a non-local operation. It is more convenient to do it in Fourier space, where a projector to filter out only the TT $dof$ of a tensor can be easily built. The $eom$ of GW in fourier space reads
\begin{eqnarray}\label{eq:GWeq}
\ddot{h}_{ij}(\bk,t)+2\mathcal{H}\dot{h}_{ij}(\mathbf{k},t)+ k^2 h_{ij}(\mathbf{k},t)=16\pi G\,\Pi_{ij}^{\rm TT}(\mathbf{k},t),
\end{eqnarray}
where $\bk$ is the comoving wave-number and $k=|\mathbf{k}|$ is the modulus. The GW source can then be written as
\begin{equation}
\Pi_{ij}^{\rm TT}(\mathbf{k},t) = \Lambda_{ij,lm}(\hat\bk)\,\Pi_{lm}(\mathbf{k},t).
\end{equation}
where $\Lambda_{ij,lm}(\hat\bk)$ is a TT-projection operator defined as\footnote{Note that for the homogeneous mode $\bk = {\bf 0}$ this projection is ill-defined. However, the transversality condition is automatically verified by a homogeneous mode, so it is then sufficient to redefine the projector for this case as $\Lambda_{ij,lm}({\bf 0}) \equiv \delta_{il}\delta_{jm}-{1\over3}\delta_{ij}\delta_{lm}$. This guarantees the tracelessness. For a discussion on how to define the analogous TT-projection on a discrete space (a lattice grid in numerical simulations), see~\cite{GWpreheatingTT}.}
\begin{eqnarray}\label{projector}
\indent \Lambda_{ij,lm}(\mathbf{\hat k}) \equiv P_{il}(\hat\bk)
P_{jm}(\hat\bk) - {1\over2} P_{ij}(\hat\bk)
P_{lm}(\hat\bk),\,\\
P_{ij} = \delta_{ij} - \hat k_i \hat k_j\,,\hV \hat k_i = k_i/k\hI\hV
\end{eqnarray}
One can easily see that the transverse-traceless conditions in Fourier space, $k_i\Pi_{ij}^{\rm TT}(\hat\bk,t) = \Pi_{ii}^{\rm TT}(\hat\bk,t) = 0$, are fulfilled at any time, thanks to the fact that $P_{ij}\hat k_j = 0$ and $P_{ij}P_{jm} = P_{im}$. 

The anisotropic stress tensor $\Pi_{\mn}$ describes the deviation of the energy momentum tensor $T_{\mn}$ with respect that of a 
perfect fluid. The spatial-spatial components read
\begin{equation}
\Pi_{ij} \equiv T_{ij} - p\,g_{ij},
\end{equation}
with $p$ the homogeneous background pressure and $g_{ij}=a^2(t)(\delta_{ij}+h_{ij})$ the spatial-spatial metric tensor.
In the scenarios we consider in this paper the energy budget is dominated by a homogeneous and isotropic background fluid, which causes the universe to expand effectively either as radiation- (RD) or matter-dominated (MD). The corresponding energy-momentum of this background is that of a perfect fluid, with spatial-spatial components $T^{\rm pf}_{ij} = p\,g_{ij}$. On top of this there is a subdominant contribution from fermions, which have their own energy-momentum tensor $T_{ij}^{\rm F}$. Hence, in these scenarios, the (spatial-spatial components of the) total energy-momentum tensor are given by $T_{ij} = T^{\rm pf}_{ij} + T^{\rm F}_{ij}$. It is clear then that $\Pi_{ij} = T_{ij}^{\rm F}$, what implies that it is the TT-part of the fermions energy-momentum tensor that will act as a source of GW.

Rescaling the tensor perturbations as
\begin{equation}
\bar{h}_{ij}\left(\mathbf{k},t\right)=a(t)h_{ij}\left(\mathbf{k},t\right)\,, \end{equation}
the GW $eom$ become
\begin{equation}
\ddot{\bar{h}}_{ij}\left(\mathbf{k},t\right)+\left(k^{2}-\frac{\ddot{a}(t)}{a(t)}\right)\bar{h}_{ij}\left(\mathbf{k},t\right)=16\pi G\,a(t)\Pi_{ij}^{\rm TT}\left(\mathbf{k},t\right).
\end{equation}
Either in RD or MD, and in general for a scale factor with any power law behavior in time, $\ddot{a}/{a} \sim \mathcal{H}^{2}$. Thus the term $\ddot{a}/{a}$ is negligible at sub-horizon scales $k \gg \mathcal{H}$, and therefore we will drop it from now on. The GW $eom$ then looks like
\begin{equation}\label{eq:h_e.o.m.}
\ddot{\bar{h}}_{ij}\left(\mathbf{k},t\right)+k^{2}\bar{h}_{ij}\left(\mathbf{k},t\right)=16\pi G\,a(t)\Pi_{ij}^{\rm TT}\left(\mathbf{k},t\right).
\end{equation}
The solution of eq.~(\ref{eq:h_e.o.m.}) is given by a convolution with the Green function associated to a free wave-operator in Minkowski, $G(k,t-t') = {1\over k}\sin(k(t-t'))$. That is, at times $t > t_I$, with $t_{I}$ the initial time with no gravitational waves, i.\,e.~$h_{ij}\left(\mathbf{k},t_{I}\right) = \dot{h}_{ij}\left(\mathbf{k},t_{I}\right) = 0$, we obtain
\begin{equation}\label{eq:h(Pi)}
{h}_{ij}\left(\mathbf{k},t\right) = {\bar{h}_{ij}(\bk,t)\over a(t)} = \frac{16\pi G}{ a(t)k}\int_{t_{I}}^{t}dt'\sin\left[k\left(t-t'\right)\right]a(t')\,\Pi_{lm}^\TT\left(\mathbf{k},t'\right)\,.\\
\end{equation}
\vspace{.1cm}

\subsection{The Spectrum of Gravitational Waves}

Expanding the Einstein equations to second order in the tensor perturbations, one recognizes that the energy density of a GW background is given by~\cite{Maggiore}
\begin{eqnarray}\label{eq:GWrhoCont}
\rho_{\GW}(t) &=& \frac{1}{32\pi G a^2(t)}\left\langle \dot{h}_{ij}(\bx,t)\dot{h}_{ij}(\bx,t)\right\rangle_V \nn\\
&\equiv & \frac{1}{32\pi G a^2(t)}\frac{1}{V}\int_V d\bx\, \dot h_{ij}(\bx,t)\dot h_{ij}(\bx,t) \nn\\
&=& \frac{1}{32\pi G a^2(t)}\int\frac{d\bk}{(2\pi)^3}\frac{d\bk'}{(2\pi)^3}~\dot h_{ij}(\bk,t)\dot h_{ij}^*(\bk',t)~\frac{1}{V}\int_V\hspace*{-1mm} d\bx~e^{-i\bx(\bk-\bk')}
\end{eqnarray} 
with $\langle...\rangle_V$ a spatial average over a sufficiently large volume $V$ encompassing all the relevant wavelengths $\lambda_*$ of the $h_{ij}$ perturbations. In the limit $V^{1/3} \gg \lambda_*$, \\
$$\int_{V\gg\lambda_*^3}\hspace*{-1mm} d\bx~e^{-i\bx(\bk-\bk')} \,\hV\rightarrow\hV\, (2\pi)^3\delta^{(3)}(\bk-\bk'),$$
and then,
\begin{equation}
 \rho_{\GW}(t) = \frac{1}{32\pi G a^2(t) V}\int\frac{d\bk}{(2\pi)^3}~\dot h_{ij}(\bk,t)\dot h_{ij}^*(\bk,t)
\end{equation} 
The GW energy density spectrum (per logarithmic interval), defined as
\begin{equation}
 \rho_{\GW}(t) = \int\frac{d\rho_{\GW}}{d\log k}\,d\log k\,,
\end{equation}
is then found to be
\begin{eqnarray}\label{eq:GWrhoContSpectrum}
\frac{d\rho_{\GW}}{d\log k} = 
\frac{k^3}{(4\pi)^3 G\,a^2(t)V} \int \frac{d\Omega_k}{4\pi}\,\dot h_{ij}(\bk,t)\dot h_{ij}^*(\bk,t)\,,
\end{eqnarray}
where $d\Omega_k$ represents a solid angle element in $\bk$-space. 

In the early Universe, however, we often encounter situations where GW are created from a stochastic source. Hence the spatial distribution of the tensor perturbations is assumed to be stochastic, following the random distribution of the source. In such cases we can apply the {\tt Ergodic hypothesis}, which amounts to replace $\langle...\rangle_V$ by an ensemble average $\langle...\rangle$ over realizations. A stochastic background of GW can then be described by
\begin{eqnarray}\label{eq:GW_energyDensity}
\rho_{\mathrm{\GW}} &=& \frac{1}{32\pi G a^2(t)}\left\langle \dot{h}_{ij}(\bx,t)\dot{h}_{ij}(\bx,t)\right\rangle \nn\\
&=& \frac{1}{32\pi G a^2(t)}\int\frac{d\mathbf{k}}{\left(2\pi\right)^{3}}\frac{d\mathbf{k}'}{\left(2\pi\right)^{3}}~e^{i\bx(\mathbf{k-k'})}\left\langle\dot{{h}}_{ij}\left(\mathbf{k},t\right)\dot{{h}}_{ij}^{*}\left(\mathbf{k'},t\right)\right\rangle 
\end{eqnarray}
The expectation value in the second line of eq.~(\ref{eq:GW_energyDensity}), assuming statistical homogeneity and isotropy, defines the power spectrum of the tensor perturbation first derivatives,
\begin{equation}
\left\langle\dot{{h}}_{ij}\left(\mathbf{k},t\right)\dot{{h}}_{ij}^{*}\left(\mathbf{k'},t\right)\right\rangle \equiv (2\pi)^3\,\mathcal{P}_{\dot h}(k,t)\delta^{(3)}(\bk-\bk')\,.
\end{equation}
Thus
\begin{equation}\label{eq:GWspectByPhdot}
\rho_{\mathrm{\GW}}(t) = \frac{1}{(4\pi)^3 G a^2(t)}\int\hspace*{0mm} {dk\over k}~k^3\,\mathcal{P}_{\dot h}(k,t)  
\end{equation}
and from here, the GW energy density spectrum reads
\begin{eqnarray}\label{eq:GWrhoStochaSpectrum}
\frac{d\rho_{\GW}}{d\log k}(k,t) = \frac{1}{(4\pi)^3 G\,a^2(t)}\,k^3\,\mathcal{P}_{\dot h}(k,t)
\end{eqnarray}

\noindent Obtaining $\mathcal{P}_{\dot h}(k,t)$ is simple. With the help of eq.~(\ref{eq:h(Pi)}), first we write
\begin{equation}
\dot{h}_{ij}(\mathbf{k},t) = \frac{16\pi G}{ka(t)}\int_{t_{I}}^{t}dt'a(t')\,\mathcal{G}(k(t-t'))\,\Pi_{ij}^\TT(\mathbf{k},t'),\label{eq:Dh(Pi)}
\end{equation}
where
\begin{equation}
\mathcal{G}(k(t-t')) \equiv \left(k\cos[k(t-t')]-\mathcal{H}\sin[k(t-t')]\right)
\end{equation}
From here we obtain
\begin{eqnarray}\label{eq:P_doth_expanded}
&& \left\langle\dot{{h}}_{ij}\left(\mathbf{k},t\right)\dot{{h}}_{ij}^{*}\left(\mathbf{k'},t\right)\right\rangle\\
&& \hspace{0.3cm} = (2\pi)^3\,\frac{(16\pi G)^2}{k^2a^2(t)}\int_{t_{I}}^{t}dt'\int_{t_{I}}^{t}dt''a(t')a(t'')\,\mathcal{G}(k(t-t'))\,\mathcal{G}(k(t-t''))\,\Pi^2(k,t',t'')\,\delta^{(3)}(\bk-\bk'),\nn
\end{eqnarray} 
where we have introduced the {\tt unequal-time-correlator} (UTC), ${\Pi}^2(k,t,t')$, of the TT-part of the anisotropic-stress $\Pi_{ij}^{\TT}$, 
\begin{eqnarray}\label{eq:UTC}
\left\langle {\Pi}_{ij}^\TT(\bk,t)\,{{\Pi}_{ij}^{\TT}}(\bk',t')\right\rangle \equiv (2\pi)^3\,{\Pi}^2(k,t,t')\,\delta^{(3)}(\bk-\bk')
\end{eqnarray}

Once GW production ends GW propagate as free waves, each mode oscillating with period $T_k = {2\pi\over k}$. In order to define correctly the energy density spectrum of GW we need to perform a time average over those oscillations. Strictly speaking, at the moment when GW production ends, we should match the solution~(\ref{eq:h(Pi)}) with the freely propagating wave solution\footnote{The freely propagating waves are described by a superposition of the linearly independent homogeneous solutions to the free-source GW $eom$.}. From there we should build the GW energy density spectrum with the free waves, and only then perform the time average over oscillations. It is however mathematically equivalent to take the time average over the product of $\mathcal{G}(\bk,t,t')$ functions. We obtain
\begin{eqnarray}
\la\mathcal{G}(\bk,t,t')\mathcal{G}(\bk,t,t'')\ra_{T_k} \equiv {1\over T_k}\int_t^{t+T_k}d\tilde t~\mathcal{G}(\bk,\tilde t,t')\mathcal{G}(\bk,\tilde t,t'') = {1\over2}(k^2+\mathcal{H}^2(t))\cos[k(t'-t'')]\nn\\
\end{eqnarray}
Replacing $\mathcal{G}(\bk,t,t')\,\mathcal{G}(\bk,t,t'')$ by $\la\mathcal{G}(\bk,t,t')\mathcal{G}(\bk,t,t'')\ra_{T_k}$ in eq.~(\ref{eq:P_doth_expanded}), and taking into account that at subhorizon scales $(k^2+\mathcal{H}^2(t)) \approx k^2$, we then find $\mathcal{P}_{\dot h}$ as 
\begin{eqnarray}\label{eq:PShdot}
\mathcal{P}_{\dot h} = \frac{(16\pi G)^2}{2a^2(t)}\int_{t_{I}}^{t}dt'\int_{t_{I}}^{t}dt''a(t')a(t'')\,\cos[k(t'-t'')]\,\Pi^2(k,t',t'')
\end{eqnarray}
Plugging eq.~(\ref{eq:PShdot}) into~eq.~(\ref{eq:GWrhoStochaSpectrum}), we finally find the energy density spectrum of a stochastic background of GW (at subhorizon scales) as
\begin{equation}\label{eq:GW_spectra(Pi)}
\frac{d\rho_{\mathrm{\GW}}}{d\log k}\left(k,t\right) =  \frac{2}{\pi}\,{G\,k^3\over a^4(t)}\int_{t_{I}}^{t}dt'\int_{t_{I}}^{t}dt''\,a(t')\,a(t'')\cos[k(t'-t'')]\,\Pi^2(k,t',t''),
\end{equation}
\vspace{.2cm}

\subsection{The Spectrum Today}

Besides $t_{I}$ as the initial time of GW generation, it is convenient to coin also the following times: the end of GW production, $t_{*}$, the first moment when the Universe become RD, $t_{_{\rm RD}}$, and finally today, as $t_o$. Gravitational waves decouple immediately after production, so we can evaluate the GW energy density spectrum today from the spectrum computed at the time of production, simply by redshifting the frequency and amplitude correspondingly. In order to do so we need to relate the scale factor today $a_o$, with that at the beginning of GW production $a_I$. Denoting $T_o$ and $\rho_o$ the temperature and energy density of the CMB today, we obtain
\begin{eqnarray}
{a_I\over a_o} &=& \left({a_I\over a_{_{\rm RD}}}\right)\left(g_{s,o}\over g_{s,_{\rm RD}}\right)^{1\over3}\left({T_o\over T_{_{\rm RD}}}\right)\nn\\
&=& \left({a_I\over a_{*}}\right)\left({a_*\over a_{_{\rm RD}}}\right)\left(g_{s,o}\over g_{s,_{\rm RD}}\right)^{1\over3}\left(g_{o}\over g_{_{\rm RD}}\right)^{-{1\over4}}\left({\rho_o\over\rho_{_{\rm RD}}}\right)^{1\over4}\nn\\
&=& \left({a_I\over a_{*}}\right)\left(g_{s,o}\over g_{s,_{\rm RD}}\right)^{1\over3}\left(g_{o}\over g_{_{\rm RD}}\right)^{-{1\over4}}\left({\rho_o\over\rho_{*}}\right)^{1\over4}\left(a_* \over a_{_{\rm RD}}\right)^{(1-3w)\over4}\,,
\end{eqnarray}
where in the first line we have used the entropy conservation law in a thermal expanding background $a(t)T(t) \propto g_{s,t}^{-1/3}$, with $T$ and $g_{s,t}$ the background temperature and the entropic degrees of freedom at time $t$; in the second line we used the temperature-energy density relation of a relativistic thermal fluid $\rho(t) \propto g_t T^4(t)$, with $g_t$ the relativistic degrees of freedom at time $t$; and in the third line we used the energy density evolution law $\rho(t) \propto a(t)^{-3(1+w)}$ in an expanding background dominated by a fluid with effective equation of state (pressure-to-density ratio) $p/\rho = w$. Assuming that the effective degrees of freedom do not change from $t_*$ to $t_{_{\rm RD}}$, i.e.~$g_{s,*} = g_{s,_{\rm RD}}$ and $g_{*} = g_{_{\rm RD}}$, and taking into account that $g_{s,t} \sim g_{t}$, we then have $\left(g_{s,o}/g_{s,_{\rm RD}}\right)^{1/3}\left(g_{o}/g_{_{\rm RD}}\right)^{-{1/4}} \sim \left(g_{o}/g_{*}\right)^{1/12} \sim \mathcal{O}(1)$ [$\approx 1.77$ if 
$g_{o}/g_{*} = 
10^3$, $\approx 1.47$ if $g_{o}/g_{*} = 10^2$]. With all this, and taking into account the value today of the energy density of relativistic species $\rho_o \approx 2\cdot 10^{-15} eV^4$, the frequency today reads
\begin{equation}
f \equiv \left(a_I\over a_o\right){k\over2\pi} \approx \epsilon^{1/4}\left(\frac{a_{I}}{a_{*}}\right)\left(\frac{k}{\rho_{*}^{1/4}}\right)\times5\cdot10^{10}\mathrm{Hz},\label{eq:ftoday}
\end{equation}
where we have introduced the factor
\be{eq:epsilonParameter}\epsilon \equiv \left({a_*\over a_{_{\rm RD}}}\right)^{(1-3w)},\ee
to quantify our ignorance about the expansion rate between $t_*$ and $t_{_{\rm RD}}$. If at $t_*$ the Universe is already in a RD phase with $w = 1/3$, i.e.~$t_{_{\rm RD}} < t_*$, then we have $\epsilon = 1$. If on the contrary, the Universe is in an expanding phase with $w \neq 1/3$ between $t_*$ and $t_{_{\rm RD}}$, then $a_*/a_{_{\rm RD}} < 1$. In general, unless in a RD background since $t_*$, there is always a frequency shift by a factor $\epsilon^{1/4} < 1$.

The spectral amplitude of the GW background today, normalized to the actual critical energy density $\rho_c$, can be obtained as
\begin{eqnarray}
h^{2}\Omega_{\GW} &\equiv& {h^2\over \rho_c}\left(d\rho_{\GW}\over d\log k\right)_o =
h^2\Omega_{\rm rad}{1\over\rho_o}\left(a_{*}\over a_o\right)^4\left(d\rho_{\GW}\over d\log k\right)_*
\nn\\
&=& h^2\Omega_{\rm rad}\left(a_{*}\over a_{_{\rm RD}}\right)^4\left(g_{s,o}\over g_{s,_{\rm RD}}\right)^{4\over3}\left(g_{_{\rm RD}}\over g_{o}\right){1\over\rho_{_{\rm RD}}}\left(d\rho_{\GW}\over d\log k\right)_* \nn\\
&=& h^2\Omega_{\rm rad}\left(a_{*}\over a_{_{\rm RD}}\right)^{1-3w}\left(g_{s,o}\over g_{s,_{\rm RD}}\right)^{4\over3}\left(g_{_{\rm RD}}\over g_{o}\right){1\over\rho_*}\left(d\rho_{\GW}\over d\log k\right)_*\,,
\end{eqnarray}
where in the first line we used the fact that the energy density of freely propagating GW scales as a radiation fluid, i.e.~$\rho_{\GW} \propto 1/a(t)^4$; in the second line we used the evolution of the radiation $dof$ from $t_{_{\rm RD}}$ till today, $\rho_o = \left(g_{s,_{\rm RD}}/g_{s,o}\right)^{4/3}\left(g_{o}/g_{_{\rm RD}}\right)\rho_{_{\rm RD}}\left(a_{_{\rm RD}}/a_o\right)^4$; and in the third line we used the evolution of the total energy density from $t_*$ to $t_{_{\rm RD}}$ (assuming again that the effective $dof$ do not change), $\rho_{_{\rm RD}} = \rho_*(a_{_{\rm RD}}/a_*)^{-3(1+w)}$.

Applying the same considerations as with the frequency, the amplitude today can be finally written as
\begin{equation}
h^{2}\Omega_{\GW}\equiv\frac{h^{2}}{\rho_{c}}\left(\frac{d\rho_{\GW}}{d\log k}\right)_{o}=h^{2}\Omega_{\mathrm{rad}}\left(\frac{g_{o}}{g_{*}}\right)^{1/3}\frac{\epsilon}{\rho_{*}}\left(\frac{d\rho_{\GW}}{d\log k}\right)_{*}\label{eq:Amptoday}
\end{equation}
where $\epsilon$ is defined as before. Here $h^{2}\Omega_{\mathrm{rad}}\simeq 4\cdot10^{-5}$, and the ratio of the number of relativistic degrees of freedom today to those active at end of GW production is $\left(g_{o}/g_{*}\right)^{1/3}\sim\mathcal{O}(0.1)$, so the total prefactor is of the order $\mathcal{O}(10^{-6})$.

\section{Fermions as a source of Gravitational Waves - Theory}
\label{sec:Theory}

The previously presented formalism applies to any stochastic GW source in a (flat) FRW background, characterized by its unequal-time-correlator (UTC), $\Pi^2(k,t,t')$. In this section we develop a formalism to calculate specifically the spectrum of GW created by fermions. We discuss first, in section~\ref{subsec:FermionsFRWdynamics}, the dynamics of fermions in a FRW background. Then we derive the explicit form of the fermionic UTC in section~\ref{subsec:GWspectrumFromFermions}, followed by a discussion in section~\ref{subsec:Regularization} about the need and procedure to regularize it.

\subsection{Fermions in a FRW background}
\label{subsec:FermionsFRWdynamics}

The Dirac equation for a spin-$\frac{1}{2}$ fermion field in a flat FRW background is
\begin{equation}
\left[\frac{i}{a(t)}\gamma^{\mu}D_{\mu}-m_{\psi}(t)\right]\Psi\left(\mathbf{x},t\right)=0,\label{eq:Dirac}
\end{equation}
where the time-dependence considered in the mass is due to possible interactions of $\Psi$ with other (homogeneous) fields, and $D_{\mu}$ is the covariant derivative given by
\begin{equation}
D_{\mu}=\partial_{\mu}+\frac{1}{4}\gamma_{\alpha\beta}\omega_{\mu}^{\alpha\beta},\label{eq:CovDer}
\end{equation}
with $\gamma_{\alpha\beta}\equiv\frac{1}{2}\left(\gamma_{\alpha}\gamma_{\beta}-\gamma_{\beta}\gamma_{\alpha}\right)$, and $\omega_{\mu}^{\alpha\beta}$ the spin connection coefficients. The $\gamma^{\mu}$ are the ordinary flat-space gamma matrices verifying the anti-commutation relations
\begin{equation}
\lbrace\gamma_\mu,\gamma_\nu\rbrace = 2\eta_{\mu\nu},\hI \lbrace\gamma^\mu,\gamma^\nu\rbrace = 2\eta^{\mu\nu},
\end{equation}
with $\eta^{\mn} = \eta_{\mn} = \mathrm{diag}(-1,1,1,1)$ the Minkowski metric, and $\gamma^{\mu} = \eta^{\mu\nu}\gamma_\nu$. In the Dirac basis, these $4\times4$ matrices read
\begin{eqnarray}\label{eq:gamma:matrices}
\gamma_{0}=\left(\begin{array}{cc}
{\tt 1} & 0\\
0 & -{\tt 1}
\end{array}\right),\hV
\gamma_{i}=\left(\begin{array}{cc}
0 & -\sigma_{i}\\
\sigma_{i} & 0
\end{array}\right), 
\end{eqnarray}
with ${\tt 1}$ the $2\times2$ identity matrix and $\sigma_{i}$ ($i=\{ 1,2,3\}$) the Pauli matrices.

The spin connection coefficients $\omega_{\mu}^{\alpha\beta}$ in eq.~(\ref{eq:CovDer}) are defined \cite{SUPERGRAVITY} as
\begin{eqnarray}\label{eq:SpinCon}
\omega_{\mu}^{ab} = \frac{1}{2}e_{c\mu}\left(\Omega^{abc}-\Omega^{bca}-\Omega^{cab}\right)\,,\\
\Omega_{abc} \equiv e_{a}^{\mu}e_{b}^{\nu}\left(\partial_{\mu}e_{\nu c}-\partial_{\nu}e_{\mu c}\right)\,,\hV
\end{eqnarray} 
with $e_{\nu}^{\alpha}$ the local Lorentz frame fields, i.e.~the {\it vierbeins}, defined from the background metric $g_{\mu\nu} = a^2(t)\eta_\mn$ as
\begin{equation}
g_{\mu\nu}=\eta_{\alpha\beta}e_{\mu}^{\alpha}e_{\nu}^{\beta}.
\end{equation}

\noindent By using eq.~(\ref{eq:SpinCon}) we obtain the spin connection coefficients in a FRW background as 
\begin{eqnarray}
\omega_{\mu}^{i0}=-\omega_{\mu}^{0i} = \mathcal{H}\delta_{\mu}^{i},\hI
\omega_{\mu}^{ij}=\omega_{\mu}^{00}=0,\label{eq:spinconn.}
\end{eqnarray}
with $\mathcal{H}$ the Hubble rate in conformal time.
The Dirac equation (\ref{eq:Dirac}) becomes then
\begin{equation}
\left[i\gamma^{\mu}\partial_{\mu}+i\frac{3}{2}\mathcal{H}\gamma^{0}-a(t)m_{\psi}(t)\right]\Psi\left(\mathbf{x},t\right)=0.\label{eq:Dirac2}
\end{equation}
By an appropriate conformal redefinition of the Dirac field as
\begin{equation}
\psi(\mathbf{x},t)=a^{3/2}(t)\Psi(\mathbf{x},t),
\end{equation}
we can remove the friction term $\propto \mathcal{H}\gamma^0$, and write the Dirac equation like
\begin{equation}
\left[i\gamma^{\mu}\partial_{\mu}-a(t){m}_{\psi}(t)\right]\psi\left(\mathbf{x},t\right)=0.\label{eq:DiracPhi}
\end{equation}
Thus we have reduced the problem to the Dirac equation in Minkowski, but with an effective time-dependent mass $a(t){m}_{\psi}(t)$.

Since the effective mass is homogeneous in space, we can quantize the Dirac field as usual, like
\begin{eqnarray}
\psi(\mathbf{x},t) &=& \int\frac{d\mathbf{k}}{\left(2\pi\right)^{3}}\,e^{-i\mathbf{k\cdot x}}\left[\hat a_{\mathbf{k},r}{\tt u}_{\mathbf{k},r}(t)+\hat b_{-\mathbf{k},r}^{\dagger}{\tt v}_{\mathbf{k},r}(t)\right],
\label{eq:Phi(u,v)}
\\
\bar{\psi}(\mathbf{x},t) &=& \psi^{\dagger}(\mathbf{x},t)\gamma_{0}=\int\frac{d\mathbf{k}}{\left(2\pi\right)^{3}}\,e^{+i\mathbf{k\cdot x}}\left[
\hat a^{\dag}_{\mathbf{k},r}{\bar{\tt u}}_{\mathbf{k},r}(t) + \hat b_{-\mathbf{k},r}{\bar{\tt v}}_{\mathbf{k},r}(t)\right],
\label{eq:barPhi(u,v)}
\end{eqnarray}
with time-independent creation/annihilation operators satisfying the canonical anticommutation relations 
\begin{equation}\label{eq:anticom}
\left\{ \hat a_{\mathbf{k},r},\hat a_{\mathbf{k}',r'}^{\dagger}\right\} =\left\{ \hat b_{\mathbf{k},r},\hat b_{\mathbf{k}',r'}^{\dagger}\right\} =\left(2\pi\right)^{3}\delta_{r,r'}\delta^{(3)}(\mathbf{k-k'})\,.
\end{equation}
Other anticommutators vanish and the vacuum state $|0\rangle$ is defined as usual by
\begin{equation}
\label{eq:IntVac}
\hat a_{\bk,r} |0\rangle = \hat b_{\bk,r} |0\rangle = 0, \hV\forall\,\bk,r
\end{equation} 
The four-component spinors can be
written as 
\begin{eqnarray}\label{eq: u_r=00003Du_+,-}
{\tt u}_{\mathbf{k},r}(t) \equiv 
\left(\begin{array}{c}
\vspace{0.4cm}u_{\mathbf{k},+}(t)\,S_{r}\\ u_{\mathbf{k},-}(t)\,S_{r}
\end{array}\right),\hI
{\tt v}_{\mathbf{k},r}(t)\equiv
\left(\begin{array}{c}
\vspace{0.4cm}v_{\mathbf{k},+}(t)\,S_{-r}\\ v_{\mathbf{k},-}(t)\,S_{-r}
\end{array}\right),
\end{eqnarray}
with $\{S_{r}\}$ 2-component spinors, eigenvectors of the helicity operator, normalized as $S_{r}^{\dagger}S_{r}=2$, and $S_{-r} \equiv -i\sigma_2S_r^*$ \cite{Chaicherdsakul}. Choosing eigenstates of the Pauli matrix $\sigma_{3}$ to be the basis of the spinors $S_{r}$, these read
\begin{eqnarray}
S_{1}=-S_{-2}=\left(\begin{array}{c}
\vspace{0.2cm}1 \\ 0\end{array}\right),\hI
S_{2}=S_{-1}=\left(\begin{array}{c}
\vspace{0.2cm}0 \\ 1\end{array}\right)
\end{eqnarray}
The spinors ${\tt v}_{\mathbf{k},r}(t)$ and ${\tt u}_{\mathbf{k},r}(t)$ are not independent. The former are related to the latter by charge conjugation as ${\tt v}_{\mathbf{k},r}(t) = \mathcal{C}\,\bar{\tt u}^{\rm T}_{\bk,r}(t)$, where $\mathcal{C} = i\gamma^0\gamma^2$~\cite{PeskinSchroeder}. It follows from here that the mode functions satisfy the relation 
\begin{equation}\label{eq:u_v_relation}
v_{\mathbf{k},\pm}(t)= \pm\,u_{\mathbf{k},\mp}^{*}(t).
\end{equation}
Thus, from now on we can work with only one of the mode functions. We choose $u_{\mathbf{k},\pm}(t)$.

By introducing the decomposition of eq.~(\ref{eq:Phi(u,v)}) into the Dirac equation eq.~(\ref{eq:DiracPhi}), and differentiating the latter with respect conformal time, it follows that the $eom$ for the complex mode functions $u_{\mathbf{k},+}(t)$ and $u_{\mathbf{k},-}(t)$ decouple. The $eom$ correspond to that of an oscillator with a complex time-dependent frequency,
\begin{eqnarray}\label{eq:DiracSimple}
{d^2\over dt^2}{u}_{\mathbf{k},\pm}(t)+\left(\omega_{\mathbf{k}}^{2}(t) \pm i{d\over dt}(am_\psi)\right) u_{\mathbf{k},\pm}(t)=0\,,\qquad
\omega_{\mathbf{k}}^{2}(t) =  k^{2}+a^{2}(t)m_{\psi}^{2}(t)
\end{eqnarray}

In order to obtain the time evolution of $u_{\bk,\pm}(t)$ we need to specify the initial conditions at time $t = t_I$. These should correspond to vanishing initial fermion number density or, equivalently, vanishing energy density. To find out the $\bk$-dependence of $u_{\bk,\pm}^{(I)} \equiv u_{\bk,\pm}(t_I)$ and $\dot{u}_{\bk,\pm}^{(I)} \equiv {\dot u}_{\bk,\pm}(t_I)$ we need to introduce the fermion Hamiltonian density, and show how to diagonalize it by means of canonical Bogoliubov transformations. The Hamiltonian of a spin-$\frac{1}{2}$ fermion field in a FRW background is given by
\begin{equation}
H(t)=\frac{1}{2a}\int d\mathbf{x}\,\psi^{\dagger}(\mathbf{x},t)i\partial_{t}\psi(\mathbf{x},t)\label{eq:H(Phi)}
\end{equation}
By substituting eqs.~(\ref{eq:Phi(u,v)}) and~(\ref{eq: u_r=00003Du_+,-}) in eq.~(\ref{eq:H(Phi)}), we obtain a non-diagonal form for the Hamiltonian as
\begin{equation}
H(t)=\frac{1}{a}\int\frac{d\mathbf{k}}{\left(2\pi\right)^{3}}\left\{ \left[\hat a_{\mathbf{k},r}^{\dagger}\hat a_{\mathbf{k},r}-\hat b_{-\mathbf{k},r}\hat b_{-\mathbf{k},r}^{\dagger}\right]E_{\mathbf{k}}(t)+\hat b_{-\mathbf{k},r}\hat a_{\mathbf{k},r}F_{\mathbf{k}}(t)+\hat a_{\mathbf{k},r}^{\dagger}\hat b_{-\mathbf{k},r}^{\dagger}F_{\mathbf{k}}^{*}(t)\right\} ,\label{eq:H(E,F)}
\end{equation}
with (real) diagonal coefficient $E_{\mathbf{k}}(t)$
and (complex) non-diagonal coefficient $F_{\mathbf{k}}(t)$ given as
\begin{eqnarray}
\label{eq:E_def}
E_{\mathbf{k}}(t) &=& \frac{i}{2}\left[u_{\mathbf{k},+}^{*}\dot{u}_{\mathbf{k},+}+u_{\mathbf{k},-}^{*}\dot{u}_{\mathbf{k},-}\right] = 2k{\rm Re}\{u_{\bk,+}^*u_{\bk,-}\} + am_\psi\left(1-2|u_{\bk,+}|^2\right)\,,
\\
\label{eq:F_def}
F_{\mathbf{k}}(t) &=& \frac{i}{2}\left[u_{\mathbf{k},-}\dot{u}_{\mathbf{k},+}-u_{\mathbf{k},+}\dot{u}_{\mathbf{k},-}\right] = k\left(u_{\mathbf{k},+}^2-u_{\mathbf{k},-}^2\right) + 2am_\psi u_{\bk,+}u_{\bk,-}\,.
\end{eqnarray}
In the last equality of both expressions, we have used the relation between ${\dot u}_{\bk,\pm}(t)$ and ${u}_{\bk,\pm}(t)$, as derived from the 'Dirac equation' $i\gamma^0\dot{\tt u}_{\bk,r} = [\gamma_jk^j + a(t)m_{\psi}(t)]{\tt u}_{\bk,r}$. From the above expressions we find that the following relation is satisfied 
\begin{equation}
E_{\mathbf{k}}^{2}+\left|F_{\mathbf{k}}\right|^{2} = \omega_{\mathbf{k}}^{2}
\end{equation}

The Hamiltonian can be brought to a diagonal form by an appropriate canonical Bogoliubov transformation of the creation and annihilation operators,
\begin{eqnarray}\label{eq:BogoliubovTrans}
\hat{\tilde a}_{\mathbf{k},r}(t) &=&\alpha_{\mathbf{k}}(t)\hat a_{\mathbf{k},r}+\beta_{\mathbf{k}}(t)\hat b_{-\mathbf{k},r}^{\dagger}\\
\label{eq:BogoliubovTrans2}
\hat{\tilde b}_{\mathbf{k},r}^{\dagger}(t) &=& \alpha_{\mathbf{k}}^{*}(t) \hat b_{-\mathbf{k},r}^{\dagger}-\beta_{\mathbf{k}}^{*}(t)\hat a_{\mathbf{k},r},
\end{eqnarray}
where $\alpha_{\mathbf{k}}(t)$ and $\beta_{\mathbf{k}}(t)$ are complex coefficients. The idea is that by choosing judiciously the latter coefficients, the Hamiltonian will read, in terms of the new creation/annihilation operators, as
\begin{equation}
H(t)=\frac{1}{a}\int\frac{d\mathbf{k}}{\left(2\pi\right)^{3}}\left[\hat{\tilde a}_{\mathbf{k},r}^{\dagger}(t)\hat{\tilde a}_{\mathbf{k},r}(t)+\hat{\tilde b}_{\mathbf{k},r}^{\dagger}(t)\hat{\tilde b}_{\mathbf{k},r}(t)\right]\omega_{\mathbf{k}}(t),\label{eq:H(Omega)}
\end{equation}
A new (''quasi-particle'') vacuum $|0_t\rangle$ is defined by
\begin{equation}
\label{eq:tVac}
\hat{\tilde a}_\bk(t)|0_t\rangle = \hat{\tilde b}_\bk(t)|0_t\rangle = 0
\end{equation} 
By demanding that $\hat{\tilde a}_{\mathbf{k}}(t)$ and $\hat{\tilde b}_{\mathbf{k}}(t)$ also satisfy the canonical anticommutation relations~(\ref{eq:anticom}), we first note that $\alpha_{\mathbf{k}}(t)$ and $\beta_{\mathbf{k}}(t)$ must verify the relation 
\begin{equation}\label{eq:BogoliubovNorm}
\left|\alpha_{\mathbf{k}}(t)\right|^{2}+\left|\beta_{\mathbf{k}}(t)\right|^{2}=1.
\end{equation}
By comparing eq.~(\ref{eq:H(E,F)}) and eq.~(\ref{eq:H(Omega)}) we find that the conditions that we must impose on the Bogoliubov coefficients for the Hamiltonian to be diagonal as in eq.~(\ref{eq:H(Omega)}), are
\begin{eqnarray}\left|\beta_{\mathbf{k}}(t)\right|^{2}&=&\frac{\omega_{\mathbf{k}}(t)-E_{\mathbf{k}}(t)}{2\omega_{\mathbf{k}}(t)}\label{eq:a/b,b^2}\\
\label{eq:a/b}
\frac{\alpha_{\mathbf{k}}(t)}{\beta_{\mathbf{k}}(t)} &=& \frac{E_{\mathbf{k}}(t)+\omega_{\mathbf{k}}(t)}{F_{\mathbf{k}}^{*}(t)}
\end{eqnarray}
Hence there are four $dof$ in the complex Bogoliubov coefficients and four constraints; two real equations, eqs.~(\ref{eq:BogoliubovNorm}), (\ref{eq:a/b,b^2}), and one complex equation, eq.~(\ref{eq:a/b}). These are, however, not independent constraints, and in reality only three $dof$ are independently constrained from eqs.~(\ref{eq:BogoliubovNorm}),~ (\ref{eq:a/b,b^2}) and~(\ref{eq:a/b}). Thus, there is a certain degree of arbitrariness in the Bogoliubov coefficients, which allow us to choose $\alpha_{\mathbf{k}}(t)$ to be real-valued\footnote{This corresponds to a choice of the phase of $\beta_\bk = |\beta_\bk|e^{i\varphi_\beta}$ opposite to the phase of $F_\bk = |F_\bk|e^{i\varphi_F}$, $\varphi_\beta = -\varphi_F$}.

The number density of produced particles up to time $t$ (equal to the number of produced antiparticles) is given by the vacuum expectation value of the particle number operator $\tilde n_\bk \equiv \hat{\tilde a}^\dag_{\bk}\hat{\tilde a}_{\bk}$,
\begin{equation}\label{eq:OccuNum}
n(t) = {1\over 2\pi^2 a^3(t)}\int dk\,k^2\,\langle 0| \tilde n_\bk|0 \rangle = {1\over 2\pi^2 a^3(t)}\int dk\,k^2\,|\beta_\bk(t)|^2
\end{equation} 
Thus $\left|\beta_{\mathbf{k}}(t)\right|^{2}$ represents the occupation number of fermions with momentum $\mathbf{k}$ at time $t$. In agreement with the Pauli exclusion principle the occupation number cannot exceed unity $\left|\beta_{\mathbf{k}}(t)\right|^{2}\leq1$. The
initial conditions then, corresponding to vanishing initial occupation number, $\big|\beta_\bk^{(I)}\big|^2 = 0$ (and thus $\alpha_{\mathbf{k}}^{(I)} = 1$), are determined by $E_{\mathbf{k}}^{(I)} \equiv \omega_{\mathbf{k}}^{(I)}$
and $F_{\mathbf{k}}^{(I)}=0$. From the definitions of $E_\bk$ and $F_\bk$ by eqs.~(\ref{eq:E_def}), (\ref{eq:F_def}), we finally find that the initial amplitudes are
\begin{eqnarray}
u_{\mathbf{k},\pm}^{(I)} &=&\sqrt{1\pm\frac{a_Im_{\psi}^{(I)}}{\omega_{\mathbf{k}}^{(I)}}},\label{eq:u(0)}
\\
\dot{u}_{\mathbf{k},\pm}^{(I)} &=& -iku_{\mathbf{k},\mp}^{(I)} \mp ia_Im_{\psi}^{(I)}u_{\mathbf{k},\pm}^{(I)}\label{eq:dotu(0)}\,,
\end{eqnarray}
where $\dot{u}_{\mathbf{k},\pm}^{(I)}$ was just obtained as a function of $u_{\mathbf{k},\pm}^{(I)}$ through the Dirac equation.

We have now at hand all ingredients needed to study the dynamics of fermions in a FRW background. Starting from the initial conditions given by eqs.~(\ref{eq:u(0)}), (\ref{eq:dotu(0)}), we can solve the Dirac eq.~(\ref{eq:DiracSimple}) to obtain the time evolution of $u_{\bk,\pm}(t)$. From there we can compute any expectation value of the fermion field. For instance the spectrum of the fermion occupation number given by eq.~(\ref{eq:a/b,b^2}), from which ultimately we can obtain the total number density of fermions at time $t$ from eq.~(\ref{eq:OccuNum}).

We can only solve the Dirac equation, of course, in the context of a specific model. In section~\ref{sec:Applications}, we will consider a fermionic field $\Psi$ coupled to a homogeneous scalar field $\varphi$ via a Yukawa coupling $\mathcal{L}_{\rm int}=h\varphi\bar{\Psi}\Psi$, with $h$ a dimensionless coupling. We will study scenarios in which the scalar field is either the inflaton, like in (p)reheating after inflation, or either a spectator field, like in the thermal era after the reheating of the Universe. In these models the scalar field oscillates around the minimum of its potential, generating an effective mass for the fermionic field as $m_{\Psi}(t)= h\varphi(t)$. The variations in time of $m_\Psi(t)$ will give rise to a non-trivial production of fermions which will source GW. See section~\ref{sec:Applications} for details.

\subsection{Spectrum of Gravitational Waves from fermions}
\label{subsec:GWspectrumFromFermions}

The spatial-spatial components of the energy-momentum tensor of a spin-$\frac{1}{2}$ fermion field are given by
\begin{equation}
T_{ij} = i\,\frac{a}{2}\left[\bar{\Psi}\gamma_{(i}\overrightarrow{D}_{j)}\Psi-\bar{\Psi}\overleftarrow{D}_{(i}\gamma_{j)}\Psi\right],\label{eq:T(Phi)}
\end{equation}
where $a$ is the scale factor, $D_*$ is the covariant derivative\footnote{$\gamma_{(i}\overrightarrow{D}_{j)} \equiv \gamma_{i}\overrightarrow{D}_{j} + \gamma_{j}\overrightarrow{D}_{i}$,  $\overleftarrow{D}_{(i}\gamma_{j)} \equiv \overleftarrow{D}_{i}\gamma_{j} + \overleftarrow{D}_{j}\gamma_{i}$.} given by eq.~(\ref{eq:CovDer}), and the overhead arrow points to the field with respect which the derivatives are taken.

As mentioned before, the source of GW is simply the TT-part of the fermions' energy-momentum tensor
\begin{equation}\label{eq:TTaniso}
\Pi_{ij}^{\rm TT}(\bk,t) = \Lambda_{ij,lm}(\hat k)T_{lm}(\bk,t)
\end{equation} 
Hence, by substituting the decomposition eqs.~(\ref{eq:Phi(u,v)})-(\ref{eq:barPhi(u,v)}) into eq.~(\ref{eq:T(Phi)}), we obtain the explicit form of the TT-part of the anisotropic stress in Fourier space, eq.~(\ref{eq:TTaniso}), like
\begin{equation}
\Pi_{ij}^{\rm TT}(\mathbf{k},t) = \int\hspace*{-1mm} \frac{d\bx}{a^{2}(t)}\frac{d\mathbf{p}\,d\bp'}{\left(2\pi\right)^{6}}\left(\hat b_{-\mathbf{p},s}{\bar{\tt v}}_{\mathbf{p},s} + \hat a_{\mathbf{p},s}^{\dagger}{\bar{\tt u}}_{\mathbf{p},s}\right)\Delta_{ij}\left(\hat a_{\mathbf{p}',r}{\tt u}_{\mathbf{\mathbf{p}'},r}+\hat b_{-\mathbf{p}',r}^{\dagger}{\tt v}_{\mathbf{p}',r}\right)e^{+i(\bk+\bp-\bp')\bx},\label{eq:T(u_r)}
\end{equation}
where $\Delta_{ij}$ is given by
\begin{equation}
\Delta_{ij} \equiv \Lambda_{ij,lm}(\hat{k})\left( {1\over2}p_{(l}\gamma_{m)} + {1\over 2}p'_{(l}\gamma_{m)} + \frac{i}{8}\left[\gamma_{(l}\gamma_{\alpha\beta}\omega_{m)}^{\alpha\beta}-\gamma_{\alpha\beta}\omega_{(l}^{\alpha\beta}\gamma_{m)}\right]\right).\label{eq:Nabla_lm}
\end{equation}
Since $\int\hspace*{-1mm}d\bx\,e^{i(\bk+\bp-\bp')\bx} = (2\pi)^3\delta^{(3)}(\bk+\bp-\bp')$, we can substitute $\bp'$ by $\bk+\bp$, and eliminate all integrations in eq.~(\ref{eq:T(u_r)}) but the one over $\bp$. By doing so and introducing the explicit form of the spin connection, eq.~(\ref{eq:spinconn.}), we find
\begin{equation}
\Delta_{ij}= \Lambda_{ij,lm}(\hat k)\left({1\over2}p_{(l}\gamma_{m)} + {1\over2}(k+p)_{(l}\gamma_{m)}-\frac{i}{2}\mathcal{H}\gamma_{0}\delta_{lm}\right) = \Lambda_{ij,lm}p_{(l}\gamma_{m)}\,,
\end{equation}
where in the last equality we used that $\Lambda_{ij,lm}(\hat{k})k_{l} = 0$ (transversality), and we eliminated the part from the spin connection $\propto \delta_{lm}$, thanks to $\Lambda_{ij,lm}(\hat{k})\delta_{lm}=0$ (tracelessness). With all this, eq.~(\ref{eq:T(u_r)}) finally reads
\begin{equation}
\Pi_{ij}^{\rm TT}(\mathbf{k},t) = \frac{\Lambda_{ij,lm}(\hat k)}{a^{2}(t)}\int\hspace*{-1mm}\frac{d\mathbf{p}}{\left(2\pi\right)^{3}}\left(\hat b_{-\mathbf{p},s}{\bar{\tt v}}_{\mathbf{p},s} + \hat a_{\mathbf{p},s}^{\dagger}{\bar{\tt u}}_{\mathbf{p},s}\right)p_{(l}\gamma_{m)}\left(\hat a_{\mathbf{k+p},r}{\tt u}_{\mathbf{\mathbf{k+p}},r}+\hat b_{-(\mathbf{k+p}),r}^{\dagger}{\tt v}_{\mathbf{k+p},r}\right),\label{eq:T(u_r)-1}
\end{equation} 
Analogously, we also obtain
\begin{equation}
{\Pi_{ij}^{\rm TT}}^{\hspace*{-0.2mm}*}\hspace*{-0.4mm}(\mathbf{k}',t') = \frac{\Lambda_{ij,lm}(\hat k')}{a^{2}(t')}\int\hspace*{-1mm}\frac{d\mathbf{q}}{\left(2\pi\right)^{3}}\left(\hat b_{-\mathbf{q},s}{\bar{\tt v}}_{\mathbf{q},s} + \hat a_{\mathbf{q},s}^{\dagger}{\bar{\tt u}}_{\mathbf{q},s}\right)q_{(l}\gamma_{m)}\left(\hat a_{\mathbf{q-k'},r}{\tt u}_{\mathbf{\mathbf{q-k'}},r}+\hat b_{(\mathbf{k'-q}),r}^{\dagger}{\tt v}_{\mathbf{q-k'},r}\right),\label{eq:T(u_r)-1*}
\end{equation} 
We need now to compute the UTC, i.e.~the $\Pi^2(k,t,t')$ function defined in eq.~(\ref{eq:UTC}), characterizing the expectation value $\langle 0|\Pi_{ij}^{\rm TT}(\mathbf{k},t){\Pi_{ij}^{\rm TT}}^{\hspace*{-0.2mm}*}(\mathbf{k'},t')|0\rangle$. From eqs.~(\ref{eq:T(u_r)-1}), (\ref{eq:T(u_r)-1*}) we infer that out of the 16 different quadrilinear combinations of creation/annihilation operators in $\Pi_{ij}^{\rm TT}(\mathbf{k},t){\Pi_{ij}^{\rm TT}}^{\hspace*{-0.2mm}*}(\mathbf{k'},t')$, only the following two have non-zero expectation value, 
\begin{eqnarray}
\langle 0|\hat b_{-\mathbf{p},s}\hat a_{\mathbf{k+p},r}\hat a_{\mathbf{q},s'}^{\dagger}b_{\mathbf{k'-q},r'}^{\dagger}|0\rangle = (2\pi)^{6}\delta^{(3)}(\mathbf{k+p-q})\delta^{(3)}(\mathbf{k-k'})\delta_{s,r'}\delta_{r,s'},\label{eq:baab}\\
\langle 0|b_{-\mathbf{p},s}b_{-(\mathbf{k+p}),r}^{\dagger}b_{-\mathbf{q},s'}b_{\mathbf{k'-q},r'}^{\dagger}|0\rangle  = (2\pi)^{6}\delta^{(3)}(\mathbf{k})\delta^{(3)}(\mathbf{k}-\bk')\delta_{s,r}\delta_{s',r'}.\label{eq:bbbb}
\end{eqnarray}
The second expectation value, eq.~(\ref{eq:bbbb}), corresponds to the zero-mode $\bk = \bk' = 0$ of the anisotropic-stress, and must vanish due to isotropy\footnote{Such term is proportional to $\delta^{3}(\mathbf{0})$ and thus represents a non-physical divergence, corresponding to the zero-point fluctuations, that must be removed.}. Therefore only the first term, eq.~(\ref{eq:baab}), will contribute to the UTC.

Now we are ready to obtain $\Pi^{2}(k,t,t')$ as a function of the 4-spinors ${\tt u}_{\mathbf{p},r}$ and ${\tt v}_{\mathbf{p},r}$. From the definition of the UTC by eq.~(\ref{eq:UTC}), and using eqs.~(\ref{eq:T(u_r)-1}), (\ref{eq:T(u_r)-1*}) and (\ref{eq:baab}), we finally obtain
\begin{equation}
\Pi^{2}(k,t,t')=\frac{1}{a^{2}(t)a^{2}(t')}\int\hspace*{-1mm}\frac{d\mathbf{p}}{\left(2\pi\right)^{3}}\,\left({\bar{\tt v}}_{\mathbf{p},s}(t)p_{(i}\gamma_{j)}{\tt u}_{\mathbf{k-p},r}(t)\Lambda_{ij,lm}(\hat{k}){\bar{\tt u}}_{\mathbf{k-p},r}(t')p_{(l}\gamma_{m)}{\tt v}_{\mathbf{p},s}(t')\right),\label{eq:Pi^2(u_r)}
\end{equation}
where we have used the property $\Lambda_{ij,pq}(\hat k)\Lambda_{pq,lm}(\hat k) = \Lambda_{ij,lm}(\hat k)$.
By representing the 4-spinors as in eq.~(\ref{eq: u_r=00003Du_+,-}) we can calculate the integrand in eq.~(\ref{eq:Pi^2(u_r)}), which corresponds to a trace over spinorial-indices. We obtain, see Appendix~\ref{app:Traceology} for details,
\begin{eqnarray}\label{eq:Trace}
\mathrm{Tr}\left\{{\bar{\tt v}}_{\mathbf{p},s}(t)p_{(i}\gamma_{j)}{\tt u}_{\mathbf{k-p},r}(t)\Lambda_{ij,lm}(\hat{k}){\bar{\tt u}}_{\mathbf{k-p},r}(t')p_{(l}\gamma_{m)}{\tt v}_{\mathbf{p},s}(t')\right\} \hspace*{4.2cm}\\
= 2p^2\sin^2\theta\,\left[u_{\mathbf{k-p},+}(t)u_{\mathbf{p},+}(t)u_{\mathbf{k-p},+}^*(t')u_{\mathbf{p},+}^*(t') + u_{\mathbf{k-p},-}(t)u_{\mathbf{p},-}(t)u_{\mathbf{k-p},-}^*(t')u_{\mathbf{p},-}^*(t') \right.\hV\nonumber\\
\left. -\,u_{\mathbf{k-p},+}(t)u_{\mathbf{p},+}(t)u_{\mathbf{k-p},-}^*(t')u_{\mathbf{p},-}^*(t') - u_{\mathbf{k-p},-}(t)u_{\mathbf{p},-}(t)u_{\mathbf{k-p},+}^*(t')u_{\mathbf{p},+}^*(t')\right],\nonumber
\end{eqnarray}
where $\theta$ is the angle between $\mathbf{k}$ and $\mathbf{p}$. 
Note that the r.h.s.~of eq.~(\ref{eq:Trace}) is expressed as a function only of the mode functions $u_{\mathbf{k},\pm}$ and not $v_{\mathbf{k},\pm}(t)$, since the latter are related to the former by eq.~(\ref{eq:u_v_relation}). 
We can then write eq.~(\ref{eq:Pi^2(u_r)}) like
\begin{equation}\label{eq:Pi2_spectra(u_+,-)}
\Pi^{2}(k,t,t')=\frac{1}{2\pi^{2}a^{2}(t)a^{2}(t')}\int dp\,d\theta\,p^{4}\sin^{3}\hspace*{-1mm}\theta\,W_{\bk,\bp}(t)W_{\bk,\bp}^*(t'),
\end{equation}
where we have defined
\begin{equation}\label{eq:Pi^2(u_+,-)}
W_{\bk,\bp}^{}(t) \equiv u_{\mathbf{k-p},+}(t)u_{\mathbf{p},+}(t) - u_{\mathbf{k-p},-}(t)u_{\mathbf{p},-}(t).
\end{equation}
From here, substituting eq.~(\ref{eq:Pi2_spectra(u_+,-)}) into eq.~(\ref{eq:GW_spectra(Pi)}), we find that the spectrum of GW produced by fermions is given by
\begin{eqnarray}
\frac{d\rho_{\GW}}{d\log k}(k,t) = \frac{Gk^{3}}{\pi^{3}a^{4}(t)}\int dp\,d\theta\, p^{4}\sin^{3}\hspace*{-1mm}\theta \,\left(\left|I_{(c)}(k,p,\theta,t)\right|^{2} + \,\left|I_{(s)}(k,p,\theta,t)\right|^{2}\right),
\label{eq:GW_spectra(u_+,-)}
\end{eqnarray} 
where
\begin{eqnarray}
I_{(c)}(k,p,\theta,t) \equiv \int_{t_{i}}^{t}\frac{dt'}{a(t')}\cos(kt')W_{\bk,\bp}(t'), \hV I_{(s)}(k,p,\theta,t) \equiv \int_{t_{i}}^{t}\frac{dt'}{a(t')}\sin(kt')W_{\bk,\bp}(t')\nn\\\label{eq:F_and_I_functions(u_+,-)}
\end{eqnarray}

Eqs.~(\ref{eq:Pi^2(u_+,-)})-(\ref{eq:F_and_I_functions(u_+,-)}) are the set of master formulas describing the energy-density spectrum of GW generated at sub-horizon scales by some fermionic field. They were first presented in~\cite{KariDaniTuukka}. For any process in the early Universe at which fermions develop an anisotropic stress, the spectrum of the GW generated by such fermions can be just found by plugging the corresponding mode functions $u_{\mathbf{k},\pm}(t)$ into eqs.~(\ref{eq:Pi^2(u_+,-)}) and (\ref{eq:F_and_I_functions(u_+,-)}), and then calculating eq.~(\ref{eq:GW_spectra(u_+,-)}). However, as we will show in what follows, such computation does not yet give the final answer. The UTC needs to be regularized in order to remove the (otherwise infinite) contribution from the zero-point fermion fluctuations. Thus, the set of master formulas eqs.~(\ref{eq:Pi^2(u_+,-)})-(\ref{eq:F_and_I_functions(u_+,-)}) need to be revisited, as shall be explained in the next section.
Note that the structure of the formula in eq.~(\ref{eq:GW_spectra(u_+,-)}) resembles that of scalar fields sourcing GW, see Appendix~\ref{app:GWbosons}.


\subsection{Regularized spectrum of Gravitational Waves from fermions}
\label{subsec:Regularization}

The calculation presented so far leads to an ultraviolet (UV) divergence in the momentum integral in eq.~(\ref{eq:GW_spectra(u_+,-)}). This is due to the contribution from the fermionic vacuum fluctuations. In order to obtain a physical amplitude of the GW background generated by fermions, we need to regularize this divergence. We need a regularization procedure which subtracts the contribution from the large momentum fermion vacuum fluctuations. We can do indeed a very similar procedure to what it is normally done in flat-space. That is, we can impose a {\tt normal~ordering} in the creation and annihilation operator products. The difference with respect the case in Minkowski is that now the background is time-dependent. Therefore we will need a {\tt time-dependent normal-ordering} (tNO) procedure. 

Our aim is to regularize the unequal-time correlator $\Pi^2(k,t,t')$, which is a quadrilinear combination of the Dirac field. However note that the vacuum expectation value (VEV) of the GW source, the TT-part of anisotropic stress tensor, $\Pi_{ij}^\TT$, also needs to be regularized. Therefore we will begin by discussing how to regularize an operator $\mathcal{O} \sim \bar\Psi \Psi$ bilinear in the Dirac field, like $\Pi_{ij}^{\rm TT}$. In the Heisenberg picture VEVs are considered with respect to the initial vacuum $|0\rangle$. If the expectation value $\langle0|\mathcal{O}|0\rangle$ of some operator $\mathcal{O}$ diverges, we can regularize it by simply subtracting the contribution from the zero-point fluctuations at time $t$. However the state $|0\rangle$ is not regarded anymore as the physical vacuum at time $t$; rather there is a new state $|0_t\rangle$ that has the properties of vacuum at that time. That is, the physical vacuum changes in time. Thus the vacuum-fluctuations that are to be removed \footnote{This is the reason why the tNO procedure is written as $\langle 0 |\mathcal{O}(t)|0\rangle -\langle 0_{t}|\mathcal{O}(t)|0_{t}\rangle$ and not as $\langle 0_{t}|\mathcal{O}(t)|0_{t}\rangle - \langle 0 |\mathcal{O}(t)|0\rangle$.} 
at every time $t$, should depend on the vacuum at that time, i.e.~$|0_t\rangle$. Therefore the tNO prescription amounts to compute expectation values as
\begin{equation}
\langle \mathcal{O}(t)\rangle_{\rm reg} \equiv \langle 0 |\mathcal{O}(t)|0\rangle -\langle 0_{t}|\mathcal{O}(t)|0_{t}\rangle
\label{eq:tNO}
\end{equation}
By its very definition, the operation of $tNO$ gives the VEV of $\mathcal{O}$ only for the existing particles at time $t$, since the VEV of the vacuum quantum fluctuations (as opposed to particles) at time $t$ is subtracted.


To perform this regularization we note that for an operator of the form $\mathcal{O} \sim \bar\Psi \Psi$, it is possible to define another operator $\tilde{\mathcal{O}}$ such that
\begin{equation}
\langle 0|\tilde{\mathcal{O}}(t)|0\rangle = \langle 0_{t}|\mathcal{O}(t)|0_{t}\rangle,
\end{equation}
see Appendix~\ref{app:AuxFermField} for details. The new operator is written as $\tilde{\mathcal{O}}(t) \sim \bar{\Phi}\Phi$, with\footnote{The ${1\over a^{3/2}}$ prefactors are simply taken by analogy with the conformal relation $\Psi = {1\over a^{3/2}}\psi$}
\begin{eqnarray}
\Phi(\bx,t) &=& {1\over a^{3/2}}\int \frac{d\bk}{\left(2\pi\right)^{3}}\,e^{-i\bk\bx}\left[\hat a_{\bk,r}\,\mathcal{U}_{\bk,r}(t) + \hat b_{-\bk,r}^\dag \mathcal{V}_{\bk,r}(t)\right]\\
\bar\Phi(\bx,t) &=& \Phi^\dag(\bx,t)\gamma_0 = {1\over a^{3/2}}\int \frac{d\bk}{\left(2\pi\right)^{3}}\, e^{+i\bk\bx}\left[\hat a_{\bk,r}^\dag\,\bar{\mathcal{U}}_{\bk,r}(t) + \hat b_{-\bk,r}\bar{\mathcal{V}}_{\bk,r}(t)\right],
\end{eqnarray} 
where the new four-spinors are given as
\begin{eqnarray}\label{eq:U}
\mathcal{U}_{\mathbf{k},r}(t) &=& \alpha_{\mathbf{k}}(t){\tt u}_{\mathbf{k},r}(t)+\beta_{\mathbf{k}}^{*}(t){\tt v}_{\mathbf{k},r}(t), \\
\label{eq:V}
\mathcal{V}_{\mathbf{k},r}(t) &=& \alpha_{\mathbf{k}}(t){\tt v}_{\mathbf{k},r}(t)-\beta_{\mathbf{k}}(t){\tt u}_{\mathbf{k},r}(t), 
\end{eqnarray} 
with $\alpha$ and $\beta$ the same Bogoliubov coefficients quoted before. Note that indeed eqs.~(\ref{eq:U}) and (\ref{eq:V}), relate the 4-spinors $\mathcal{U}_{\bk,r},\mathcal{V}_{\bk,r}$ with ${\tt u}_{\bk,r},{\tt v}_{\bk,r}$, similarly to how the Bogoliubov transformation eqs.~(\ref{eq:BogoliubovTrans}) and (\ref{eq:BogoliubovTrans2}), relate the operators $\hat {\tilde a}_{\bk,r}, \hat {\tilde b}_{\bk,r}$ with $\hat{a}_{\bk,r}, \hat{b}_{\bk,r}$.

The regularized VEV of $\mathcal{O}$ can then be written as
\begin{equation}
\langle \mathcal{O}(t)\rangle_{\rm reg} = \langle 0|{\mathcal{O}}(t)-\tilde{\mathcal{O}}(t)|0\rangle \equiv \langle 0|\mathcal{O}_{\rm reg}(t)|0\rangle ,\label{eq:Pi_reg.}
\end{equation}
with
\begin{equation}
\mathcal{O}_{\rm reg}(t) \sim \left(\bar\Psi\Psi-\bar\Phi\Phi\right)(t)
\end{equation}



We can apply this procedure to any operator bilinear in the Dirac field. Thus the regularized VEV of the TT-part of the anisotropic stress-tensor, i.e.~the GW source, could be obtained as
\begin{equation}
\left\langle\Pi_{ij}^{\rm TT}(\bk,t)\right\rangle_{\rm reg} = \big\langle 0\big|\,\Pi_{ij}^{\rm TT}({\tt u},{\tt v})
-{\tilde \Pi}_{ij}^{\rm TT}(\mathcal{U},\mathcal{V})\,\big|0\big\rangle \equiv \left\langle 0 \left|\,\Pi_{ij,\rm{reg}}^{\rm TT}\,\right|0 \right\rangle,\label{eq:Pi_reg.-1}
\end{equation}
with ${\Pi}_{ij,{\rm reg}}^{\TT}(\mathbf{k},t)$ a function of all spinors ${\tt v}, {\tt u}, {\mathcal{V}}, \mathcal{U}$. We will not provide however the explicit expression for ${\Pi}_{ij,{\rm reg}}^{\TT}(\mathbf{k},t)$, since in general grounds it is expected that $\big\langle\Pi_{ij}^{\rm TT}(\bk,t)\big\rangle = 0$, and thus the regularization of $\Pi_{ij}^\TT$ is unnecessary. In particular, due to statistical homogeneity and isotropy, $\langle 0|h_{ij}(\bx,t)|0\rangle = 0, \forall\,\bx, t$. Then, from the linearized GW eom, we are forced to conclude that $\langle 0 |\,\Pi_{ij}^{\rm TT}(\bx,t)\,|0 \rangle = 0,\forall\,\bx, t$, and correspondingly $\langle 0 |\,\Pi_{ij}^{\rm TT}(\bk,t)\,|0 \rangle = 0, \forall\,\bk, t$. Therefore, regularizing the expectation value of $\Pi_{ij}^{\TT}(\bk,t)$ as in eq.~(\ref{eq:Pi_reg.-1}) does not change this fact, and the regularized VEV still vanishes (see Appendix~\ref{subsec:2point_corr} for more details). We will then rather focus on the regularization of a 
bilinear operator with a non-vanishing VEV. The simplest one is the scalar quantity $\mathcal{O}_2(\bx,t) = \bar\Psi(\bx,t)\Psi(\bx,t)$, which has a Fourier transform as
\begin{eqnarray}\label{eq:FT_Os}
\mathcal{O}_{2}(\mathbf{k},t) &=& \int\hspace*{-1mm} \frac{d\mathbf{p}}{(2\pi)^{3}}\left[\,\hat b_{-\mathbf{p},s}{\bar{\tt v}}_{\mathbf{p},s}{\tt u}_{\mathbf{\mathbf{k+p}},r}\hat a_{\mathbf{k+p},r} + \hat b_{-\mathbf{p},s}{\bar{\tt v}}_{\mathbf{p},s}{\tt v}_{\mathbf{\mathbf{k+p}},r}\hat b_{-(\mathbf{k+p}),r}^\dag \right.\\ 
&& ~~~~~~~~~~~~~~+ \left.\hat a_{\mathbf{p},s}^\dag{\bar{\tt u}}_{\mathbf{p},s}{\tt u}_{\mathbf{\mathbf{k+p}},r}\hat a_{\mathbf{k+p},r} + \hat a_{\mathbf{p},s}^\dag{\bar{\tt u}}_{\mathbf{p},s}{\tt v}_{\mathbf{\mathbf{k+p}},r}\hat b_{-(\mathbf{k+p}),r}^\dag\,\right],\nn
\end{eqnarray}
The regularized VEV of this quantify is then
\begin{equation}
\left\langle \mathcal{O}_2(\bk,t)\right\rangle_{\rm reg} \equiv \big\langle 0\big|\,\mathcal{O}_2({\tt u},{\tt v})-\tilde{\mathcal{O}}_2(\mathcal{U},\mathcal{V})\,\big|0\big\rangle \equiv \big\langle 0\big|\,\mathcal{O}_{2,{\rm reg}}(\bk,t)\,\big|0 \big\rangle,\label{eq:RegO_s}
\end{equation}
where the explicit form of $\mathcal{O}_{2,{\rm reg}}(\mathbf{k},t)$ is found as (see Appendix~\ref{app:Traceology})
\begin{eqnarray}\label{eq:Os_reg}
\mathcal{O}_{2,{\rm reg}}(\mathbf{k},t) &=& \int\hspace*{-1mm} \frac{d\mathbf{p}}{(2\pi)^{3}}\left[\,\hat b_{-\mathbf{p},s}\left({\bar{\tt v}}_{\mathbf{p},s}{\tt u}_{\mathbf{\mathbf{k+p}},r}-\bar{\mathcal{V}}_{\mathbf{p},s}\mathcal{U}_{\mathbf{\mathbf{k+p}},r}\right)\hat a_{\mathbf{k+p},r}\right.\\ 
&& ~~~~~~~~~~~~~~+~ \hat b_{-\mathbf{p},s}\left({\bar{\tt v}}_{\mathbf{p},s}{\tt v}_{\mathbf{\mathbf{k+p}},r}-\bar{\mathcal{V}}_{\mathbf{p},s}\mathcal{V}_{\mathbf{\mathbf{k+p}},r}\right)\hat b_{-(\mathbf{k+p}),r}^\dag\nn\\
&& ~~~~~~~~~~~~~~~~+~ \hat a_{\mathbf{p},s}^\dag\left({\bar{\tt u}}_{\mathbf{p},s}{\tt u}_{\mathbf{\mathbf{k+p}},r}-\bar{\mathcal{U}}_{\mathbf{p},s}\mathcal{U}_{\mathbf{\mathbf{k+p}},r}\right)\hat a_{\mathbf{k+p},r} \nn\\ 
&& ~~~~~~~~~~~~~~~~~~~+\, \left.\hat a_{\mathbf{p},s}^\dag\left({\bar{\tt u}}_{\mathbf{p},s}{\tt v}_{\mathbf{\mathbf{k+p}},r}-\bar{\mathcal{U}}_{\mathbf{p},s}\mathcal{V}_{\mathbf{\mathbf{k+p}},r}\right)\hat b_{-(\mathbf{k+p}),r}^\dag\,\right]\nn
\end{eqnarray}
All the information we need to regularize the VEV of any bilinear operator in the Dirac field is contained in eq.~(\ref{eq:Os_reg}). The latter expression is indeed identical to that in eq.~(\ref{eq:FT_Os}), but with every bilinear product of spinors, $\bar{{\tt v}}{\tt u}, \bar{{\tt v}}{\tt v}, \bar{{\tt u}}{\tt u}$ and $\bar{{\tt u}}{\tt v}$, replaced respectively by its regularized version, $(\bar {{\tt v}}{\tt u}-\bar{\mathcal{V}}\mathcal{U})$, $(\bar {{\tt v}}{\tt v}-\bar{\mathcal{V}}\mathcal{V})$, $(\bar {{\tt u}}{\tt u}-\bar{\mathcal{U}}\mathcal{U})$ and $(\bar {{\tt u}}{\tt v}-\bar{\mathcal{U}}\mathcal{V})$. Such replacements constitute precisely our prescription for regularizing the VEV of any operator formed by an even product of fermionic fields. For instance, out of the four bilinear operators in eqs.~(\ref{eq:FT_Os}) and (\ref{eq:Os_reg}), $\hat b \hat a, \hat b \hat b^\dag, \hat a^\dag \hat b, \hat a^\dag \hat b^\dag$, only $\hat b \hat b^\dag$ has a non-zero VEV as $\langle0| b_{\mathbf{p},s} 
b_{\mathbf{k},r}^\dag |0\rangle = (2\pi)^3\delta_{rs}\delta_{\rm D}(\bp-\bk)$. Thus, we obtain the regularized VEV as
\begin{equation}\label{eq:Os_regII}
\big\langle 0 \big|\,\mathcal{O}_{2}(\bx,t)\,\big|0 \big\rangle_{\rm{reg}} \equiv \big\langle 0 \big|\,\mathcal{O}_{2,\rm{reg}}(\bx,t)\,\big|0 \big\rangle = \int\hspace*{-1mm} {d\bp\over(2\pi)^3} \,\left({\bar{\tt v}}_{\mathbf{p},s}{\tt v}_{\mathbf{\mathbf{p}},s}-\bar{\mathcal{V}}_{\mathbf{p},s}\mathcal{V}_{\mathbf{\mathbf{p}},s}\right),
\end{equation}
which is independent of $\bx$. The non-regularized VEV reads
\begin{equation}\label{eq:Os_NonReg}
\big\langle 0 \big|\,\mathcal{O}_2(\bx,t)\,\big|0 \big\rangle = \int\hspace*{-1mm} {d\bp\over(2\pi)^3}~{\bar{\tt v}}_{\mathbf{p},s}{\tt v}_{\mathbf{\mathbf{p}},s}\,,
\end{equation}
also independent of $\bx$, being simply the same as in eq.~(\ref{eq:Os_regII}), but with ${\bar{\tt v}}_{\mathbf{p},s}{\tt v}_{\mathbf{\mathbf{p}},s}$ in the integrand, instead of $\left({\bar{\tt v}}_{\mathbf{p},s}{\tt v}_{\mathbf{\mathbf{p}},s}-\bar{\mathcal{V}}_{\mathbf{p},s}\mathcal{V}_{\mathbf{\mathbf{p}},s}\right)$. More explicitly, computing the trace in eqs.~(\ref{eq:Os_NonReg}), (\ref{eq:Os_regII}), the details of which can be found in Appendix \ref{app:Traceology}, we find
\begin{eqnarray}\label{eq:Os_reg3_I}
\big\langle 0 \big|\,\mathcal{O}_{2}(\bx,t)\,\big|0 \big\rangle = 2\int\hspace*{-1mm} {d\bp\over(2\pi)^3} \,\left(|u_{\bp,+}|^2-|u_{\bp,-}|^2\right)\hV\,\\\label{eq:Os_reg3_II}
\big\langle 0 \big|\,\mathcal{O}_{2}(\bx,t)\,\big|0 \big\rangle_{\rm{reg}} = 4\int\hspace*{-1mm} {d\bp\over(2\pi)^3} \,|\beta_\bp|^2\left(|u_{\bp,+}|^2-|u_{\bp,-}|^2\right)
\end{eqnarray}
Defining the 'regularized' mode functions\footnote{Note that in our original letter~\cite{KariDaniTuukka}, we proposed a different 'dressing' for the regularized mode functions $\tilde u_{\bp,\pm}(t)$, based on a different approach than (though very similar to) the one carefully detailed in this section~\ref{subsec:Regularization}, and complemented by  Appendices~\ref{app:Traceology},~\ref{app:AuxFermField}.
} by
\begin{eqnarray}\label{eq:RegModeFunc}
\tilde{u}_{\bp,\pm} \equiv \sqrt{2}\,|\beta_\bp|\, u_{\bp,\pm}\,,
\end{eqnarray}
we can then write the regularized VEV like
\begin{equation}\label{eq:Os_reg4}
\big\langle 0 \big|\,\mathcal{O}_2(\bx,t)\,\big|0 \big\rangle_{\rm reg} = 2\int\hspace*{-1mm} {d\bp\over(2\pi)^3}\,\left(\big|\tilde{u}_{\bp,+}\big|^2-\big|\tilde{u}_{\bp,-}\big|^2\right)
\end{equation} 
Note that this expression is actually the same as that of the non-regularized VEV, but as a function of the regularized mode functions. In other words, the regularized VEV, eq.~(\ref{eq:Os_reg4}), can be obtained under the prescription that it is equal to the non-regularized VEV, eq.~(\ref{eq:Os_reg3_I}), but substituting $u_{\bp,\pm}$ by $\tilde{u}_{\bp,\pm}$. Therefore, we can write
\begin{eqnarray}
\big\langle 0 \big|\,\mathcal{O}_2[u_+,u_-]\,\big|0 \big\rangle_{\rm reg} \equiv \big\langle 0 \big|\,\mathcal{O}_2\big[u_\pm \rightarrow \tilde{u}_\pm\big]\,\big|0 \big\rangle
\end{eqnarray}
We can easily appreciate now how the regularization is working at removing the UV divergence: In any scenario under consideration there will always be a characteristic UV cut-off scale $k_*$, such that fermions with larger momenta $k > k_*$ are not excited. Thus the occupation number of the high momenta modes should vanish (often it dies away exponentially), $n_{k > k_*} \rightarrow 0$. In other words, for UV modes $k > k_*$, there are only fermion vacuum fluctuations, which are precisely responsible for the divergence. Since $n_k \equiv |\beta_k|^2$, then $\beta_{k > k_*} \rightarrow 0$, and hence the regularized mode functions die also away for $k > k_*$, as $\tilde{u}_{\bk,\pm} \propto |\beta_k| \rightarrow 0$. Therefore the integral in eq.~(\ref{eq:Os_reg3_II}) [or equivalently that in eq.~(\ref{eq:Os_reg4})] is naturally regularized at the scale $k = k_*$, which separates the excited fermionic modes from the vacuum fluctuations. The short wave-length vacuum fluctuations with $k > k_*$, responsible of 
the UV divergence in the original integration in eq.~(\ref{eq:Os_reg3_I}), are therefore removed (typically they are exponentially suppressed, depending on the scenario), and the divergence is then regularized.

We can generalize this prescription to regularize the VEV of any operator formed by an arbitrary product of pairs of fermionic fields, each evaluated at a given time. We can write this symbolically, in Fourier space, as
\begin{eqnarray}
\mathcal{O}_{2N}(t_1,t_2,...,t_N) \sim \prod_{n=1}^{N} \mathcal{O}_2(t_n) \sim \prod_{n=1}^{N} \bar\Psi[u_\pm(\bp_n,t_n)]\Psi[u_\pm(\bk_n,t_n)]\,,
\end{eqnarray}
where each $\mathcal{O}_2$ is a characteristic operator evaluated at a given time, bilinear in the fermion fields and possibly including matrices or momentum components (from the fourier transform of derivatives) multiplying any of the fields within the pair. 
The regularized VEV of $\mathcal{O}_{2N}$ is then obtained as
\begin{eqnarray}
\big\langle 0 \big|\,\mathcal{O}_{2N}(t_1,t_2,...,t_N)\,\big|0 \big\rangle_{\rm reg} \equiv \big\langle 0 \big|\,\prod_{n=1}^{N} \mathcal{O}_2\big[u_\pm(t_n) \rightarrow \tilde{u}_\pm(t_n)\big]\,\big|0 \big\rangle
\end{eqnarray}

Following this prescription, the regularization of the source of GW, the unequal time correlator
$\Pi^2(k,t,t')$, can now be achieved straight ahead. We just have to substitute the mode functions within each pair of spinors evaluated at the same time, by their regularized version. That is
\begin{eqnarray}\label{eq:Pi_reg_Definition}
\big{\langle} \Pi_{ij}^{\rm TT}(\mathbf{k},t)\Pi_{ij}^{\rm TT^*}\hspace*{-0.6mm}(\mathbf{k'},t')\big{\rangle}_{\rm reg} &\equiv& \big{\langle} 0\big{|}\,\Pi_{ij}^{\rm TT}[u_\pm \rightarrow \tilde{u}_\pm](\bk,t)\,\Pi_{ij}^{\rm TT^*}[u_\pm \rightarrow \tilde{u}_\pm](\bk',t')\,\big{|}0\big{\rangle}\nn\\
 &\equiv& (2\pi)^{3}\Pi_{\rm reg}^{2}(k,t,t')\delta^{3}(\mathbf{k}-\mathbf{k'}).
\end{eqnarray}
The regularized UTC, $\Pi_{\rm reg}^{2}(k,t,t')$, like the expression of the non-regularized UTC, eq.~(\ref{eq:Pi^2(u_r)}), will be a function of quadrilinear combinations of spinors. 
It is actually obtained by the same trace calculation as in eq.~(\ref{eq:Trace}), but substituting every $u_\pm$ by $\tilde{u}_\pm$. We then have
\begin{equation}\label{eq:Pi2_regularized}
\Pi^{2}_{\rm reg}(k,t,t') = \frac{1}{2\pi^{2}a^{2}(t)a^{2}(t')}\int dp\,d\theta\,p^{4}\sin^{3}\hspace*{-1mm}\theta\,\tilde W_{\bk,\bp}(t)\tilde W_{\bk,\bp}^*(t'),
\end{equation} 
with $\tilde W_{\bk,\bp}$ defined analogously to $W_{\bk,\bp}$ in eq.~(\ref{eq:Pi^2(u_+,-)}), but in terms of $\tilde{u}_{\pm}$'s as
\begin{eqnarray}\label{eq:W_reg}
\tilde W_{\bk,\bp}(t) &\equiv& \tilde{u}_{\mathbf{k-p},+}(t)\tilde{u}_{\mathbf{p},+}(t) - \tilde{u}_{\mathbf{k-p},-}(t)\tilde{u}_{\mathbf{p},-}(t) \\
&=& 2|\beta_{\mathbf{p}}(t)||\beta_{\mathbf{k}-\bp}(t)|{W}_{\mathbf{k},\mathbf{p}}(t) \nonumber
\end{eqnarray}
Substituting eq.~(\ref{eq:W_reg}) in eq.~(\ref{eq:GW_spectra(Pi)}), we find the regularized spectrum of GW produced by fermions like
\begin{eqnarray}
\frac{d\rho_{\GW}}{d\log k}(k,t) = \frac{Gk^{3}}{\pi^{3}a^{4}(t)}\int dp\,d\theta\, p^{4}\sin^{3}\hspace*{-1mm}\theta \,\left(\left|\tilde I_{(c)}(k,p,\theta,t)\right|^{2} + \,\left|\tilde I_{(s)}(k,p,\theta,t)\right|^{2}\right),
\label{eq:GW_spectra_Reg}
\end{eqnarray} 
where
\begin{eqnarray}
\tilde I_{(c)}(k,p,\theta,t) \equiv \int_{t_{i}}^{t}\frac{dt'}{a(t')}\cos(kt')\tilde W_{\bk,\bp}(t'), \hV \tilde I_{(s)}(k,p,\theta,t) \equiv \int_{t_{i}}^{t}\frac{dt'}{a(t')}\sin(kt')\tilde W_{\bk,\bp}(t')\nn\\\label{eq:I_reg}
\end{eqnarray}
Denoting again as $k_*$ the UV cut-off momentum scale above which only fermion vacuum fluctuations are present, we expect that the fermionic occupation number should then vanish for modes $k > k_*$, i.e.~$n_{p > k_*} \equiv |\beta_{p > k_*}|^2 = 0$. Consequently $\tilde{W}^{\pm}_{\mathbf{k},\mathbf{p}}(t) = 2|\beta_{\mathbf{p}}(t)||\beta_{\mathbf{k}-\bp}(t)|{W}_{\mathbf{k},\mathbf{p}}(t) \rightarrow 0$ for when $p \gg k_*$ and/or $k \gg k_*$. That is, the would be UV-divergence in eq.~(\ref{eq:Pi2_regularized}) is regularized by the suppression of the large-momentum tail of $\tilde{W}^{\pm}_{\mathbf{k},\mathbf{p}}(t)$, and the convergence of the GW spectrum at large $k$ is also guaranteed. Note that eqs.~(\ref{eq:GW_spectra_Reg}), (\ref{eq:I_reg}) are actually identical to the ones before we considered regularization, eqs.~(\ref{eq:GW_spectra(u_+,-)}), (\ref{eq:F_and_I_functions(u_+,-)}), but simply evaluated at the regularized mode functions $\tilde u_{\pm}$. Eqs.~(\ref{eq:GW_spectra_Reg}), (\ref{eq:I_reg}),
 together with eq.~(\ref{eq:W_reg}), constitute therefore the master set of equations that should be really used to determine the spectral amplitude of the GW emitted by an ensemble of fermions.

\section{Fermions as a source of Gravitational Waves - Applications}
\label{sec:Applications}

\indent In this section we will apply the previous formalism to early Universe scenarios where out-of-equilibrium fermions are created through parametric excitation interactions with a scalar field. Such fermions develop naturally an anisotropic stress which sources GW. This is the case of fermions produced during (p)reheating after inflation, but also during the post-reheating thermal era previous to BBN. All along section~\ref{sec:Applications} we present the numerical results for the spectrum of the GW generated by fermions excited from a Yukawa type interaction with a homogeneous oscillatory scalar field. We have characterized how the GW spectrum depends on the parameters involved in the various models considered. In particular, we have investigated two possible general scenarios: 1) fermion particle creation during the preheating period following the end of inflation, considering both the case of a massless or a massive inflaton, and 2) fermion production from some spectator field during the thermal era 
following reheating, before BBN, like e.g.~in curvaton-like scenarios. We will refer to the first case as fermionic preheating, and to the second as post-reheating thermal scenarios. We begin in section~\ref{subsec:YukawaAndScalarDynamics} with considerations common to all scenarios studied, discussing general aspects of the dynamics of homogeneous oscillating scalar fields, and their coupling to fermions through a Yukawa term. In section~\ref{subsec:ParametricAntsazs}, we complete the discussion on the common aspects of the scenarios considered, by providing a general parametric estimation of the amplitude and frequency of the expected background of GW. sections~\ref{subsec:FermionicMasslessPreheating} and~\ref{subsec:FermionicMassivePreheating} are dedicated respectively to the cases of fermionic preheating from a massless and a massive inflaton. Finally, section~\ref{subsec:ThermalEraScenarios} is devoted to the post-reheating fermion production scenarios in the thermal era. 

\subsection{Scalar field dynamics and the Yukawa interaction}
\label{subsec:YukawaAndScalarDynamics}

In all scenarios we have always considered a coupling of a scalar field $\varphi$ to some fermionic species $\Psi$, via a Yukawa interaction as
\begin{equation}\label{ref:YukawaCoupling}
\mathcal{L}_{\rm int} = h\varphi\bar{\Psi}\Psi,
\end{equation}
with $h$ a dimensionless Yukawa coupling. If the scalar field forms a homogeneous condensate, this interaction induces an effective time-dependent mass in the fermionic field like $m_{\Psi} = h\varphi(t)$. Depending on the dynamics of $\varphi$, the fermion modes can be excited, so it is important to characterize well the behavior of $\varphi$. All along sections~\ref{subsec:YukawaAndScalarDynamics}-\ref{subsec:ThermalEraScenarios} we will assume that the only mass that fermions have is the dynamical one from their interaction with $\varphi$. We leave for Appendix~\ref{app:FermionsMass} the consideration of fermions having an additional time-independent bare mass.

The $eom$ of a homogeneous scalar field in a FRW background is given by
\begin{equation}
\ddot{\varphi}+2\mathcal{H}\dot{\varphi}+a^{2}\frac{dV}{d\varphi}=0
\label{eq:eomPhi}
\end{equation}
where derivatives are taken with respect to conformal time $t$. In principle, since the scalar field is coupled to fermions, one should then include somehow the Yukawa interaction in eq.~(\ref{eq:eomPhi}). However, in most situations the excitation of fermions from the scalar field represents a marginal transfer of energy out of the scalar condensate. Hence one can ignore the effect of the fermions into the scalar field dynamics. Obviously this depends on the exact excitation of the fermions and on the strength of the Yukawa coupling $h$. For reasonable coupling amplitudes, the interaction with fermions is never an issue for the scalar field dynamics. It is therefore important to characterize the behavior of $\varphi$ just as dictated from its own potential $V(\varphi)$, ignoring the Yukawa coupling. The condensate $\varphi$ will simply act as a source for the fermions, while the 'backreaction' of the latter into the former will be negligible. 

We will consider two type of potentials, that of a massive field, $V(\varphi) = {1\over2}m_\varphi^2\varphi^2$, and that of a self-interacting field, $V(\varphi) = {\lambda\over4}\varphi^4$. In the scenarios that we have studied the condensate starts with some initial amplitude, $\varphi(t_I) = \Phi_I$, and then oscillates around the minimum of $V(\varphi)$, located at $\varphi = 0$ in the two potentials considered. Due to the expansion of the universe the amplitude of the field will decrease with time as the condensate oscillates. Soon after the field begins to roll towards the 'bottom' of the potential for the first time, the solution of eq.~(\ref{eq:eomPhi}) can be written in the form
\begin{equation}\label{eq:ApproxVarPhi}
\varphi(t) \approx \Phi(t)F(t),
\end{equation}
with $\Phi(t)$ a decreasing amplitude and $F(t+2\pi/\omega) = F(t)$
an oscillatory periodic function with angular frequency $\omega$ and amplitude equal to unity. This is a common aspect shared by the two potentials that we have considered\footnote{The periodicity in the case of a massive scalar field is actually exact only in cosmic time, $y(t) = \int_0^t dt' a(t')$, not in conformal one $t$.}, quadratic and quartic, though the details differ. We will discuss in each section~\ref{subsec:FermionicMasslessPreheating},~\ref{subsec:FermionicMassivePreheating} and~\ref{subsec:ThermalEraScenarios}, some of the details of the analytical solutions for $\varphi(t)$ corresponding to each different scenario. However, in practice, we will not rely in such analytical expressions (which are only approximate), and rather we will solve numerically the $eom$ of the scalar field, together with the Fermions' mode equation.

In the light of the coupling in eq.~(\ref{ref:YukawaCoupling}), the $eom$ for the fermionic mode functions $u_{\mathbf{\boldsymbol\kappa},\pm}(\tau)$ can be written in the form
\begin{equation}\label{eq:FermEOM}
\frac{d^{2}}{d\tau^{2}}u_{\mathbf{\boldsymbol\kappa},\pm}(\tau)+\left[\boldsymbol\kappa^{2}+q\, a^{2}(\tau)\tilde{\varphi}^{2}(\tau)\pm i\sqrt{q}\frac{d}{d\tau}(a(\tau)\tilde{\varphi}(\tau))\right]u_{\mathbf{\boldsymbol\kappa},\pm}(t)=0,
\end{equation}
where we have defined a dimensionless time $\tau$ and momentum $\boldsymbol\kappa$, as well as a dimensionless scalar field amplitude  (normalized to unity at $t_I = 0$), as
\begin{eqnarray}
\tau\equiv \omega t\,,~~~~~~~~~~~~~~~~\boldsymbol\kappa\equiv \mathbf{k}/\omega\,, ~~~~~~~~~~~~~~~\tilde{\varphi}(\tau)\equiv \varphi/\Phi_{I}
\label{eq:NaturalUnits}
\end{eqnarray}
We have also introduced a resonance parameter
\begin{equation}
\label{ResPar}
q \equiv {h^2\Phi_I^2\over\omega^2},
\end{equation}
which becomes $q = h^2/\lambda$ in the case of a quartic potential, and $q = h^2\Phi_I^2/m_\varphi^2$ in the case of a quadratic potential. 

For each scenario, we have solved numerically eq.~(\ref{eq:eomPhi}) together with eq.~(\ref{eq:FermEOM}). The initial conditions for the fermions correspond simply to an initial zero number density, eqs.~(\ref{eq:u(0)}), (\ref{eq:dotu(0)}), which we already discussed in section~\ref{subsec:FermionsFRWdynamics}. The initial conditions of the scalar field vary from scenario to scenario. In the case of preheating, the two potentials considered, quadratic and quartic, do not necessarily represent the inflationary potential (as it would be the case in chaotic inflation models), but rather the effective potential for the inflaton after inflation. In the particular circumstance in which the preheating potential represents as well the inflationary potential, then the initial conditions for the inflaton are simply determined by the breaking of the slow-roll regime. That is, considering the slow-roll parameters (each subscript corresponding to a derivative with respect $\varphi$)
\begin{equation}
\varepsilon \equiv \frac{1}{16 \pi G}\left(\frac{V_{\varphi}}{V}\right)^{2}, \qquad \eta \equiv \frac{1}{8 \pi G} \left|\frac{V_{\varphi \varphi}}{V}\right|,
\end{equation}
the initial amplitude $\Phi_I$ can be obtained from the moment when either of the slow-roll parameters becomes unity, $\varepsilon=1$ or $\eta=1$. The initial velocity is then read from the slow-roll eom $\dot\Phi_I \approx -V(\Phi_I)/2\mathcal{H}_I$. However, in general, as said, the shape of the potential during preheating is not necessarily linked to the potential during inflation (which could be, for instance, a plateau). Thus the initial amplitude and velocity for the inflaton in such case are free parameters, though bounded to be smaller than the ones from the breaking of the slow-roll condition in the chaotic inflationary regime. Finally, in the case of a scalar field oscillating during the thermal era after the completion of reheating, clearly its potential is not related to the inflationary one, since the scalar field is not identified with the inflaton. Therefore the initial conditions in this case can be freely chosen, with the only constraint that the energy stored in the scalar condensate should 
correspond to only a marginal fraction of the energy in the thermal bath, which dominates the energy budget of the universe. 

\subsection{Parametrization of the Gravitational Waves amplitude and frequency}
\label{subsec:ParametricAntsazs}

Let us note that since we will solve the fermion mode equation in the natural units defined in eq.~(\ref{eq:NaturalUnits}), it is also convenient to express the GW spectrum in such units. Thus, eqs.~(\ref{eq:GW_spectra_Reg}), (\ref{eq:I_reg}) can be rewritten as a function of dimensionless variables, such that the total fraction of GW energy density when production ends at $t = t_*$, reads
\begin{eqnarray}\label{eq:AmpProd}
\Omega_\GW^{(*)} \equiv {1\over \rho_*}\left({d\rho_\GW\over d\log k}\right)_* = \left(\omega\over M_p\right)^2\left(a_I\over a_*\right)^{1-3w}{\kappa^3 \mathcal{F}_*(\kappa)\over \pi^3 \tilde \rho_I} 
~~~~~~~~\\
\label{eq:AmpProd2}
\mathcal{F}_{*}(\kappa) \equiv \int d\tilde p\, d\theta\, {\tilde p}^{4}\sin^{3}\theta \,\left(\left|\tilde I_{(c)}(\kappa,\tilde p,\theta,\tau_*)\right|^{2} + \,\left|\tilde I_{(s)}(\kappa,\tilde p,\theta,\tau_*)\right|^{2}\right)\,,
\end{eqnarray}
where $\kappa \equiv k/\omega$, $\tilde p \equiv p/\omega$, $\tau_* \equiv \omega t_*$, $\tilde \rho_I \equiv \rho_I/\omega^4$, and the $\tilde I_{(x)}$ functions are just given by the same formulas as in eq.~(\ref{eq:I_reg}), but expressed in terms of the dimensionless variables. Note that we have also introduced the Planck mass $M_p \equiv 1/\sqrt{G} \simeq 1.22\cdot10^{19}$ GeV, and used the fact that the energy density $\tilde\rho_*$ at the end of GW production, can be expressed as a function of the initial energy density $\tilde \rho_I \equiv \rho_I/\omega^4$, as $\tilde \rho_* = \tilde \rho_I (a_I/a_*)^{3(1+w)}$, where $w$ is the equation of state parameter. In the case of a RD universe with $w = 1/3$, then $(a_I/a_*)^{1-3w} = 1$ in eq.~(\ref{eq:AmpProd}), since the GW energy density scales exactly as the background one, i.e.~both behave as relativistic $dof$.

According to eq.~(\ref{eq:ftoday}) and eq.~(\ref{eq:Amptoday}), today's frequency and amplitude of the GW, redshifted from the time of production, are given by
\begin{equation}
h^{2}\Omega_{\GW} = h^{2}\Omega_{\mathrm{rad}}\left(\frac{g_{o}}{g_{*}}\right)^{1\over3}\times\epsilon\,\Omega_\GW^{(*)},
\label{eq:AmpToday}
\end{equation}
\begin{equation}
f = \kappa\,\left(\frac{a_{I}}{a_{*}}\right)\left(\frac{\epsilon}{\tilde\rho_{*}}\right)^{1\over4}\times5\cdot10^{10}\mathrm{Hz},
\label{eq:fToday}
\end{equation}
where we recall that $\epsilon = ({a_*/a_{_{\rm RD}}})^{(1-3w)}$ as defined in eq.~(\ref{eq:epsilonParameter}).

In general, fermions excited from a homogeneous oscillating scalar field $\varphi$ are expected to fill up a "Fermi-sphere" with a comoving radius~\cite{Fermions2}
\begin{equation}
\label{eq:kF}
\begin{cases}
\kappa_{F}\sim q^{1/4} &, V(\varphi)\propto\varphi^{4}\vspace{0.3cm}\\
\kappa_{F}\sim(a/a_{I})^{1/4}q^{1/4} &, V(\varphi)\propto\varphi^{2}
\end{cases}
\end{equation}
where $q$ is the resonance parameter. Outside the Fermi-sphere, for $\kappa > \kappa_F$, the fermion occupation number vanishes and hence the GW production is strongly suppressed for those modes. Only inside the Fermi-sphere, i.e.~for $\kappa < \kappa_F$, fermions are excited. Let us now look carefully at the integrand of $\mathcal{F}_*$ in eq.~(\ref{eq:AmpProd2}). We see that there is an angular modulation $\sin^3\theta$, and more importantly, that its amplitude grows with the internal momenta as $\tilde p^4$. Besides, given the structure of the $\tilde I_{(x)}$ functions, these are not expected to depend significantly on $\tilde p$ for $\tilde p < \kappa_F$, though they should drop abruptly for $\tilde p > \kappa_F$. Thus we expect a growth of the integrand with $\tilde p$ until we hit the Fermi-radius scale $\kappa_F$. From there on, i.e.~for $\tilde p > \kappa_F$, the integrand amplitude will be suppressed (typically exponentially), since the modes are outside of the Fermi-sphere. Therefore, on general 
grounds, we can expect that there will be a peak at some scale $k = k_p$ in the spectrum of GW, located roughly at the Fermi-radius scale, i.e.~$k_p \sim \kappa_F\,\omega$. In other words, the position of the highest amplitude of the spectrum of GW from fermions, is expected to be around the maximum momentum excited in the fermions. This feature can actually be nicely seen in the plots in figure~\ref{fig:FermPhi4Spectra}, figure~\ref{fig:FermPhi2Spectra} and figure~\ref{fig:FermPhi2SpectraCurvaton}, which we present in sections~\ref{subsec:FermionicMasslessPreheating}, \ref{subsec:FermionicMassivePreheating} and \ref{subsec:ThermalEraScenarios}, respectively.

Let us try to find an analytic estimate for the $q$-dependence of the peak amplitude of the GW background produced by fermions. For this we need to figure out an analytic estimate for $\mathcal{F}_{*}(\kappa_{p})$, where $\kappa_p = k_p/\omega$. Ignoring the angular dependence in the integrand of $\mathcal{F}_{*}(\kappa_{p})$, which will contribute only to a $\mathcal{O}(1)$-modulation factor, we can actually write
\begin{equation}
\mathcal{F}_{*}(k_{p}) \sim \int d\tilde p\, {\tilde p}^{4} \,\left|\tilde I(\kappa,\tilde p,\tau_*)\right|^{2},
\end{equation}
where $\tilde{I}$ represents any of the $\tilde{I}_{(x)}$ functions, whose amplitude is expected to be of the same order, $|\tilde{I}_{(c)}| \sim |\tilde {I}_{(s)}|$. Only the modes within the Fermi-sphere are excited, so we can make the following $Antsaz$
\begin{equation}\label{eq:Antsaz}
\big|\tilde{I}\big| = A\,{\tilde p^{\,n}}\,\theta(\kappa_{p}-\tilde p),
\end{equation}
where $A$ is a dimensionless amplitude, $n$ is an effective spectral index characterizing the dependence of $\tilde I$ with $\tilde p$, and $\theta(x)$ is the step function. We expect that $\big|\tilde I\big|$ should decrease with growing $\tilde p$, or in other words that $n < 0$. This is because the higher the value of $\tilde p$, the faster the fermion mode functions in the integrand of $\big|\tilde I_{(x)}\big|$ oscillate, and hence the smaller the amplitude of $\big|\tilde{I}_{(x)}\big|$ should be due to a 'phase erasing' effect. Before writing the $\tilde I_{(x)}$ functions in terms of the natural dimensionless variables, we see that $\tilde I_{(x)}$ has dimensions of inverse energy. Therefore, the natural value for the spectral index, just based on dimensional considerations, is $n = -1$. Using eq.~(\ref{eq:kF}), we expect then
\begin{equation}\label{eq:Antsaz_F*}
\mathcal{F}_{*}(\kappa_{p}) \sim A^{2}\kappa_{F}^{3+2\delta} \,\simeq\, A^{2}q^{\frac{3}{4}+\frac{\delta}{2}}\times\begin{cases}
 ~~1 &, ~V(\varphi)\propto\varphi^{4}\vspace{0.3cm}\\
\left({a_{*}\over a_{I}}\right)^{\frac{3+2\delta}{4}} &, ~V(\varphi)\propto\varphi^{2}\,,
\end{cases}
\end{equation}
where we have introduced the parameter
\begin{equation}
\delta \equiv n+1,
\end{equation}
to account for possible deviations with respect our educated guess of $n = -1$. From here, we can infer that the final GW peak amplitude at $\tau = \tau_*$, $\Omega_\GW^{(p)} \equiv \Omega_\GW^{(*)}(\kappa_p)$, should scale as 
\begin{eqnarray}\label{eq:PeakAmplitude}
\Omega_{\GW}^{(p)} &=& \left(\frac{\omega}{M_{p}}\right)^{2}\left(\frac{a_{I}}{a_{*}}\right)^{1-3w}\frac{\kappa_{p}^{3}\mathcal{F}_{*}(\kappa_{p})}{\pi^{3}\tilde{\rho}_{I}}\nonumber\\
\nonumber\\
&\sim& \frac{A^{2}}{\pi^{3}}\frac{\omega^{6}}{\rho_IM_{p}^{2}}\,q^{\frac{3+\delta}{2}}\times
\begin{cases}
\left(\frac{a_{*}}{a_{I}}\right)^{3w-1} &, ~V(\varphi)\propto\varphi^{4}\vspace{0.3cm}\\
\left(\frac{a_{*}}{a_{I}}\right)^{3w+\frac{1+\delta}{2}} &, ~V(\varphi)\propto\varphi^{2}
\end{cases}
\end{eqnarray}
Let us note that in the case $V(\varphi)\propto\varphi^{4}$, we expect in most scenarios that the factor in eq.~(\ref{eq:PeakAmplitude}) is $\left({a_{*}}/{a_{I}}\right)^{3w-1} = 1$, since typically the scalar field $\phi$ is either oscillating in a relativistic thermal background or dictating itself the expansion of the Universe, in both of which cases $w = 1/3$. Only in the case that $V(\varphi)\propto\varphi^{4}$ does not dominate the energy budget and the latter is dominated by non-relativistic $dof$, i.e.~$w \neq 1/3$, then does $\left({a_{*}}/{a_{I}}\right)^{3w-1}$ may have any weight different than unity. However that corresponds to uncommon situations, unlikely to have happened in the early Universe. 

In sections~\ref{subsec:FermionicMasslessPreheating}-\ref{subsec:ThermalEraScenarios} we will quantify the goodness of this parametrization of the peak amplitude as a function of $q$, measuring $\delta$ and $A^2$ from actual numerical results varying $q$ within each scenario considered. Anticipating our results, for a massless scalar field with $q>1$, we find that the $({\delta/2})$-correction in eq.~(\ref{eq:PeakAmplitude}) amounts to a $\sim$ 7\% relative deviation with respect the 'expected' ${3\over2}$ power index of $q$ (which would correspond to $n = -1$ exactly), whereas the correction for a massive scalar field with $q>1$, goes up to $\sim 17\%$. 

The corresponding amplitude of the GW peak today is simply
\begin{equation}
\label{eq:AmppToday}
h^{2}\Omega_{\GW}\left(f_{p}\right)\simeq h^{2}\Omega_{\mathrm{rad}}\left(\frac{g_{o}}{g_{*}}\right)^{\frac{1}{3}}\times\epsilon\,\Omega_{\GW}^{(p)}\,,
\end{equation}
whereas the present frequency of the GW peak, using eqs.~(\ref{eq:fToday})-(\ref{eq:kF}), is given by
\begin{eqnarray}
\label{eq:fpToday}
f_{p} &\simeq& 5\cdot10^{10}\left(\frac{a_{I}}{a_{*}}\right)\left(\frac{\epsilon}{\tilde{\rho}_{*}}\right)^{\frac{1}{4}}\kappa_{F}~{\rm Hz}\nonumber\\
&\sim& 5\cdot10^{10}\left(\frac{\omega}{{\rho}_{I}^{1/4}}\right)\,\epsilon^{\frac{1}{4}}\,q^{\frac{1}{4}}~{\rm Hz}
\times
\begin{cases}
\left(\frac{a_{*}}{a_{I}}\right)^{\frac{3w-1}{4}} &, ~V(\varphi)\propto\varphi^{4}\vspace{0.3cm}\\
\left(\frac{a_{*}}{a_{I}}\right)^{\frac{3w}{4}} &, ~V(\varphi)\propto\varphi^{2}
\end{cases}
\end{eqnarray}
Again in the case $V(\varphi)\propto\varphi^{4}$, typically the expansion rate is determined by either $\phi$ or a relativistic thermal background, so that $w = 1/3$ and, correspondingly, the volume factor in eq.~(\ref{eq:fpToday}) is $\left({a_{*}}/{a_{I}}\right)^{(3w-1)/4} = 1$. In the case $V(\varphi)\propto\varphi^{2}$, if the expansion rate is either determined by $\phi$ or by a non-relativistic background, so that $w = 0$, the corresponding volume factor in eq.~(\ref{eq:fpToday}) is also $\left({a_{*}}/{a_{I}}\right)^{{3w/4}} = 1$.

Note that the above reasoning assumed implicitly that $q > 1$, which justifies the assumption that $\big|\tilde I\big|$ should decrease smoothly as $\tilde p$ grows (until we hit the sharp cut-off at $\tilde p = \kappa_F$). When $q < 1$ the structure of the excited momenta below the Fermi radius is complicate, and only specific set of modes $\tilde p_1 < \tilde p_2 < ... < \kappa_F$ are excited~\cite{Fermions2}. We thus expect a different behavior of $\tilde I$ with respect $\tilde p$, as compared to the previous case $q > 1$. In the case $q < 1$, while increasing $\tilde p$, more excited modes enter within the sphere of radius equal to such $\tilde p$. Thus, we may expect that $\big|\tilde I\big|$ grows with $\tilde p$ or, equivalently, that $n > 0$ in eq.~(\ref{eq:Antsaz}). In this case the $\delta$-parametrization of the deviations with respect $n=-1$ looses somehow its meaning, since in general $\delta$ will not be a small correction to $n = -1$. Therefore, in the case of $q < 1$, eq.~(\ref{eq:PeakAmplitude}) is still expected to be valid, but it is better to consider $\delta$ simply as a fitting parameter, and not as a small correction.

Now we have at hand everything needed to make an estimate of the spectral peak amplitude of the GW background produced by a fermionic source excited by a coherent scalar field through a Yukawa interaction. We will calibrate our formulas above confronting them with the actual numerical results from the several scenarios that we have studied in sections~\ref{subsec:FermionicMasslessPreheating} - \ref{subsec:ThermalEraScenarios}. 

\subsection{Fermionic preheating, part I: the case of a self-interacting inflaton}
\label{subsec:FermionicMasslessPreheating}

In this section we study the GW production from fermions produced by of a self-interacting scalar field oscillating with potential $V=\frac{1}{4}\lambda\varphi^{4}$. This is the situation during preheating after inflation, with the massless scalar field playing the role of the inflaton, and hence dominating the energy budget of the Universe. The expansion of the Universe behaves in this case, effectively as a radiation dominated (RD) FRW universe, since the trace of the inflaton energy-momentum is zero. The scale factor evolves correspondingly as $a(t) \approx a_I\left[1 + \mathcal{H}_I(t-t_I)\right]$, with $a_I$ and $\mathcal{H}_I$ the scale factor and the (comoving) Hubble rate at $t = t_I$. We take $t_I = 0$ at the end of inflation. The Hubble rate during preheating is then $\mathcal{H}(t) \approx {\mathcal{H}_I/\left[1 + \mathcal{H}_I t\right]} \sim {1/t}$. The solution to eq.~(\ref{eq:eomPhi}) is given by an elliptic cosine~\cite{Preheating1}
\be{eq:masslessInflatonSol}
\varphi(t) = {\Phi}(t)\,{\rm cn}(x(t), 1/2)\,,~~~~~~~~~~~ x(t) \equiv \sqrt{\lambda}\,\Phi_I\,t
\ee
with a decreasing amplitude $\Phi(t) \propto 1/x(t)$. Note that this expression indeed matches the form anticipated  in eq.~(\ref{eq:ApproxVarPhi}). The natural frequency of oscillations is then 
\begin{equation}
\omega = \sqrt{\lambda}\,\Phi_I\,,
\end{equation}
whereas the period in between inflaton zero crossings is given by $\omega T = \pi^{-1/2}\Gamma^2(1/4) \approx 7.416$~\cite{Preheating1}, different than the usual harmonic $2\pi \approx 6.283...$. 
Note that we could also write the behavior of $\varphi(t)$ in terms of the number of oscillations $N$ after inflation. In particular the amplitude scales as $\Phi \propto 1/N$. 

\begin{figure}
\centering\includegraphics[width=10.5cm]{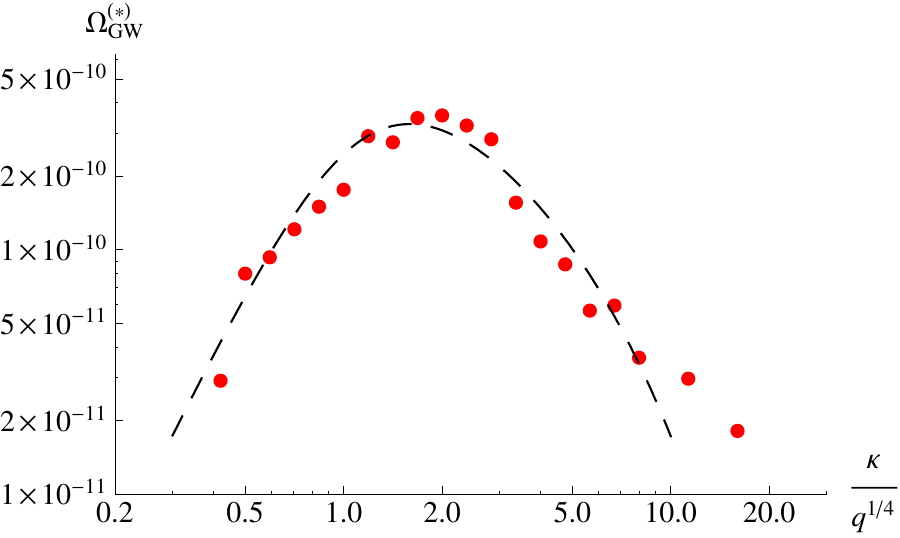}
\caption{Spectrum of GW just at the end of production, see eq.~(\ref{eq:AmpPhi4Prod}), obtained for parameters $q=10^6$, $h=0.1$ and $\Phi_{I}=\sqrt{{3}/{2 \pi}}M_{p}$, corresponding to an initial energy scale $E_I\approx 6.0\cdot 10^{16} \mathrm{GeV}$. The dashed line has been added by hand, just to help guiding the eye through the scattered points. A realistic spectral shape should be a smooth curve, but due to the rapid oscillations of some of the functions involved, the numerical calculations are not accurate enough. This affects particularly the modes in the UV tail, where the calculated amplitude decrease rather slow. The dashed line plotted lies within the estimated errors from our numerical integration (which we do not show just not to obscure the plot). The 'picture' that one should really take clear from this plot is that there is a well-defined peak in the spectrum, just as expected located at $\kappa \sim q^{1/4}$.}
\label{fig:FermPhi4Spectra}
\end{figure}

For a RD background we have an equation of state parameter $w=1/3$, and hence the parameter defined in eq.~(\ref{eq:epsilonParameter}) is $\epsilon = 1$. This implies that the energy density of the GW scales exactly as the energy of the inflaton, which drives the expansion of the Universe\footnote{This fact is actually advantageous, since there is no extra suppression of the GW energy density relative to the total one, as it will be the case when the expansion is matter dominated (MD), for when $\epsilon < 1$.}. The total background energy density of the Universe at the end of the production of GW is then given by
\begin{equation}
\rho_{*}=\rho_{I}\left(\frac{a_{I}}{a_{{\rm *}}}\right)^{4}=\frac{\lambda}{4}\Phi_{I}^{4}\left(\frac{a_{I}}{a_{{\rm *}}}\right)^{4},
\label{eq:TotEDensPhi4}
\end{equation}
where $\rho_{I}=\frac{\lambda}{4}\Phi_{I}^{4}$ is the initial energy density at the end of inflation. The resonance parameter in this scenario is given by
\begin{equation}
q=\frac{h^{2}}{\lambda}
\label{eq:qPhi4}
\end{equation}
Using eq.~(\ref{eq:AmpProd}) we obtain that the spectrum of GW right after production ends is given by
\begin{equation}
\Omega_\GW^{(*)}(\kappa) = {4\lambda^2\over\pi^3}\left(\frac{\Phi_{I}}{M_{p}}\right)^{2}\kappa^{3}\mathcal{F}_{*}(\kappa;q).
\label{eq:AmpPhi4Prod}
\end{equation}
We see from eq.~(\ref{eq:AmpPhi4Prod}) that, as expected, the initial amplitude $\Phi_I$, the self coupling of the inflaton $\lambda$, and the resonance parameter $q$ (or equivalently the Yukawa coupling $h$ for a fixed $\lambda$), determine completely the final shape of the GW spectrum. In principle, for this model, the amplitude of the CMB fluctuations fix the inflaton self-coupling to $\lambda \sim 10^{-13}$, whereas the radiative corrections require, in order not to spoil the flatness of the potential, that $h \lesssim 10^{-3}$~\cite{LindeBook}. However this is only true if the potential $V(\phi) = {\lambda\over4}\,\phi^4$ describes the inflationary period, corresponding to a chaotic inflation scenario. However, if that was the case, this model would be rule out in the light of the latest analysis of the CMB fluctuations by the Planck Collaboration~\cite{Planck}. It is clear then, that the chaotic inflation scenario in which $V \propto \phi^4$ describes both the inflation and reheating periods, is not an 
acceptable model anymore. We are therefore forced to conclude that, for our purposes, the inflationary potential should be described by some unspecified shape, and the quartic form $V \propto \phi^4$ becomes valid only at (p)reheating, after inflation. In general we will then consider $\lambda$ and $h$ as free parameters (within reasonable margins), whereas the initial amplitude $\Phi_I$ will not need to be determined from the breaking of the slow-roll condition. Yet the latter, $i.e.$ the moment when the slow-roll parameter $\varepsilon$ or $\eta$ would become unity in the chaotic scenario, determines an upper bound for the initial amplitude $\Phi_I$. In this scenario $\eta$ becomes unity first, and thus
\begin{equation}
\eta \geq 1~~~~~~\Rightarrow ~~~~~~ \Phi_{I} \leq \Phi_I^{\rm sr} = \sqrt{\frac{3}{2 \pi}}M_{p}
\end{equation}
In the case that $V \propto \phi^4$ is only valid after inflation, the initial field amplitude is simply restricted as $\Phi_I \leqslant \Phi_I^{\rm sr}$. Equivalently, the initial energy scale would be $E_I = (\lambda \Phi_I^4/4)^{1/4} < E_I^{\rm sr} \equiv \sqrt{3/(4\pi)}\lambda^{1/4}M_{p}$, with realistic situations corresponding to $E_I \ll E_I^{\rm sr}$.

In figure~\ref{fig:FermPhi4Spectra} we show an example of a GW spectrum, right after production ends, in an scenario $V(\phi) = {\lambda\over4}\phi^4$ with $\lambda=10^{-8}$ and $h=0.1$. The resonance parameter is then $q=10^{6}$. The initial energy scale in the example is $E_{I} = \lambda^{1/4}\Phi_I/\sqrt{2} \approx 6.0\cdot10^{16} \mathrm{GeV}$, corresponding to an initial amplitude $\Phi_I = \Phi_I^{\rm sr}$. This energy scale is too big to be acceptable, and was chosen just to show how big, in an extreme case, the fraction of energy in the GW could be with respect that in the background. To rescale the amplitude of this GW spectrum to more realistic initial smaller amplitudes $\Phi_I < \Phi_I^{\rm sr}$, one just need to multiply the spectral amplitude by an overall constant $({\Phi_I/\Phi_I^{\rm sr}})^2$. Of course the physical scale at which the spectrum peaks would also shift as $k_p \rightarrow (\Phi_I/\Phi_I^{\rm sr})k_p$. However, since in the plot we show the GW spectral amplitude as a function of 
the natural units for the peak position, $\sim q^{1/4}\sqrt{\lambda}\Phi_I$, such shift would not be apparent.

In figure~\ref{fig:FermPhi4Spectra} we observe that, indeed as expected, the position of the maximum amplitude ('peak' hereafter) of the GW spectrum is located at $k_{p} \sim \kappa_p\omega \sim q^{1/4}\omega$. 
The actual peak position, measured in the numerics as
\begin{equation}
k_{p}^{\rm (num)} \simeq 2\,q^{1/4}\sqrt{\lambda}\Phi_{I},
\end{equation}
matches very well 
the expectation.

\begin{figure}
\centering\includegraphics[width=10.5cm]{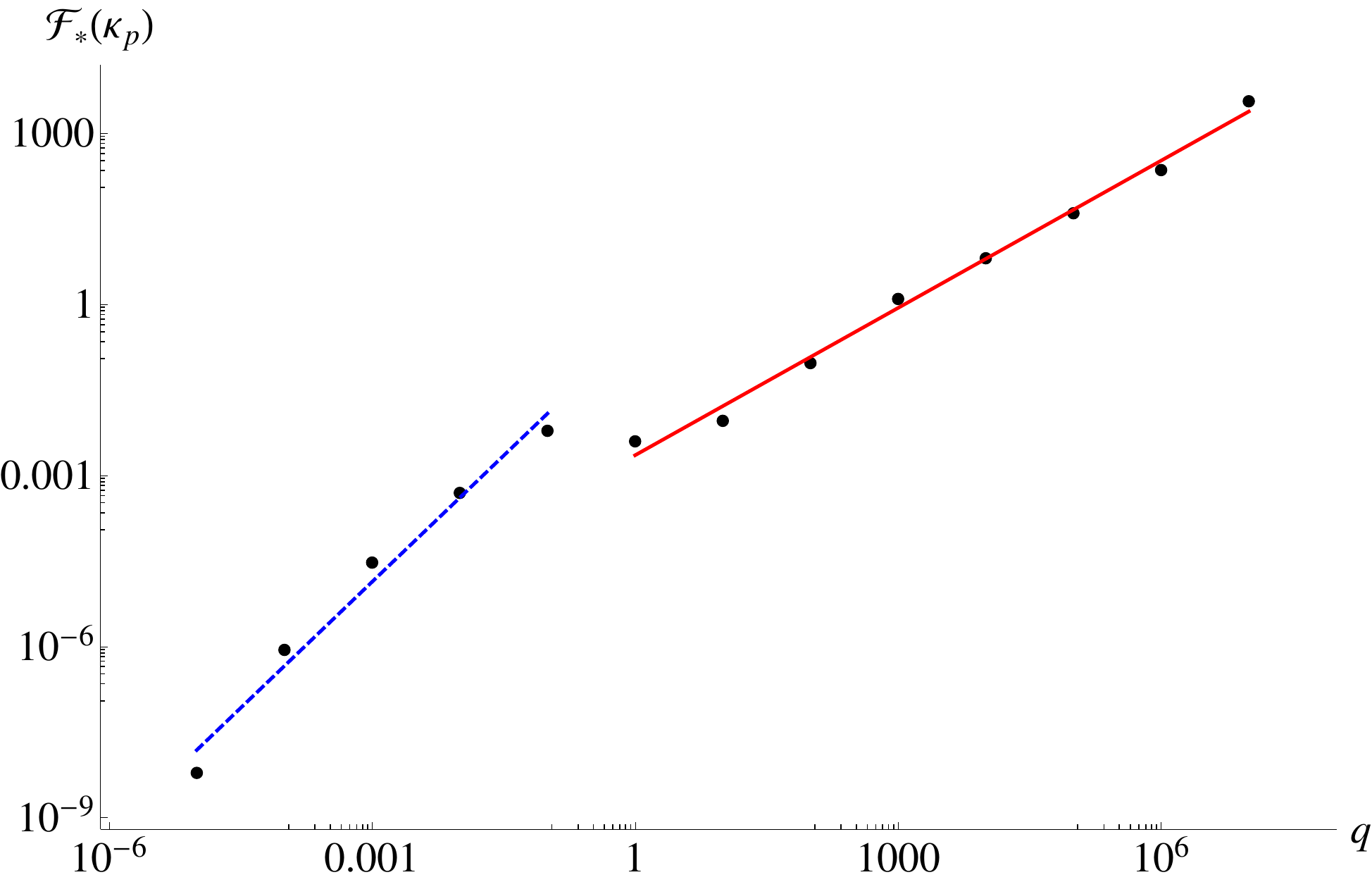}
\caption{$q$-dependence of  $\mathcal{F}_{*}(\kappa_p)$. The fits correspond to $\mathcal{F}_*(\kappa_p) \propto q^{0.86}$ for $q \geq 1$, and $\mathcal{F}_*(\kappa_p) \propto q^{1.48}$ for $q \ll 1$, see eq. (\ref{qdepmassless}).}
\label{fig:FermPhi4HighqDep}
\end{figure}

In figure~\ref{fig:FermPhi4HighqDep} we show the numerical dependence of $\mathcal{F}_*(\kappa_p)$ at the peak amplitude, with the resonance parameter $q = h^2/\lambda$. That is, we have obtained numerically the spectral GW amplitude at $k = k_p$, for a fixed $\Phi_I$, while varying $q$. Let us recall that given our Antsaz eq.~(\ref{eq:Antsaz}), which leads to eq.~(\ref{eq:Antsaz_F*}), we expect that $\mathcal{F}_*$ evaluated at $\kappa_p$, scales as
\begin{equation}
\mathcal{F}_*(\kappa_p,\, q) \simeq A^2 \cdot q^{\frac{3}{4}+\frac{\delta}{2}} 
\end{equation}
When confronting these expressions with our numerical results, see figure~\ref{fig:FermPhi4HighqDep}, we obtain the following results
\begin{eqnarray}
\label{qdepmassless}
\left\lbrace
\begin{array}{l}
q > 1: \mathcal{F}_*(\kappa_p) \approx 2.3\cdot10^{-3}\times q^{0.86} ~~~\Rightarrow~~~ A^2\approx2.3\cdot10^{-3}\,,\quad\delta \approx 0.22\,\vspace{0.3cm}\\
q < 1: \mathcal{F}_*(\kappa_p) \approx 0.36\times q^{1.48} ~~~~~~~~~~ \Rightarrow~~~~~~~ A^2\approx0.36\,,\quad\delta \approx 1.46\,
\end{array}
\right.
\end{eqnarray}

In the case of $q > 1$, we see that the correction ${\delta\over2}$ to the ${3\over2}$ exponent 
in $q$, is of the order of $\sim 7\%$. Note that $\delta$ was introduced originally as a correction to the spectral index of $\big|\tilde I\big|$ in eq.~(\ref{eq:Antsaz_F*}) for $q>1$. If we then approximate this correction to $\delta \simeq 0.25$, we see that the scaling for $q > 1$ in our original Antsaz eq.~(\ref{eq:Antsaz}), really goes as
\begin{equation}
\big|\tilde I\big| \propto \tilde p^{(-1+\delta)} \approx  {1\over{\tilde p}^{\,3/4}}\,,
\end{equation}
In the opposite regime, when $q < 1$, $\delta$ is not considered as a small correction. Approximating the numerical fit in this case to $\delta \approx 1.5$, we then find that our Antsaz scales as 
\begin{equation}
\big|\tilde I\big| \propto \tilde p^{(-1+\delta)}  \approx \tilde p^{1/2}\,,
\end{equation}

The GW production in the example in figure~\ref{fig:FermPhi4Spectra} lasts for several oscillations of the inflaton, with the scale factor growing as $a_*/a_I \simeq 100$.  The approximated amplitude of the spectral peak given by our parametrization in eq.~(\ref{eq:PeakAmplitude}), is however independent of $a_*/a_I$, and reads
\begin{eqnarray}
\Omega_{\GW}^{(p)}\simeq\frac{4A^{2}}{\pi^{3}}\lambda^{2}\left(\frac{\Phi_{I}}{M_{p}}\right)^{2}q^{\frac{3+\delta}{2}}\simeq\begin{cases}
3.0\cdot10^{-4}\lambda^{2}\left(\frac{\Phi_{I}}{M_{p}}\right)^{2}q^{1.61}, & q \geq 1\vspace*{0.3cm}\\
4.6\cdot10^{-2}\lambda^{2}\left(\frac{\Phi_{I}}{M_{p}}\right)^{2}q^{2.23}, & q \ll 1
\end{cases}
\end{eqnarray}

We can easily translate our results from the time of GW production to the present frequency and amplitude of the GW spectrum today. We just need to use eqs.~(\ref{eq:PeakAmplitude})-(\ref{eq:fpToday}). In particular, after having calibrated our parametrization schemes against the numerical results, today's amplitude and characteristic frequency of the peak 
$$h^{2}\Omega_{\GW}(f_p) \equiv h^{2}\Omega_{\mathrm{rad}}\left({g_{o}}\over{g_{*}}\right)^{1/3}\Omega_\GW^{(p)}\,, ~~~~~{\rm and} ~~~~~ f_p \simeq 5\cdot10^{10}{\kappa_p\over\tilde\rho_{I}^{1/4}}\,\mathrm{Hz},$$ 
can be written as
\begin{equation}
\label{eq:FinalAmpTodayPhi4}
h^{2}\Omega_{\GW}\left(f_{p}\right)\simeq\begin{cases}
1.2\cdot10^{-9}\lambda^{2}\left(\frac{\Phi_{I}}{M_{p}}\right)^{2}q^{1.61}, & q \geq 1\vspace*{0.3cm}\\
1.8\cdot10^{-7}\lambda^{2}\left(\frac{\Phi_{I}}{M_{p}}\right)^{2}q^{2.23}, & q \ll 1
\end{cases}
\end{equation}
\begin{equation}
\label{eq:FinalFreqTodayPhi4}
~~~~~~~~~~~~~~f_p \simeq 7\cdot10^{10}q^{1\over4} \lambda^{1\over4}\,\mathrm{Hz} = 7\cdot10^{10}\,h^{1\over2}\,\mathrm{Hz}\,,~~~~~~~
\end{equation}
As it was expected, the typical frequencies are very high. For instance, the spectrum in the example in figure~\ref{fig:FermPhi4Spectra} would be peaked today at $f_p \sim 10^{10}$ Hz, and it would have an amplitude of $h^2\Omega_\GW(f_p) \sim 10^{-15}$. Although the amplitude is not that small, the frequency is however simply too large as compared to the frequency range for which the future GW direct observatories such as BBO or DECIGO will be designed. Can we reduce the frequency? This is actually possible by having a very small Yukawa coupling. Let us actually take a very tiny value, say $h = 10^{-20}$. Then the prefactor in eq.~(\ref{eq:FinalFreqTodayPhi4}) is $q^{1/4}\lambda^{1/4} = h^{1/2} = 10^{-10}$, and thus the peak frequency $f_p$ would be in the $\sim 10$ Hz regime. Let us now take a big amplitude for $\Phi_I$, say $\sim M_p$ (even if this is too big to be acceptable), such that the prefactor $(\Phi_I/M_p)^2$ in eq.~(\ref{eq:FinalAmpTodayPhi4}) is $\mathcal{O}(1)$. We are then left with a 
remaining prefactor $\sim (10^{-9})10^{-7}\lambda^2q^{(3+\delta)/2} \lesssim (10^{-89})10^{-87}q^{(\delta-1)/2}$, which is ridiculously small, independently of the resonance parameter $q$. Actually $q^{(\delta-1)/2}$ is always smaller than unity, because for $q > 1$, $\delta < 1$, whereas for $q < 1$, $\delta > 1$. This contributes to make even smaller the overall amplitude, as $h^2\Omega_\GW(f_p) \ll (10^{-89})10^{-87}$.

In conclusion, the GW background from fermions excited from a Yukawa interaction with a coherent oscillating field -- the inflaton -- with potential\footnote{Let us note that in this case of a quartic potential, the homogeneous inflaton also decays through a self-resonance. This self-resonant process is however very inefficient, since it happens analogously as in a theory with an extra scalar field $\chi$ coupled to the inflaton as ${1\over2}g^2\phi^2\chi^2$, with fixed coupling ratio $g^2/\lambda = 3$. For such ratio there is no 'principal' resonant band at long wavelengths, so only short wavelengths are mildly excited (the 'Floquet' index is small) and the resonance is not very broad. Thus, there is always a long period of inflaton oscillations until the moment when self-resonance has finally any impact. During that period, GW from the fermionic decay are expected to be produced exactly as we have described in this section, unless the resonant parameter $q = h^2/\lambda$ Yukawa is ridiculously small. It might be interesting to study the regime at which both decay processes (fermionic and self-resonant) influence each other, but most likely this would only affect the cases where, from the simple analysis ignoring the self-resonance, the GW amplitude is already expected to be extremelly small, and therefore uninteresting.} $\propto \phi^4$, can have a non-negligible amplitude, but it is naturally peaked at high frequencies.

\subsection{Fermionic preheating, part II: the case of a massive Inflaton}
\label{subsec:FermionicMassivePreheating}

In this section we study the GW production from fermions excited by a massive homogeneous
scalar field oscillating in a potential $V=\frac{1}{2}m_\varphi^{2}\varphi^{2}$, in the context of preheating after inflation. The massive scalar field is again identified with the inflaton, so we consider that dominates the energy budget of the Universe. The approximated solution to eq.~(\ref{eq:eomPhi}) is~\cite{Preheating1}
\be{eq:massiveInflatonSol}
\varphi(t) \simeq  {\Phi_I\over m_\varphi y(t)} \,\sin(m_\varphi y(t))
\ee
which, expressed in terms of cosmic time $y(t) \equiv \int_0^t dt' a(t')$, matches the form in eq.~(\ref{eq:ApproxVarPhi}). The frequency of oscillations in this case is $\omega = m_\varphi$, and we see that the inflaton crosses zero every time $m_\varphi y(t) = \pi, 2\pi, 3\pi, ...$.

The energy density of the inflaton in a quadratic potential scales like in a non-relativistic fluid~\cite{Preheating1}. So the Universe is effectively MD for as long as the inflaton condensate dominates the energy budget. Therefore, the scale factor behaves (back to conformal time) approximately as $a(t) \approx a_I\left[1 + {1\over2} \mathcal{H}_I(t-t_I)\right]^{2}$, with $a_I$ and $\mathcal{H}_I$ the scale factor and the Hubble rate at $t = t_I$ (again we take $t_I = 0$ at the end of inflation). The Hubble rate during preheating is then $\mathcal{H}(t) \approx {\mathcal{H}_I/\left[1 + {1\over2}\mathcal{H}_I t\right]} \sim {2/t}$. In a MD background the equation of state parameter is $w=0$, and hence the parameter defined in
eq.~(\ref{eq:epsilonParameter}) is $\epsilon=({a_{*}}/{a_{\rm RD}}) < 1$. This implies that the radiation-like energy density of the GW will decrease faster than the matter-like background energy density.
 
The total energy density of the Universe right at the end of the GW production, at $t = t_*$, is given by
\begin{equation}
\rho_{*}=\rho_{I}\left(\frac{a_{I}}{a_{{\rm *}}}\right)^{3}=\frac{1}{2}m_\varphi^{2}\Phi_{I}^{2}\left(\frac{a_{I}}{a_{{\rm *}}}\right)^{3},
\label{eq:TotEDensPhi2}
\end{equation}
where $\rho_{I}=\frac{1}{2}m_\varphi^{2}\Phi_{I}^{2}$ is the initial energy density of the Universe at the start of preheating. The resonance parameter in this case is given by
\begin{equation}
q=\frac{h^{2}\Phi_{I}^{2}}{m_\varphi^{2}},
\label{eq:qPhi2}
\end{equation}
From eq.~(\ref{eq:AmpProd}) we obtain that the spectrum of GW at $t = t_*$ is given by
\begin{equation} 
\Omega_\GW^{(*)} = 
{2\over\pi^3}\left(\frac{m_{\varphi}^2}{\Phi_{I}M_p}\right)^{2}\left(a_I\over a_*\right)\kappa^{3}\mathcal{F}_{*}(\kappa;q),
\label{eq:AmpPhi2Prod} 
\end{equation}
We see from eq.~(\ref{eq:AmpPhi2Prod}) that the initial amplitude $\Phi_I$, the inflaton mass $m_\varphi$ and the resonance parameter $q$ (or equivalently the Yukawa coupling $h$ for fixed $\Phi_I$ and $m_\varphi$) determine completely the final amplitude of the GW spectrum. The potential $V(\phi) = {1\over2}m_\varphi^2\phi^2$ describes, in principle, only the (p)reheating stage, not the inflationary period. But if it did, as in a chaotic inflation scenario, the amplitude of the CMB fluctuations fixes the inflaton mass to $m_\varphi \sim 10^{13}$ GeV, whereas avoiding that radiative corrections spoil the flatness of the potential requires $h \lesssim 10^{-3}$~\cite{LindeBook}. In the case of chaotic inflation, therefore, $V \propto \phi^2$ is valid both at inflation and (p)reheating, and contrary to the case of chaotic inflation with $V \propto \phi^4$, this scenario is not ruled out by the Planck Results~\cite{Planck}. The chaotic inflation scenario with $V \propto \phi^2$ is still viable, though it is true 
that it is not preferred by the data, since it is $2\sigma$ "away" from the preferred value in the (tensor-to-scalar)-(spectral index) plane~\cite{Planck}. For our purposes, we can either consider the chaotic inflation case, or rather that the inflationary potential is described by some unspecified shape with the quadratic form $V \propto \phi^2$ only valid during (p)reheating. Therefore, in analogous situation to the quartic case, we can consider $m_\varphi$ and $h$ as free parameters within reasonable margins. The initial amplitude $\Phi_I$, on the other hand, is simply restricted to not being bigger than the amplitude inferred from the breaking of the slow-roll regime in the chaotic scenario. As before, we can determine such upper bound for the initial amplitude $\Phi_I$ when any of the slow-roll parameters becomes unity in the chaotic scenario. In this model both parameters are equal and thus become unity at the same time, so from $ \varepsilon = \eta \geq 1$ we find that
\begin{equation}
\Phi_{I} \leq \Phi_I^{\rm sr} ={M_{p}\over2\sqrt{\pi}}
\end{equation}
In general $\Phi_I \leqslant \Phi_I^{\rm sr}$ or, equivalently, the initial energy at the end of inflation should be restricted as $E_I = ({1\over2} m_\varphi^2 \Phi_I^2)^{1/4} < E_I^{\rm sr} \equiv (m_\varphi M_p)^{1/2}/(8\pi)^{1/4}$.

\begin{figure}
\centering\includegraphics[width=10.5cm]{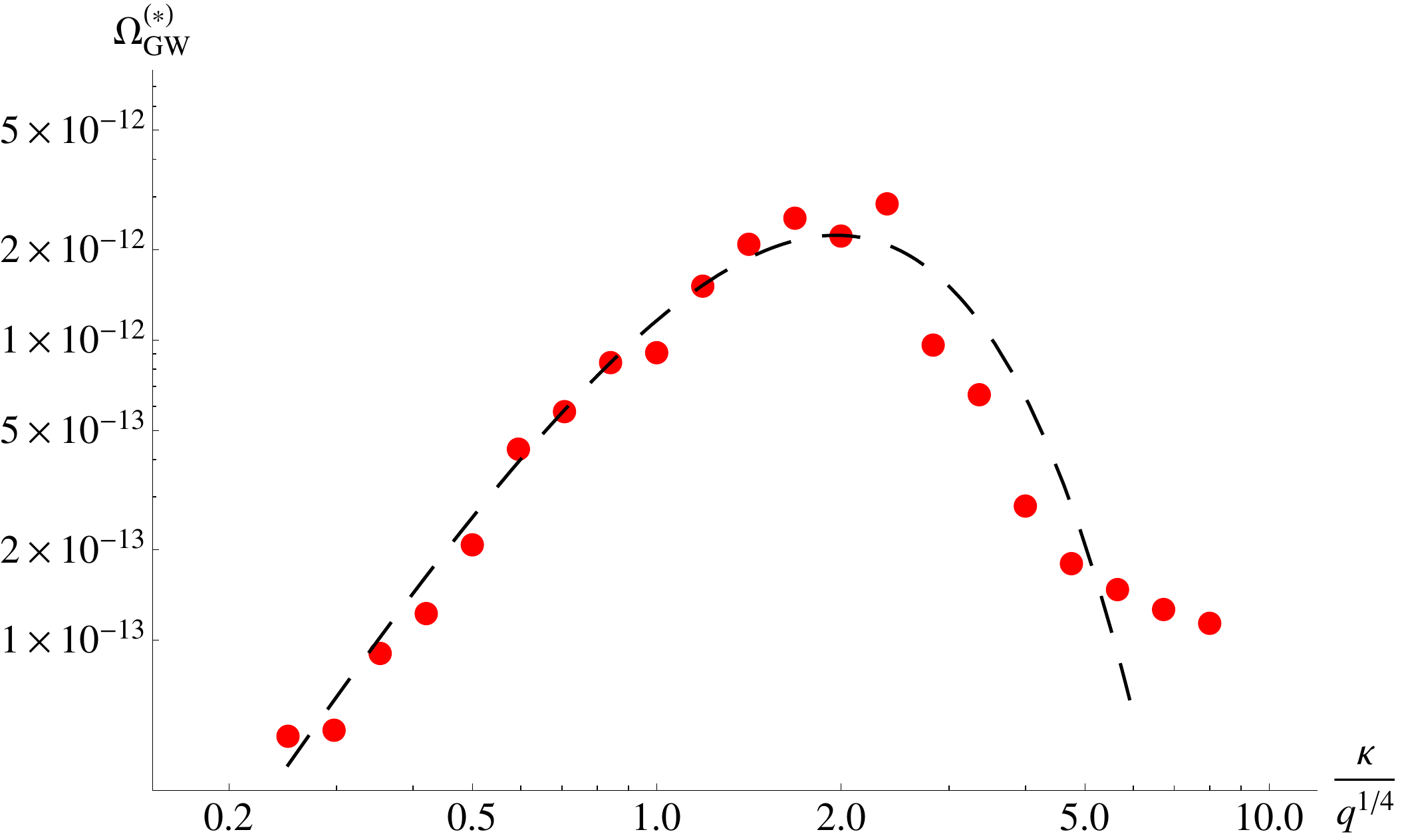}
\caption{The spectrum of GW right after production, see eq.~(\ref{eq:AmpPhi2Prod}), for parameters $q=10^{6}$, $h=0.1$ and $\Phi_{I}=M_{p}/(2\sqrt{\pi})$, and thus correspondingly for an inflaton mass $m_\varphi=3.4\cdot10^{14} \mathrm{GeV}$ and initial energy density $E_{I}=3\cdot10^{16} \mathrm{GeV}$. Again, the dashed line is simply a guide for the eye over the scattered points. Similar considerations to the ones discussed in the previous section (see the caption of figure~\ref{fig:FermPhi4Spectra}) apply here as well, in particular what concerns the amplitude of the spectrum at the most UV scales ($\kappa > 5q^{1/4}$).}
\label{fig:FermPhi2Spectra}
\end{figure}

In figure~\ref{fig:FermPhi2Spectra} we show an example of a GW spectrum obtained right at the end of production in an scenario
$V(\phi) = {1\over2}m_\varphi^2\phi^2$ with $m_\varphi = 3\cdot10^{13}$ GeV, inflaton initial amplitude $\Phi_I = \Phi_I^{\rm sr}$, and Yukawa coupling $h=0.1$. The resonance parameter and initial energy in this example are, correspondingly, $q=10^{6}$ and $E_{I} \simeq 3\cdot10^{15} \mathrm{GeV}$. To rescale the amplitude of this GW spectrum for initial smaller amplitudes $\Phi_I < \Phi_I^{\rm sr}$, we need to multiply the spectral amplitude by an overall constant $({\Phi_I^{\rm sr}/\Phi_I})^{2} < 1$, at the same time that we re-evaluate $\mathcal{F}_*$ with a new (smaller) resonance parameter $q \rightarrow ({\Phi_I/\Phi_I^{\rm sr}})^{2}q$. Note that indeed this is substantially different from the quartic potential case, since in the latter the resonance parameter is independent of $\Phi_I$, and hence there is no need for re-evaluating $\mathcal{F}_*$. The scale at which the spectrum peaks would also shift, as $k_p \rightarrow (\Phi_I/\Phi_I^{\rm sr})^{1/2}k_p$.

If we maintain the inflaton initial amplitude fixed but change its mass as $m_\varphi \rightarrow m_{\varphi}'$, we would then have to multiply the spectrum by a factor $(m_\varphi'/m_\varphi)^4$, at the same time that we re-evaluate $\mathcal{F}_*$ at a new resonance parameter $q \rightarrow (m_\varphi/m_\varphi')^2q$. The physical scale of the peak would shift as $k_p \rightarrow (m_\varphi'/m_\varphi)^{1/2}$, but again, due to the units used in the plot, such shift would not be apparent in figure~\ref{fig:FermPhi2Spectra}.

In figure~\ref{fig:FermPhi2Spectra} we can also see very clearly that, as expected (analogously to the case of a quartic potential), the position of the maximum amplitude of the GW spectrum is located at $k_{p} \sim \kappa_p\omega$. In the example chosen the GW production lasts for $\ln(a_*/a_I) \approx \ln 11 \approx 2.4$ e-foldings, so we expect $\kappa_p \simeq (a_*/a_I)^{1/4}q^{1/4} \simeq
1.8\, q^{1/4}$. We see therefore that the actual peak position obtained from the numerical outcome, \begin{equation} k_{p}^{\rm (num)}
\simeq 2\,q^{1/4}m_\varphi, \end{equation} matches very well the expectation.

\begin{figure}
\centering\includegraphics[width=10.5cm]{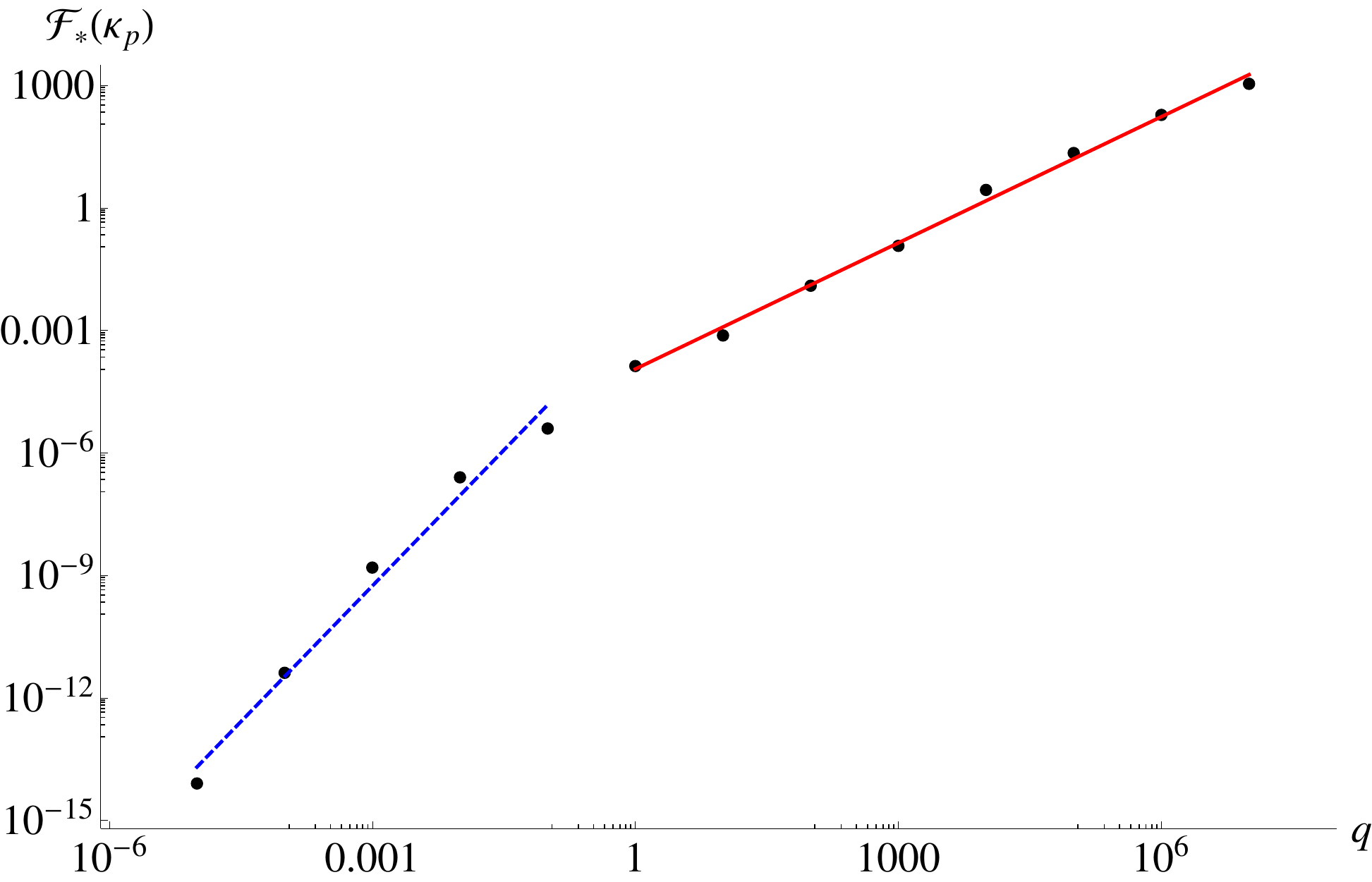}
\caption{$q$-dependence of  $\mathcal{F}_{*}(\kappa_p)$. The fits, see eq.(\ref{qdepmassive}), correspond to $\mathcal{F}_*(\kappa_p) \propto q^{1.03}$ for $q > 1$, and $\mathcal{F}_*(\kappa_p) \propto q^{2.22}$ for $q < 1$.}
\label{fig:FermPhi2HighqDep}
\end{figure}

In figure~\ref{fig:FermPhi2HighqDep} we show the numerical dependence of the peak amplitude $\mathcal{F}_*(\kappa_p;q)$ against the resonance parameter $q = (h\Phi_I/m_\varphi)^2$. From our Antsaz in eqs.~(\ref{eq:Antsaz}),~(\ref{eq:Antsaz_F*}) we expect at the peak scale
\begin{equation}
\mathcal{F}_*(\kappa_p,\, q) \simeq A^2 \cdot (a_{*}/a_I)^{\frac{3}{4}+\frac{\delta}{2}}  q^{\frac{3}{4}+\frac{\delta}{2}} 
\end{equation}
From our numerical outcome, see figure~\ref{fig:FermPhi2HighqDep}, we obtain that
\begin{eqnarray}
\label{qdepmassive}
\left\lbrace
\begin{array}{l}
q > 1: \mathcal{F}_*(\kappa_p) \approx 1.1\cdot10^{-4}\times q^{1.03} ~~~\Rightarrow~~~ A^2\approx9.6\cdot10^{-6}\,,\quad\delta \approx 0.56\,,\vspace{0.3cm}\\
q < 1: \mathcal{F}_*(\kappa_p) \approx 2.5\cdot 10^{-3}\times q^{2.22} ~~~ \Rightarrow~~~ A^2\approx1.2\cdot10^{-5}\,,\quad\delta \approx 2.94\,,
\end{array}
\right.
\end{eqnarray}

So for $q > 1$, we see that the ${\delta\over2}$-correction to the ${3\over2}$ slope (expected for $n = -1$) in eq.~(\ref{eq:PeakAmplitude}), is of the order $\sim 17\%$. If we approximate $\delta \simeq 0.5$, we see that the scaling for $q > 1$ in eq.~(\ref{eq:Antsaz}), goes as 
\begin{equation} \big|\tilde I\big| \propto \tilde p^{(-1+\delta)} \approx {1\over{\tilde p}^{\,1/2}}\,, \end{equation} 
In the opposite regime, when $q < 1$, approximating the numerical fit to $\delta \approx 3.0$, we find that our Antsaz scales as 
\begin{equation}
\big|\tilde I\big| \propto \tilde p^{(-1+\delta)}  \approx {\tilde p}^2\,, 
\end{equation}

The GW production in the example in figure~\ref{fig:FermPhi2Spectra} lasts only for few e-folds, with the scale factor growing as $a_*/a_I \simeq 11$. However the duration in general will depend on $q$, and thus we will maintain explicitly in the characterization of the GW peak, the dependence with $a_*/a_I$. Note that this is opposed to the case of preheating with $V \propto \phi^4$, discussed in section~\ref{subsec:FermionicMasslessPreheating}, where neither the amplitude nor the frequency of the peak depend on the duration $a_*/a_I$. From eq.~(\ref{eq:PeakAmplitude}), we see that the approximated amplitude of the spectral peak at the end of GW production is
\begin{eqnarray}
\Omega_{GW}^{(p)}\simeq\frac{2A^{2}}{\pi^{3}}\left(\frac{m_{\varphi}^{2}}{\Phi_{I}M_{p}}\right)^{2}\left(\frac{a_{*}}{a_{I}}\right)^{\frac{1+\delta}{2}}q^{\frac{3+\delta}{2}}\simeq\begin{cases}
6.2\cdot10^{-7}\left(\frac{m_{\varphi}^{2}}{\Phi_{I}M_{p}}\right)^{2}\left(\frac{a_{*}}{a_{I}}\right)^{0.78}q^{1.78}, & q>1\vspace{0.3cm}\\
7.7\cdot10^{-7}\left(\frac{m_{\varphi}^{2}}{\Phi_{I}M_{p}}\right)^{2}\left(\frac{a_{*}}{a_{I}}\right)^{1.97}q^{2.97}, & q<1
\end{cases}\nonumber\\
\end{eqnarray}
Let us now convert our results from the end of GW production to the frequency and amplitude of the GW today, for what we simply have to use eqs.~(\ref{eq:PeakAmplitude})-(\ref{eq:fpToday}). After the calibration above of our parametrization schemes against the numerical results, the present amplitude and characteristic frequency of the peak,
$$h^{2}\Omega_{\GW}(f_p) \equiv h^{2}\Omega_{\mathrm{rad}}\left({g_{o}}\over{g_{*}}\right)^{1/3}\Omega_\GW^{(p)}\,,
~~~~~~~~~~f_p \simeq 5\cdot10^{10}\left(a_I\over a_*\right)^{1\over4}{\kappa_p\over\tilde\rho_{I}^{1/4}}\,\mathrm{Hz},$$
can be parametrized as
\begin{equation}
\label{eq:FinalAmpTodayPhi2}
h^{2}\Omega_{GW}\left(f_{p}\right)\simeq\begin{cases}
2.5\cdot10^{-12}\left(\frac{m_{\varphi}^{2}}{\Phi_{I}M_{p}}\right)^{2}\left(\frac{a_{*}}{a_{I}}\right)^{0.78}q^{1.78}, & q>1\vspace{0.3cm}\\
3.1\cdot10^{-12}\left(\frac{m_{\varphi}^{2}}{\Phi_{I}M_{p}}\right)^{2}\left(\frac{a_{*}}{a_{I}}\right)^{1.97}q^{2.97}, & q<1
\end{cases}
\end{equation} 
\begin{equation}
\label{eq:FinalFreqTodayPhi2}
~~~~~f_p \simeq 6\cdot10^{10}\left(\frac{m_{\varphi}}{\Phi_{I}}\right)^{1\over2}q^{1\over4}\,\mathrm{Hz} = 6\cdot10^{10}\,h^{1\over2}\,\mathrm{Hz} \,
\end{equation}
As in preheating with a quartic potential, the typical frequencies are also very high. The spectrum in the example in figure~\ref{fig:FermPhi2Spectra}, for instance, would be peaked at $f_p \sim 10^{10}$ Hz, with an amplitude today of $h^2\Omega_\GW(f_p) \sim 10^{-17}$. This is not a big amplitude, but yet it is around the sensitivity threshold expected for detectors like BBO or DECIGO. The frequency, however, is again as in the quartic potential case, simply too large compared to the operational frequency range of such observatories. Of course, we can reduce the frequency by finding the appropriate parameter values. By having a very small Yukawa coupling, say we take again a very tiny value -- the same as in the quartic potential -- $h = 10^{-20}$. Then the peak frequency would be in the $\sim 10$ Hz regime. For $q > 1$, by using relation $\left(m_{\varphi}^{2}/(\Phi_{I}M_{p})\right)^{2}=(h^{2}/q)^{2}\left(\Phi_{I}/M_{p}\right)^{2}$ and by assuming $a_*/a_I \sim \mathcal{O}(10)$, we obtain that the amplitude 
of the GW can only be ridiculously small, $h^{2}\Omega_{GW}\left(f_{p}\right)\sim10^{-11}h^{4}\left(\Phi_{I}/M_{p}\right)^{2}q^{-0.22}\ll10^{-90}$. For the scenarios with $q < 1$ we obtain a similar result, $h^{2}\Omega_{GW}\left(f_{p}\right)\sim10^{-10}h^{4}\left(\Phi_{I}/M_{p}\right)^{2}q^{0.97}\ll10^{-90}$.

In conclusion, the GW background from fermions excited from a Yukawa interaction with a coherent oscillating massive inflaton with
potential $\propto \phi^2$, is naturally peaked at high frequencies if we want to maintain a non-negligible amplitude. Therefore, this is a common aspect shared by the preheating scenarios considered.

\subsection{Fermion production after reheating: the thermal era}
\label{subsec:ThermalEraScenarios}

Finally, in this section we will study the GW production from the fermionic decay of a massive homogeneous scalar field oscillating in a potential $V=\frac{1}{2}m_\varphi^{2}\varphi^{2}$, but this time not in the context of preheating after inflation. We want to analyze the case of a massive field during the thermal era after the completion of reheating. Given the nature of previous results, we will put particular attention to see if, contrary to the preheating scenarios, there might be a chance to produce a non-negligible GW amplitude at low frequencies in this case. 

In this situation the expansion of the Universe is dictated by the radiation background, and thus the scale factor goes as $a(t) = a_I\left[1 + \mathcal{H}_I(t-t_I)\right]$, where $a_I$ and $\mathcal{H}_I$ represent the scale factor and the Hubble rate at $t = t_I$. The initial time $t_I$ should be understood as the initial moment of GW generation from fermions, during the thermal era. If the Universe reheated at time $t = t_{\rm rh}$, then the only constraint is that $t_I \geq t_{\rm rh}$. The Hubble rate during the thermal era is then $\mathcal{H}(t) = {\mathcal{H}_I/\left[1 + \mathcal{H}_I t\right]} \sim {1/t}$. It is clear that the massive spectator scalar field should not be identified with the inflaton. Besides, its energy density should be subdominant within the energy budget of the Universe.

At the beginning the scalar field is light ($a^2m^2_\varphi \ll \mathcal{H}^2$), and it slowly rolls down its potential. Later the scalar field becomes heavy ($a^2m^2_\varphi > \mathcal{H}^2$) and it starts to oscillate. Let us say that this happens at time $t_I$ when the slow-roll parameter of the spectator field
\begin{equation}
\eta_{\varphi}\equiv\frac{a^{2}V_{\varphi\varphi}}{3\mathcal{H}^{2}}=\frac{a^{2}m_{\varphi}^{2}}{3\mathcal{H}^{2}}
\end{equation}
becomes unity. For the Hubble rate this gives
\begin{equation}
\mathcal{H}_{I}=\mathcal{H}(t_{I})=\frac{m_{\varphi}}{\sqrt{3}},
\end{equation}
where we have set $a_I=1$. The total energy density at the beginning of the oscillations is then given as
\begin{equation}
\rho_{I}=\frac{3}{8\pi a_{I}^{2}}\mathcal{H}_{I}^{2}M_{p}^{2}=\frac{1}{8\pi}m_{\varphi}^{2}M_{p}^{2}
\end{equation}

The solution to eq.~(\ref{eq:eomPhi}) in these circumstances is ~\cite{KariDaniRose}
\be{eq:massiveCurvatonSol}
\varphi(y(t)) =  {2^{1/4}\Gamma(5/4)\Phi_I \over \left[m_\varphi y(t) + {m_\varphi\over2\mathcal{H}_I}\right]^{1\over4}}\,J_{1\over4}\big(m_\varphi y(t) + {m_\varphi/(2\mathcal{H}_I)}\big)~,
\ee
where $y(t)$ is the cosmic time $y(t) \equiv \int_0^t dt' a(t')$, $J_{1/4}(x)$ is a Bessel function of order $1\over4$ and $\Gamma(5/4) \approx 0.9064$. This solution matches the form in eq.~(\ref{eq:ApproxVarPhi}), but again only when given in terms of cosmic time $y(t)$, which can be expressed as a function of the conformal time $t$ as $y(t) = {a_I\over2}\left[(1 + \mathcal{H}_It)^2-1\right]\mathcal{H}_I^{-1}$. The prefactors in eq.~(\ref{eq:massiveCurvatonSol}) just guarantee that $\varphi(0) = \Phi_I$, as can be simply checked by using the small-argument expansion of the Bessel function. For $m_\sigma y(t) \gtrsim 2$, the large-argument expansion of the Bessel function gives
\be{eq:curvatonMotionRD}
\varphi(t) \simeq  \Phi(t) \sin\left(m_\varphi y(t) + {\pi/8}\right);\;\;\;\;\Phi(t) \simeq 0.860~ \frac{\Phi_I}{(m_\varphi y(t))^{3/4}}~.
\ee
Thus, the scalar field oscillates with frequency $\omega = m_\varphi$ in cosmic time, crossing zero (i.e. $\varphi = 0$) every time $m_\varphi y(t) = \frac{7\pi}{8}, \frac{15\pi}{8}, \frac{23\pi}{8}, ...$\,.

The energy density of a scalar field in a quadratic potential scales, as mentioned before, like if it was a non-relativistic fluid~\cite{Preheating1}, i.e.~as $\rho_\varphi \propto 1/a^{3}(t)$. So although the Universe is RD at the start of the oscillations, if we would wait long enough, the energy density of the scalar condensate would end dominating the energy budget, and thus dictating the expansion of the Universe. We will however not consider such a possibility here, since the production of GW from fermions excited by a massive oscillatory field usually takes place during only few oscillations. Therefore, unless the initial energy in the condensate is close to the energy density of the radiation background, the GW production will happen while the energy of the scalar condensate is still subdominant. We will then consider that after the GW emission ceases, the scalar field decays at some latter moment into radiation $dof$, thus avoiding any MD anomalous expanding era. After the completion of reheating, 
the expansion of the Universe always remain as RD (receiving simply an injection of energy from the scalar condensate), until the matter-radiation equality moment $t_{eq}$ prior to recombination (at redshift $z_{eq} \simeq 3400$). The massive scalar field we want to consider then is simply an spectator in the radiation background.

For a RD background we already know that the parameter defined in eq.~(\ref{eq:epsilonParameter}) is $\epsilon = 1$. This implies that, as in the case of the quartic potential during preheating, the energy density of the GW scales exactly as it does the total energy of the thermal background.
 
The total energy density of the Universe just when the GW production ends at $t = t_*$, is given by
\begin{equation}
\rho_{*}=\rho_{I}\left(\frac{a_{I}}{a_{{\rm *}}}\right)^{4}=\frac{1}{8\pi}m_{\varphi}^{2}M_{p}^{2}\left(\frac{a_{I}}{a_{{\rm *}}}\right)^{4},
\label{eq:TotEDensPhi2Curvaton}
\end{equation}
as it corresponds to a relativistic fluid. The great difference in this post-reheating scenario, as compared to the (p)reheating ones, is that the initial energy density $\rho_{I}$ is actually unrelated to the the initial scalar condensate energy density $\rho_{\varphi}^{(I)} = \frac{1}{2}m_\varphi^{2}\Phi_{I}^{2}$. So $\rho_{I} \neq \rho_\varphi^{(I)}$, and consequently $\rho_I$ does not determine the initial amplitude of the scalar field $\Phi_I$. At most, the initial amplitude $\Phi_I$ is simply constrained such that its energy represents only a marginal fraction of $\rho_I$. This is precisely the crucial difference with respect the massive preheating case. Hopefully we will able to take advantage of this fact, rendering to more observable circumstances the GW background in these scenarios. 

The resonance parameter is the same as in the preheating case with a quadratic potential,
\begin{equation}
q=\frac{h^{2}\Phi_{I}^{2}}{m_\varphi^{2}},
\label{eq:qPhi2Curvaton}
\end{equation}
but the spectrum of GW at $t = t_*$ is now given, see eq.~(\ref{eq:AmpProd}), by
\begin{equation} 
\Omega_{\GW}^{(*)}=\frac{m_{\varphi}^{6}}{\pi^{3}\rho_{I}M_{p}^{2}}\kappa^{3}\mathcal{F}_{*}(\kappa;q)=\frac{8}{\pi^{2}}\left(\frac{m_{\varphi}}{M_{p}}\right)^{4}\kappa^{3}\mathcal{F}_{*}(\kappa;q),
\label{eq:AmpPhi2ProdCurvaton} 
\end{equation}
To determine the final amplitude of the GW spectrum, we see from eq.~(\ref{eq:AmpPhi2ProdCurvaton}) that we need to specify the mass $m_\varphi$ and the resonance parameter $q$. The potential $V(\phi) = {1\over2}m_\varphi^2\phi^2$ is not related in any how to the (p)reheating or inflationary periods and hence, we will consider $m_\varphi$ and $h$ as free parameters within reasonable margins. The initial amplitude $\Phi_I$, as mentioned above, should just be restricted as 
${1\over2}m_\varphi^2\Phi_I^2 \ll \rho_I$. Since we will consider only the situation when the thermal background always dominates the energy density budget of the Universe during GW production, then we should have that the energy of the scalar condensate at the end of GW production, $\rho_\varphi^* \equiv {1\over2}m_\varphi^2\Phi_I^2\left(a_I/a_*\right)^3$, should be smaller than the total energy of the thermal bath at that moment, $\rho_*$, given above in eq.~(\ref{eq:TotEDensPhi2Curvaton}). Therefore,
\begin{equation}
\rho_\varphi^* \leq \rho_* ~~~~~~~ \Rightarrow ~~~~~~~ \Phi_I \leq \Phi_I^{\rm max} \equiv \frac{1}{2\sqrt{\pi}}\left(\frac{a_{I}}{a_{{\rm *}}}\right)^{1\over2}M_p\,,
\end{equation}

\begin{figure}
\centering\includegraphics[width=10.5cm]{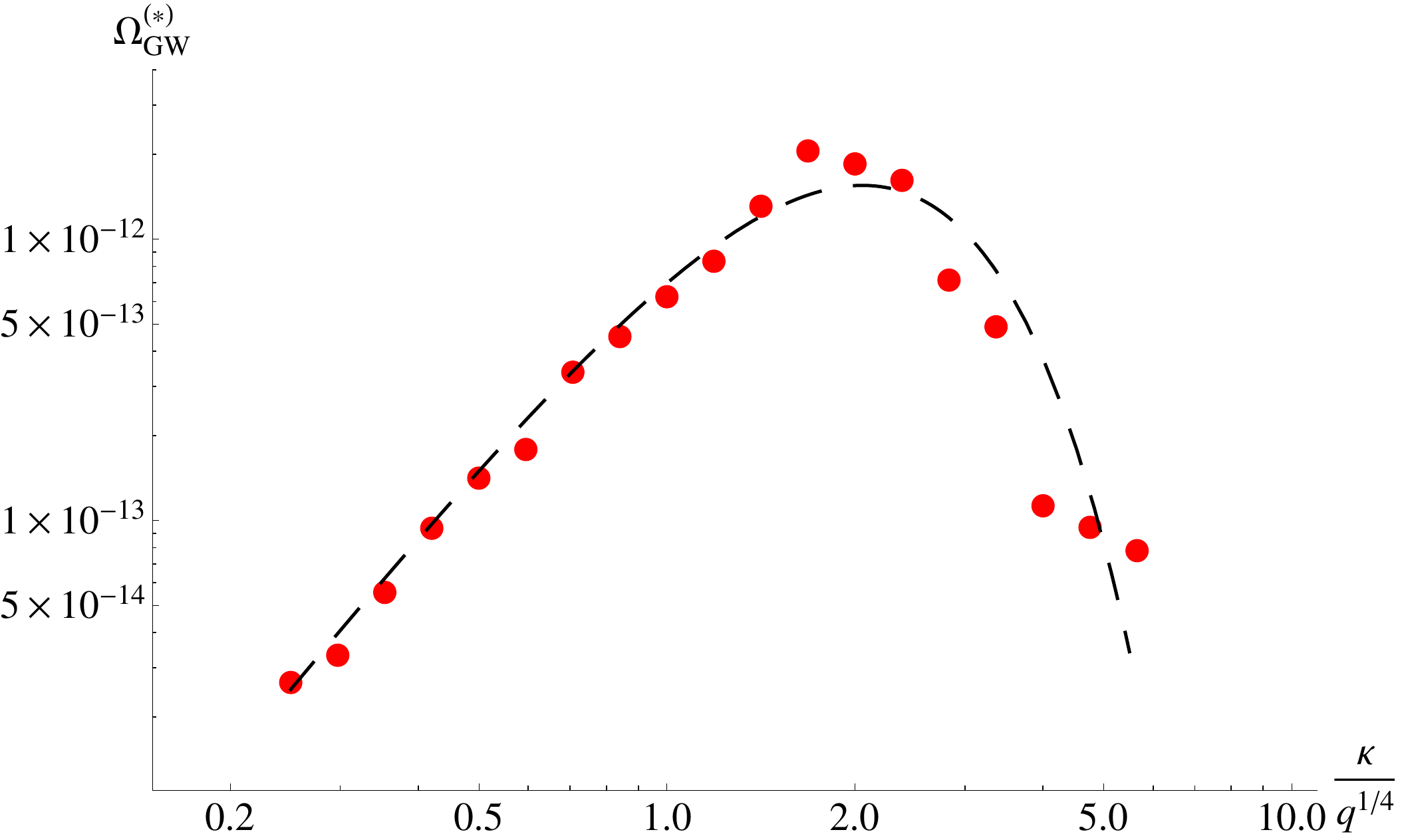}
\caption{Spectrum of GW right after the end production, for parameters $q=10^{6}$, $m_{\varphi}\approx1.2 \cdot 10^{14}\,\mathrm{GeV}$, initial energy scale $E_I\approx 1.7\cdot10^{16}\,\mathrm{GeV}$ and $\Phi_I=(10^{-2}/h)M_p$. Analogous considerations about the dashed guiding line, similar to those in figures~\ref{fig:FermPhi4Spectra} and \ref{fig:FermPhi2Spectra}, apply here.}
\label{fig:FermPhi2SpectraCurvaton}
\end{figure}

In figure~\ref{fig:FermPhi2SpectraCurvaton} we show an example of a spectrum obtained right at the end of GW production in the model under study, with parameters $m_\varphi = 10^{-5}M_p$, $q=10^{6}$, $E_{I}=\rho_I^{1/4} = 1.4\cdot10^{-3} M_p$. The initial scalar amplitude is then $\Phi_I = m_\varphi {\sqrt{q}/h} = (10^{-2}/h)M_p$, where the Yukawa coupling $h$ is not fixed. If we choose e.g.~$h = 0.1$, then $\Phi_I = 0.1 M_p$.  The point of the example, again, is not to take realistic parameter values, but rahter to show that extreme values give rise to a significant fraction of GW. To rescale the amplitude of this GW spectrum for other initial amplitudes $\Phi_I'$, we only need to re-evaluate $\mathcal{F}_*$ with the new resulting resonance parameter $q' = ({\Phi_I'/\Phi_I})^{2}q$. Note that this is very different from the situation in massive preheating, since in the latter, the expression of the GW spectral amplitude contains $\Phi_I$ explicitly as a multiplicative factor, and not only through the resonance 
parameter in the argument of $\mathcal{F}_*$. Changing $\Phi_I \rightarrow \Phi_I'$ would also shift the scale at which the spectrum peaks, as $k_p \rightarrow (\Phi_I'/\Phi_I)^{1/2}k_p$, similarly as in the case of massive preheating.

Changing instead the scalar field mass, $m_\varphi \rightarrow m_{\varphi}'$, would give rise to a rescaling of the GW spectrum by a factor $(m_\varphi'/m_\varphi)^4$, but one should at the same time re-evaluate $\mathcal{F}_*$ at a new resonance parameter $q \rightarrow (m_\varphi/m_\varphi')^2q$. The physical scale of the peak would shift again like in the massive preheating case, as $k_p \rightarrow (m_\varphi'/m_\varphi)^{1/2}k_p$.

In figure~\ref{fig:FermPhi2SpectraCurvaton} we see again that, as expected, and analogously to the preheating scenarios, the position of the maximum amplitude of the GW spectrum is located at $k_{p} \sim \kappa_p\omega$. In the example chosen the GW production lasts for $\ln(a_*/a_I) \approx \ln 11 \approx 2.4$ e-foldings, so we expect $\kappa_p \simeq (a_*/a_I)^{1/4}q^{1/4} \simeq 1.8\, q^{1/4}$. Nicely, the actual peak position seen in the numerical output, see figure~\ref{fig:FermPhi2SpectraCurvaton}, 
 \begin{equation} k_{p}^{\rm (num)}
\simeq 2\,q^{1/4}m_\varphi, \end{equation} 
matches very well (once again) the expectation.

\begin{figure}
\centering\includegraphics[width=10.5cm]{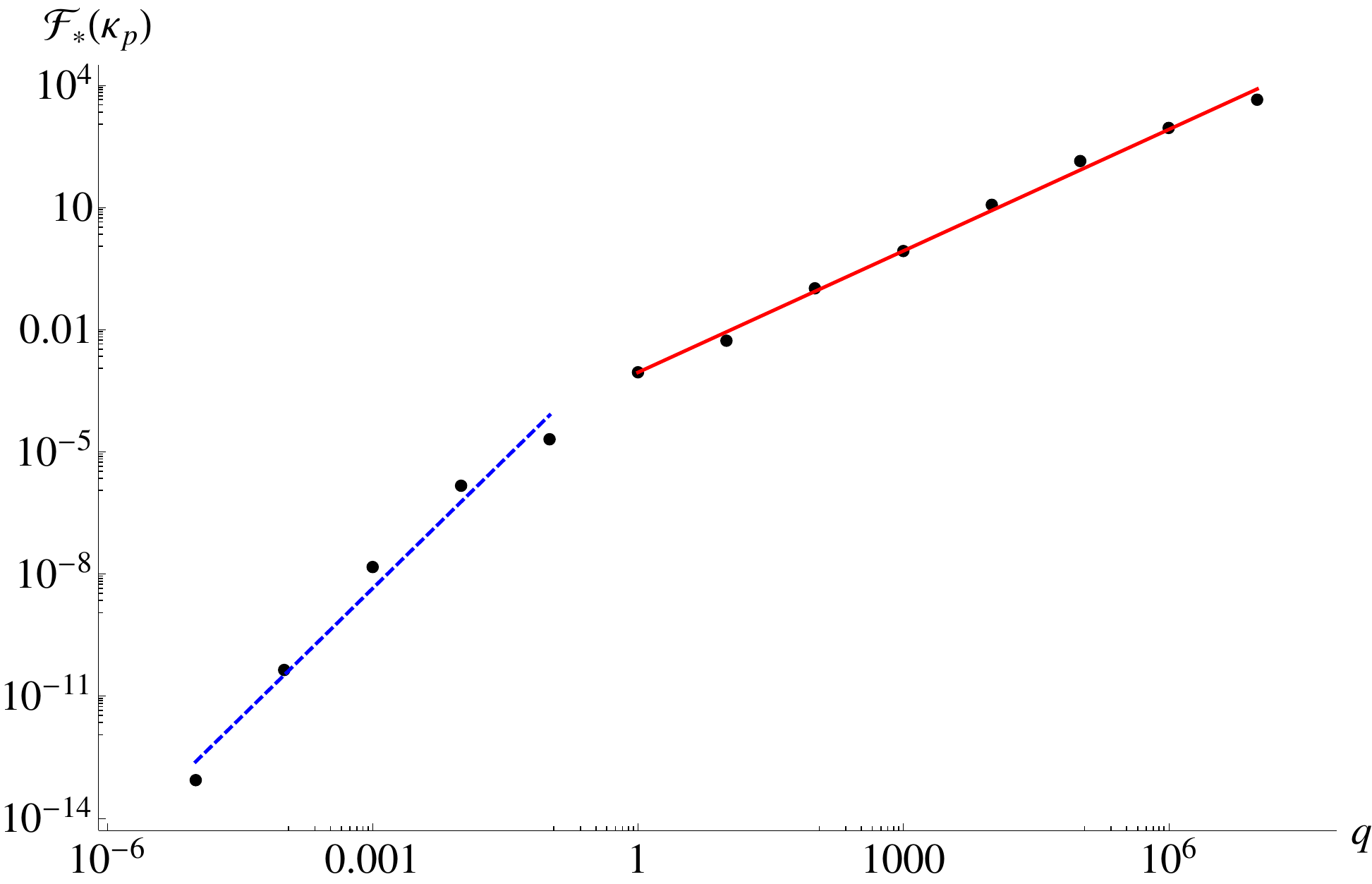}
\caption{The $q$-dependence of $\mathcal{F}_{*}(\kappa_p)$. The fittings are given in eq. (\ref{qdepCurvaton}).}
\label{fig:FermPhi2HighqDepCurvaton}
\end{figure}

In figure~\ref{fig:FermPhi2HighqDepCurvaton} we plot the function $\mathcal{F}_*(\kappa_p;q)$ versus the resonance parameter $q = (h\Phi_I/m_\varphi)^2$. Let us recall once again that our Antsaz eqs.~(\ref{eq:Antsaz}),~(\ref{eq:Antsaz_F*}) predicts that at the peak scale,
\begin{equation}
\mathcal{F}_*(\kappa_p,\, q) \simeq A^2 \cdot (a_{*}/a_I)^{\frac{3}{4}+\frac{\delta}{2}}  q^{\frac{3}{4}+\frac{\delta}{2}}\,.
\end{equation}
From our numerical results, see figure~\ref{fig:FermPhi2HighqDepCurvaton}, we learn that
\begin{eqnarray}
\label{qdepCurvaton}
\left\lbrace
\begin{array}{l}
q > 1: \mathcal{F}_*(\kappa_p) \approx 8.9\cdot10^{-4}\times q^{0.99} ~~~\Rightarrow~~~ A^2\approx8.3\cdot10^{-5}\,,\quad\delta \approx 0.48\,,\vspace{0.5cm}\\
q < 1: \mathcal{F}_*(\kappa_p) \approx 1.1\cdot10^{-2}\times q^{2.13} ~~~ \Rightarrow~~~ A^2\approx6.7\cdot10^{-5}\,,\quad\delta \approx 2.76\,, \vspace{0.3cm}
\end{array}
\right.
\end{eqnarray}
So for $q > 1$, we obtain that the ${\delta\over2}$-correction to the ${3\over2}$ slope (expected for $\delta = 0$) in eq.~(\ref{eq:PeakAmplitude}), is of the order of $\sim 16\%$. If we approximate $\delta \simeq 0.5$, we see that the scaling for $q > 1$ in eq.~(\ref{eq:Antsaz}), goes as 
\begin{equation} \big|\tilde I\big| \propto  \tilde p^{(-1+\delta)} \approx {1\over{\tilde p}^{\,1/2}}\,, \end{equation} 
In the opposite regime, when $q < 1$, approximating the numerical fit to $\delta \approx 3.0$, we find that our Antsaz scales as 
\begin{equation}
\big|\tilde I\big| \propto \tilde p^{(-1+\delta)}  \approx {\tilde p}^2\,, 
\end{equation}
From this analysis we can nicely conclude that the scaling behavior of $|\tilde{I}|$ found in the case of a scalar field with quadratic potential $V \propto \phi^2$, is universal: both at preheating (with the scalar field playing the role of the inflaton) and at thermal post-reheating stages (with the scalar field being simply an spectator), the spectral indices are the same, with $\delta \simeq 3$ for $q < 1$ and $\delta \simeq 0.5$ for $q > 1$. This could had been expected, since the physics of the GW production from fermions is only sensitive to the  shape of the potential of the scalar field acting as a source, and thus it should be independently of whether the scalar condensate dominates (inflaton in preheating) or not (spectator in thermal era) the energy budget of the Universe. The shape of the spectrum of GW is simply related to the potential of the oscillating field, and with these numerical results we nicely confirm this fact.

As in the preheating scenarios, the GW production typically lasts for few e-folds. In the example in figure~\ref{fig:FermPhi2SpectraCurvaton}, the scale factor grows as $a_*/a_I \simeq 11$. From eq.~(\ref{eq:PeakAmplitude}) we can obtain the approximated GW peak amplitude, including the duration $(a_*/a_I)$ factors, as
\begin{equation}
\label{eq:AmpPhi2PeakCurvaton}
\Omega_{\GW}^{(p)}\simeq\frac{8A^{2}}{\pi^{2}}\left(\frac{m_{\varphi}}{M_{p}}\right)^{4}\left(\frac{a_{*}}{a_{I}}\right)^{\frac{3+\delta}{2}}q^{\frac{3+\delta}{2}}\simeq\begin{cases}
6.7\cdot10^{-5}\left(\frac{m_{\varphi}}{M_{p}}\right)^{4}\left(\frac{a_{*}}{a_{I}}\right)^{1.74}q^{1.74}, & q > 1 \vspace{0.3cm}\\
5.4\cdot10^{-5}\left(\frac{m_{\varphi}}{M_{p}}\right)^{4}\left(\frac{a_{*}}{a_{I}}\right)^{2.88}q^{2.88}, & q < 1
\end{cases}
\end{equation} 

We can convert again our results from the end of GW production to the frequency and amplitude of the GW today. We simply need to use once again eqs.~(\ref{eq:PeakAmplitude})-(\ref{eq:fpToday}), together with the previous calibration of our parametrization schemes against the numerical results. We obtain that the amplitude and characteristic frequency of the peak today, 
$$h^{2}\Omega_{\GW}(f_p) \equiv h^{2}\Omega_{\mathrm{rad}}\left({g_{o}}\over{g_{*}}\right)^{1/3}\Omega_\GW^{(p)}\,,
~~~~~~~~~~f_p \simeq 5\cdot10^{10}{\kappa_p\over\tilde\rho_{I}^{1/4}}\,\mathrm{Hz},$$ can be written as 
\begin{equation}
\label{eq:FinalAmpTodayPhi2Curvaton}
h^{2}\Omega_{\GW}\left(f_{p}\right)\simeq\begin{cases}
2.7\cdot10^{-10}\left(\frac{m_{\varphi}}{M_{p}}\right)^{4}\left(\frac{a_{*}}{a_{I}}\right)^{1.74}q^{1.74}, & q>1 \vspace{0.3cm}\\
2.2\cdot10^{-10}\left(\frac{m_{\varphi}}{M_{p}}\right)^{4}\left(\frac{a_{*}}{a_{I}}\right)^{2.88}q^{2.88}, & q<1
\end{cases}
\end{equation}
\begin{equation}
\label{eq:FinalFreqTodayPhi2Curvaton}
f_{p}\simeq5\cdot10^{10}\left(\frac{a_{*}}{a_{I}}\right)^{\frac{1}{4}}\left(\frac{m_{\varphi}}{\rho_{I}^{1/4}}\right)q^{\frac{1}{4}}\,\mathrm{Hz} \simeq 10^{11}\left(\frac{a_{*}}{a_{I}}\right)^{\frac{1}{4}}\left(\frac{m_{\varphi}}{M_{p}}\right)^{\frac{1}{2}}q^{1\over4}\,\mathrm{Hz}\,.
\end{equation}
As in the preheating scenarios, the typical frequencies are very high. However now we see that the parametric dependence of the frequency and peak amplitude is not controlled by the same parameters. This is thanks to the fact that total initial energy density $\rho_{I}$ is not determined by the initial scalar field amplitude $\Phi_{I}$. This is quite a relevant difference with respect the GW produced in fermionic preheating, so we will try now to use it in our favor in order to render more observable the GW spectra generated in the post-reheating scenarios. The spectrum in the example in figure~\ref{fig:FermPhi2SpectraCurvaton} would be peaked today at $f_p \simeq10^{11}$ Hz, with an amplitude of $\Omega_\GW(f_p) \sim 10^{-16}$. This amplitude, as it happened in the examples in the preheating scenarios, it is above the sensitivity threshold of BBO or DECIGO, but its frequency is again too large to be accessible to such observatories. Let us see if we can reduce the 
characteristic frequency by finding the appropriate parameters, while at the same time we do not decrease the amplitude of the background. The frequency and amplitude of the GW spectrum peak in eqs.~(\ref{eq:FinalAmpTodayPhi2Curvaton}) and (\ref{eq:FinalFreqTodayPhi2Curvaton}) are characterized by parameters $\lbrace m_\varphi, q\rbrace$.

Let us begin by choosing $q \sim 1$. Then, in order to have a peak frequency in the $\sim 10$ Hz regime, we need to take a ratio $(m_\varphi/M_p)^{1/2} \sim 10^{-10}$. However this implies that the peak amplitude should be constrained as $h^2\Omega_\GW(f_p) \sim 10^{-10}(10^{-10})^8(a_{*}/a_I)^{1.74(2.88)} < 10^{-87}(10^{-86})$. Therefore, in the case of $q \sim 1$ it is not possible to have a non-negligible amplitude while shifting the frequency to small values. So what about the case $q \gg 1$? In such a case, we then need a ratio $(m_\varphi/M_p)^{1/2} \sim q^{-1/4}10^{-10} \ll 10^{-10}$ for the peak frequency to be in the $\sim 10$ Hz range. But, again this implies that $h^2\Omega_\GW(f_p) \sim 10^{-10}(q^{-1/4}10^{-10})^8(a_{*}/a_I)^{1.74} q^{1.74} \ll 10^{-88}$. So the case of $q \gg 1$ is even worse than when $q \sim 1$. Following that trend we could expect perhaps that in the opposite situation, i.e.~$q \ll 1$, there might be a chance to decrease the frequency but not the amplitude. Is that so? Well, 
to obtain $f_p \sim 10$ Hz, we need that $(m_\varphi/M_p)^{1/2} \sim q^{-1/4}10^{-10}$. The peak amplitude would then be  $h^2\Omega_\GW(f_p) \sim 10^{-10}(q^{-1/4}10^{-10})^8(a_{*}/a_I)^{2.88} q^{2.88} \ll 10^{-87}$, which is again way too small to be observed.

In conclusion, in these scenarios of GW production from Yukawa-excited fermions from an oscillatory spectator scalar field during the thermal era, GW backgrounds live naturally at very high frequencies, similarly to those in preheating scenarios. Note that though we only proved this for a massive scalar field, there is no particular reason to consider that this would change if we had chosen a massless scalar field.  

\section{Summary and Discussion}
\label{sec:Conclusions}

Fermions are expected to be created somewhen during the evolution of the early Universe, before BBN. Fermions could be the result of a perturbative decay from other bosonic species, but the most natural effects by which fermions can be efficiently created, correspond to non-perturbative phenomena. In the latter case the spectrum of the created fermions is non-thermal, and it develops a non-trivial anisotropic-stress, the transverse-traceless part of which acts as a source of GW. It is thus natural to consider that, non-perturbatively excited fermions in the early Universe, generate a stochastic background of GW. In this paper we have explored this possibility, characterizing the amplitude and frequency of the GW background from fermions in a variety of post-inflationary models. 

In section~\ref{subsec:GWspectrumFromFermions}, see also Appendix~\ref{app:Traceology}, we first developed a general formalism for computing the GW spectrum generated by an ensemble of fermions. In section~\ref{subsec:Regularization}, we took particular care of the UV divergences in the GW source, proposing a method to regularize the UTC of the TT $dof$ of the fermion's energy-momentum tensor. The main set of formulas encompassing our findings are eqs.~(\ref{eq:Pi^2(u_+,-)}), (\ref{eq:GW_spectra(u_+,-)}) and (\ref{eq:F_and_I_functions(u_+,-)}) [before regularization], and eqs.~(\ref{eq:W_reg}), (\ref{eq:GW_spectra_Reg}) and (\ref{eq:I_reg}) [after regularization]. See also eqs.~(\ref{eq:AmpProd}), (\ref{eq:AmpProd2}), which provide the equivalent dimensionless expressions in terms of the natural variables of the problem. All these equations are the master set of formulas in this paper, which describe the spectrum of the GW background created by any ensemble of fermions. We have also provided a parametric 
estimation of the frequency and amplitude of the spectrum of GW created by fermions, see section~\ref{subsec:ParametricAntsazs}, valid for when the fermions are excited from a Yukawa coupling with a coherent oscillatory scalar field. The main result there is given by eq.~(\ref{eq:PeakAmplitude}). 

We have applied our formalism of GW to several post-inflationary fermion-production scenarios. In section~\ref{subsec:FermionicMasslessPreheating} we have analyzed in detail the case when fermions are created during preheating from a massless inflaton. Based on our numerical outcome, we have obtained a fit to the amplitude and frequency of the GW peak in these scenarios, given by eqs.~(\ref{eq:FinalAmpTodayPhi4}), (\ref{eq:FinalFreqTodayPhi4}). Analogously, in section~\ref{subsec:FermionicMassivePreheating}, we have analyzed the case of a massive inflaton, for which we also found a fit to the amplitude and frequency of the GW peak, given by eqs.~(\ref{eq:FinalAmpTodayPhi2}), (\ref{eq:FinalFreqTodayPhi2}). Finally, the case when fermion production takes place after the completion of reheating, during the thermal era, has also been taken in consideration in section~\ref{subsec:ThermalEraScenarios}. The equivalent fit to the GW peak is captured in eqs.~(\ref{eq:FinalAmpTodayPhi2Curvaton}), (\ref{eq:
FinalFreqTodayPhi2Curvaton}).

All the scenarios studied share in common that the main mechanism of fermion excitation is due to a Yukawa coupling with a coherent oscillating scalar field; the inflaton in preheating, and a spectator field in the thermal scenario. Also common is that fermions are excited non-perturbatively, filling up a Fermi sphere of comoving radius $k_F \sim q^{1/4}\omega$, where $q$ is the resonance parameter [eq.~(\ref{ResPar})] and $\omega$ is the scalar field oscillation frequency. The modes within the Fermi sphere are continuouly excited with their occupation numbers oscillating at different rates, depending on the momentum and the resonance parameter. See~\cite{Fermions1}, \cite{Fermions2} for an extensive discussion on the details of their dynamics. The modes outside the Fermi sphere, i.e.~the UV modes of the problem, are on the contrary not excited, and remain as vacuum fluctuations. Due to this non-equilibrium structure of the fermions distribution, a non-trivial anisotropic stress appears in the system, whose 
TT part source actively GW. The final amplitude and frequency of the GW generated in each case, depends basically on the resonance parameter, and on the shape of the scalar potential. 

In figures~\ref{fig:FermPhi4Spectra}, \ref{fig:FermPhi2Spectra} and \ref{fig:FermPhi2SpectraCurvaton} we have plotted the spectrum of GW from different scenarios (having chosen the parameters in each case to maximize the amplitude). In all the cases the numerical points of the spectrum show some scattering around what it would be a physical continuous signal. We believe that the reason for this is the inaccuracy in our numerical calculations, due to the need to integrate numerically highly oscillating functions for a long period. The spectral amplitude in the UV scales seems particularly affected by this, since it tends to decrease slower that what it would be reasonably expected,  say following the initial trend of the slope on the right side of the spectral peak. We have added dashed lines to aid the visualization of how a continuous spectra would look like. With our current computational power we cannot determine the 'fine' details of the would-be continuous spectra. However, what it is clear is that we 
capture very well the 'essence' of such spectra, i.e.~the presence of a very well defined peak, accompanied by clear IR and UV tails around. The peak of the GW spectrum in all considered cases is, actually very nicely located at a scale of the order of the Fermi radius in each scenario, $\kappa_p \simeq \kappa_F$, as expected. For the peak of the spectrum today, we can write [reproducing again eqs.~(\ref{eq:PeakAmplitude})-(\ref{eq:fpToday})]
\begin{equation}
h^{2}\Omega_{\GW}\left(f_{p}\right)\simeq h^{2}\Omega_{\mathrm{rad}}\left(\frac{g_{o}}{g_{*}}\right)^{\frac{1}{3}}\,\frac{\epsilon A^{2}}{\pi^{3}}\,\frac{\omega^{6}}{\rho_IM_{p}^{2}}\,\,q^{\frac{3+\delta}{2}}\times
\begin{cases}
\left(\frac{a_{*}}{a_{I}}\right)^{3w-1} &, ~V(\varphi)\propto\varphi^{4}\vspace{0.3cm}\\
\left(\frac{a_{*}}{a_{I}}\right)^{3w+\frac{1+\delta}{2}} &, ~V(\varphi)\propto\varphi^{2}
\end{cases}
\end{equation}
and
\begin{eqnarray}
f_{p} \simeq 
5\cdot10^{10}\left(\frac{\omega}{{\rho}_{I}^{1/4}}\right)\,\epsilon^{\frac{1}{4}}\,q^{\frac{1}{4}}~{\rm Hz}
\times
\left\lbrace
\begin{array}{ll}
\left(\frac{a_{*}}{a_{I}}\right)^{\frac{3w-1}{4}} &, V(\varphi)\propto\varphi^{4}\vspace{0.3cm}\\
\left(\frac{a_{*}}{a_{I}}\right)^{\frac{3w}{4}} &, V(\varphi)\propto\varphi^{2}
\end{array}
\right.
\end{eqnarray}
valid for both $q < 1$ and $q > 1,$ as long as the right choice for the parameter $\delta$ is made. Let us recall that $w$ is the efective equation of state parameter from the start of GW production at $a_I$ till the end at $a_*$, $A^2$ is adjusted from our numerical  output (and differs from scenario to scenario, though typically is $\ll 1$), and $\epsilon \leq 1$, defined in eq.~(\ref{eq:epsilonParameter}), parametrizes the expansion history since the end of GW production till the first moment when the Universe becomes RD. From our numerical investigation, we have found that the parameter $\delta$ for each regime, $q < 1$ and $q > 1$, is determined by the shape of the scalar potential, rather than by the expansion rate (characterized by $w$). In particular we have found $\delta(q > 1) \approx 1/4$ and $\delta(q < 1) \approx 3/2$ for a massless scalar field, and $\delta(q > 1) \approx 1/2$, $\delta(q < 1) \approx 3$ for a massive scalar field (both in the case of massive preheating or of a massive spectator 
in the thermal era). 

For clarity, we reproduce again the final parametrizations found for each scenario, collecting altogether the expressions already presented in eqs.~(\ref{eq:FinalAmpTodayPhi4}), (\ref{eq:FinalFreqTodayPhi4}), (\ref{eq:FinalAmpTodayPhi2}), (\ref{eq:FinalFreqTodayPhi2}), (\ref{eq:FinalAmpTodayPhi2Curvaton}) and (\ref{eq:FinalFreqTodayPhi2Curvaton}),\\\\

\noindent Massless Preheating:
\begin{equation}
h^{2}\Omega_{\GW}\left(f_{p}\right)\simeq\begin{cases}
1.2\cdot10^{-9}\lambda^{2}\left(\frac{\Phi_{I}}{M_{p}}\right)^{2}q^{1.61}, & q>1\vspace*{0.3cm}\\
1.8\cdot10^{-7}\lambda^{2}\left(\frac{\Phi_{I}}{M_{p}}\right)^{2}q^{2.23}, & q<1
\end{cases}
\end{equation}
\begin{equation}
f_p \simeq 7\cdot10^{10}q^{1\over4} \lambda^{1\over4}\,\mathrm{Hz} = 7\cdot10^{10}\,h^{1\over2}\,\mathrm{Hz}\,,\vspace*{1.0cm}
\end{equation}

\noindent Massive Preheating:
\begin{equation}
h^{2}\Omega_{\GW}\left(f_{p}\right)\simeq\begin{cases}
2.5\cdot10^{-12}\left(\frac{m_{\varphi}^{2}}{\Phi_{I}M_{p}}\right)^{2}\left(\frac{a_{*}}{a_{I}}\right)^{0.78}q^{1.78}, & q>1\vspace{0.3cm}\\
3.1\cdot10^{-12}\left(\frac{m_{\varphi}^{2}}{\Phi_{I}M_{p}}\right)^{2}\left(\frac{a_{*}}{a_{I}}\right)^{1.97}q^{2.97}, & q<1
\end{cases}
\end{equation} 
\begin{equation}
f_p \simeq 6\cdot10^{10}\left(\frac{m_{\varphi}}{\Phi_{I}}\right)^{1\over2}q^{1\over4}\,\mathrm{Hz} = 6\cdot10^{10}\,h^{1\over2}\,\mathrm{Hz} \,, \vspace*{1.0cm}
\end{equation}

\noindent Massive Spectator in Thermal Era:
\begin{equation}
h^{2}\Omega_{\GW}\left(f_{p}\right)\simeq\begin{cases}
2.7\cdot10^{-10}\left(\frac{m_{\varphi}}{M_{p}}\right)^{4}\left(\frac{a_{*}}{a_{I}}\right)^{1.74}q^{1.74}, & q>1 \vspace{0.3cm}\\
2.2\cdot10^{-10}\left(\frac{m_{\varphi}}{M_{p}}\right)^{4}\left(\frac{a_{*}}{a_{I}}\right)^{2.88}q^{2.88}, & q<1
\end{cases}
\end{equation}
\begin{equation}
f_{p}\simeq5\cdot10^{10}\left(\frac{a_{*}}{a_{I}}\right)^{\frac{1}{4}}\left(\frac{m_{\varphi}}{\rho_{I}^{1/4}}\right)q^{\frac{1}{4}}\,\mathrm{Hz}\simeq 10^{11}\,\left(\frac{a_{*}}{a_{I}}\right)^{\frac{1}{4}}\left(\frac{m_{\varphi}}{M_{p}}\right)^{\frac{1}{2}}q^{1\over4}\,\mathrm{Hz}\,\vspace*{1.0cm}
\end{equation}

As discussed all along section~\ref{sec:Applications}, the natural frequencies for these backgrounds, $f_p \sim 10^9-10^{11}$ Hz, are too high to be reached by the frequency range that will be probed by proposed GW observatories. Tunning the parameters such that the peak frequency becomes small, for instance taking the Yukawa coupling $h$ to extremely small values, automatically implies a tremendous suppression in the amplitude of the GW background itself, to values totally undetectable as $h^2\Omega_\GW(f_p) \ll 10^{-87}$. The problem is ameliorated, as compared to preheating scenarios, in the case of the scalar spectator in the thermal era. But still remains. Therefore, we are forced to accept that the backgrounds considered in this paper peak naturally at high frequencies, where no experimental device has been envisaged with sufficient sensitivity as $10^{-20} \lesssim h^2\Omega_\GW(f_p) \lesssim 10^{-10}$. A possible scenario where one might be able to achieve the goal of having a small frequency but not 
so small amplitude, would be to consider the GW production from Fermions in Hybrid preheating ~\cite{Preheating3}. We have not explored this situation in this paper, but we will briefly comment about it later.

Let us note that in our results in sections~\ref{subsec:FermionicMasslessPreheating}-\ref{subsec:ThermalEraScenarios}, the fermions are massless, except for their interactions with the scalar field, which provides them a dynamical mass. We have considered in Appendix~\ref{app:FermionsMass} the case of fermions having a bare constant mass in the Lagrangian. We conclude there that fermions with a light mass ($m_\Phi \ll h\Phi_I$) do not affect our results from sections~\ref{subsec:FermionicMasslessPreheating}-\ref{subsec:ThermalEraScenarios}, summarized in the previous equations. Too heavy fermions ($m_\Phi \gg h\Phi_I$) on the contrary, are simply not excited, and thus GW production is completely suppressed (since it simply does not take place). Therefore, a bare mass is either irrelevant or a killing factor of the GW production, which justifies our analysis in the the bulk of the text with massless fermions.

Finally, before we conclude, we would like to raise up a number of points which could be interesting for future research in the topic that we have studied here:
\begin{itemize}
\item As mentioned already, it would be very interesting to apply our formalism from section~\ref{sec:Theory} to fermions produced in Hybrid preheating~\cite{JuanFermionicPreheating}. Our parametric formulas from section~\ref{subsec:ParametricAntsazs} would not be valid there, since the excitation of fermions in these models occurs via tachyonic effects, very differently than the oscillatory parametric effects of the scenarios analyzed in sections~\ref{subsec:FermionicMasslessPreheating}-\ref{subsec:ThermalEraScenarios}. The characteristic scale in Hybrid models (substituting the frequency scale $\omega$ in of the models considered in the paper) is given by $\omega^2 = \lambda v^2$, whereas the initial energy scale is $E_I \sim \lambda^{1/4}v$, with $\lambda$ and $v$ the self-coupling and VEV of an auxiliary (symmetry breaking) field coupled to the inflaton. The characteristic frequency would scales as $f \propto \lambda^{1/4}$~\cite{GWpreheating3}, so that for small self-couplings we could obtain a peak frequency sufficiently small. The 
question is how the amplitude of the GW background would be affected by such a small value of $\lambda$. This possibility could be explored indeed using the formalism developed in section~\ref{sec:Theory}.

\item Realistic scenarios -- rooted in particle physics -- where both our theoretical formalism and our parametric formulas could be used, include (p)reheating~\cite{RH_higgsInflation}-\cite{RH_higgsInflation3} after Higgs-inflation~\cite{HiggsInflation}, and the curvaton-Higgs model recently proposed~\cite{KariDaniRose} within the curvaton scenario~\cite{Curvaton}. In the former the SM Higgs plays the role of the inflaton, creating during preheating the gauge fields and leptons of the SM through parametric resonance. It is then expected a high frequency GW background both from the excited bosons and fermions. In the latter, the curvaton is a spectator field during the thermal era. In~\cite{KariDaniRose} it was considered that the curvaton excited non-perturbatively the SM Higgs, but one could equally consider that the SM fermions could be coupled to the curvaton, in which case these would be excited precisely as considered in section~\ref{subsec:ThermalEraScenarios}.

\item Our formalism for GW from fermions in sections~\ref{subsec:GWspectrumFromFermions}, \ref{subsec:Regularization} does not rely on solving the mode functions from the Dirac equation. Actually, it has been recently obtained the result~\cite{BergesPRL} that quantum effects might enhance the production of fermions in preheating scenarios. As a consequence fermions would approach faster to a quasistationary thermal in the IR, as compared to the rate expected in semiclassical descriptions based on the Dirac equation with a homogeneous background field. The impact of this effect goes beyond our applied results in this paper in sections~\ref{subsec:FermionicMasslessPreheating}-\ref{subsec:ThermalEraScenarios}, where we have simply solved the Dirac equation with the fermion mode functions stimulated by the classical scalar field to which they are coupled to. We speculate that this next-to-leading order effects considered in~\cite{BergesPRL}, could impact our results in two manners: on one hand it could enhance the production of GW since more fermions are stimulated faster, but on the other hand it could decrease, at the same time, the efficiency of GW production, since the IR band of momenta tends faster to a quasi-equilibrium thermal state. It should be interesting to explore this, perhaps with the recent lattice techniques proposed to simulate non-equilibrium quantum fermions coupled to classical bosonic fields~\cite{SaffinTranberg}. We expect that the overall result of taking into account these corrections in the fermion dynamics, will amount for a rescaling of our parametric formulas through a renormalization of the dimensionless constant $A^2$, whereas the functional parametrization depending on the resonance parameter $q$ should (most likely) remain.

\item In our considerations about fermion dynamics all through section~\ref{sec:Applications}, we have not considered explicitly nonlinear effects, such as the backreaction from the created fermions into the oscillating scalar field. A proper study of these aspects would enlarge the complexity of the calculations substantially. Such non-linear effects will never prevent in any case the GW production from taking place during the initial fermionic parametric excitaction, when backreation is still negligible. Only when the energy transferred into the fermions is comparable to the energy in the scalar oscillatory field, does the backreaction matters. But in most cases the fraction of energy transferred into fermions is negligible, and the transfer of energy ceases before it might become relevant (there is no more phase space and fermions of higher momenta cannot be created anymore). In~\cite{Fermions3} it was found that in fermionic preheating with $q < 10^8$, backreaction is never relevant, since the energy transferred into the fermions is simply too small. The case with a resonant parameter bigger than $q = 10^8$ is actually (most-likely) unrealistic, since it would typically require unreasonably big Yukawa couplings. Yet, if one would want to consider such situation, probably the best way to study it would be with the aforementioned lattice techniques proposed in~\cite{SaffinTranberg}.

\item In our discussion about the regularization of the fermions' GW source in section~\ref{subsec:Regularization}, we have not compared our results with other time-dependent regularization schemes, like for instance the one proposed in~\cite{BarnabyPelosoEtAl}. This would be an interesting exercise to do, though not easy: most of the existing techniques, often focused in bosonic species and not in fermions, are suitable for 2-point functions, but not necessarily for 4-point functions. In our case, given the structure of the UTC $\Pi^2(k,t,t')$, we necessarily need to deal with a 4-point fermionic operator. What it should be really clear is that our prescription for regularizing the UTC, captures well all the relevant expected physical effects. Mainly, that the UV modes which are not really excited by the scalar field, are filterred out, eliminating this way their otherwise divergent vacuum fluctuations' contribution. Only the IR modes survive naturally, thanks to the Bogoliubov coefficients $|\beta_k(t)|$ in front of the 'regularized' mode functions. Thus only the relevant modes -- those physically excited  within the Fermi sphere -- are captured, and allowed to contribute into the final GW spectrum.

\item A final, perhaps more speculative consideration about our results, would be to consider the possibility of anisotropies arising in the GW produced by fermions. It has been recently shown~\cite{LauraDaniArttu} that anisotropies emerge naturally in the GW background from preheating with bosonic fields, whenever the preheat fields were light during inflation (i.e.~its effective mass smaller than the Hubble rate). It is not clear how these type of results could be extended to the case of fermions, but it is certainly intriguing, and probably worth exploring, to consider the possibility that, indeed, anisotropies could arise in the amplitude of the GW background from fermions that we have just described in this paper.

\end{itemize}

To conclude, as a final remark, we would like to emphasize once again, that the natural frequency of the GW backgrounds that we have described in this paper, is in the $f_p \sim 10^9-10^{11}$ Hz range. There is unfortunately no experimental device planned to work at those high frequencies, with sufficiently good sensitivity. Nevertheless, we would like to see this fact not only as a caveat, which it is, but also as a reason to urge experts on the field to start considering (more actively) the possibility of developing ultra-high frequency GW detectors. There is little doubt that pursuing this endeavor is certainly worthy. We have shown here that fermions can indeed act as very efficient generators of GW. The natural frequencies of the backgrounds that we predict lie in a high-frequency window, inaccessible to currently planned GW detectors. However, a detection of such a GW background would open a direct window to the physics of the very early Universe.

\section*{Acknowledgements}

We would like to thank very specially K. Enqvist, with whom it was a pleasure to work with at the first stages of this project. We are also grateful to D. Blas, R. Durrer, J. Garc\'ia-Bellido, G. P\'erez-Nadal, M. Peloso, S. Sibiryakov, L. Sorbo and A. Riotto for useful comments. DGF would also like to thank the members of the Institute of Nuclear Research (INR) in Moscow, for their very nice hospitality during a visit in 2013, during which a significant fraction of the writing of this paper took place. DGF is supported by the Swiss National Science Fundation (SNSF), and TM is supported by the Magnus Ehrnrooth Foundation.

\appendix


\section{"Traceology" in a time-dependent background}
\label{app:Traceology}

In section~\ref{sec:Theory} we presented the result of several VEVs, from the non-regularized UTC in eq.~(\ref{eq:Pi^2(u_r)}) to the regularized one in eq.~(\ref{eq:Pi2_regularized}), passing through the 2-point correlation functions in eqs.~(\ref{eq:Pi_reg.-1}), (\ref{eq:Os_regII}) and (\ref{eq:Os_NonReg}). All these calculations involved the computation of traces over spinorial contractions. In this appendix we show the details of such calculations. First we set up a compact notation for spinorial fields, then show the computation of the 2-point correlation functions from the main text, and finally we show the calculation of the 4-point UTCs, both with and without regularization.

\subsection{Compact Notation}
\label{subsec:CompactNotation}

For the sake of clarity, let us recall that we write a Dirac field $\psi$ as
\begin{eqnarray}
\psi(\mathbf{x},t) &=& \int\frac{d\mathbf{k}}{\left(2\pi\right)^{3}}\,e^{-i\mathbf{k\cdot x}}\left[\hat a_{\mathbf{k},r}{\tt u}_{\mathbf{k},r}(t)+\hat b_{-\mathbf{k},r}^{\dagger}{\tt v}_{\mathbf{k},r}(t)\right]
\end{eqnarray}
with $a_{\bk,r}, b_{\bk,r}$ satisfying the anti-commutation relations of eq.~(\ref{eq:anticom}), and where the 4-component spinors are expressed as
\begin{eqnarray}\label{eq:u,v_Appendix}
{\tt u}_{\mathbf{k},r}(t) \equiv 
\left(\begin{array}{c}
\vspace{0.4cm}u_{\mathbf{k},+}(t)\,S_{r}\\ u_{\mathbf{k},-}(t)\,S_{r}
\end{array}\right),\hI
{\tt v}_{\mathbf{k},r}(t)\equiv
\left(\begin{array}{c}
\vspace{0.4cm}v_{\mathbf{k},+}(t)\,S_{-r}\\ v_{\mathbf{k},-}(t)\,S_{-r}
\end{array}\right),
\end{eqnarray}
with $\{S_{r}\}$ 2-component spinors given by 
\begin{eqnarray}\label{eq:Sr_Appendix}
S_{1}=-S_{-2}=\left(\begin{array}{c}
\vspace{0.2cm}1 \\ 0\end{array}\right),\hI
S_{2}=S_{-1}=\left(\begin{array}{c}
\vspace{0.2cm}0 \\ 1\end{array}\right)
\end{eqnarray}

Given the structure of the 4-component spinors shown in eqs.~(\ref{eq:u,v_Appendix}), (\ref{eq:Sr_Appendix}), we can rewrite them as
\begin{eqnarray}\label{eq:u,P+,P-_Appendix}
{\tt u}_{\mathbf{k},r}(t) &=& u_{\mathbf{k},+}(t)S_r \otimes P_+ + u_{\mathbf{k},-}(t)S_r \otimes P_-\,,\\
\label{eq:v,P+,P-_Appendix}
{\tt v}_{\mathbf{k},r}(t) &=& v_{\mathbf{k},+}(t) S_{-r}\otimes P_+ + v_{\mathbf{k},-}(t)S_{-r}\otimes P_-\,,
\end{eqnarray}
where we have defined the projectors
\begin{eqnarray}\label{eq:Pa_Appendix}
P_{+} = \left(\begin{array}{c}
\vspace{0.2cm}1 \\ 0\end{array}\right),\hI
P_{-} = \left(\begin{array}{c}
\vspace{0.2cm}0 \\ 1\end{array}\right),
\end{eqnarray}
and introduced the multiplication rule
\begin{eqnarray}
A \otimes B \equiv \left(\begin{array}{ccc}
A\hspace*{-0.1cm}\cdot\hspace*{-0.1cm}B_{11} & \cdots & A\hspace*{-0.1cm}\cdot\hspace*{-0.1cm} B_{1Q}\\
\vdots & \ddots & \vdots \\
A\hspace*{-0.1cm}\cdot\hspace*{-0.1cm} B_{P1} & \cdots & A\hspace*{-0.1cm}\cdot\hspace*{-0.1cm} B_{PQ}\end{array}\right) 
\equiv \left(\begin{array}{cccc}
{\begin{array}{ccc}
A_{11} B_{11} & \cdots & A_{1M} B_{11}\\
\vdots & \ddots & \vdots \\
A_{N1} B_{11} & \cdots & A_{NM} B_{11}\end{array}} & \cdots & \cdots & {\begin{array}{ccc}
A_{11} B_{1Q} & \cdots & A_{1M} B_{1Q}\\
\vdots & \ddots & \vdots \\
A_{N1} B_{1Q} & \cdots & A_{NM} B_{1Q}\end{array}}\\
\vdots & \ddots & & \vdots \\
\vdots & &\ddots & \vdots \\
{\begin{array}{ccc}
A_{11} B_{P1} & \cdots & A_{1M} B_{P1}\\
\vdots & \ddots & \vdots \\
A_{N1} B_{P1} & \cdots & A_{NM} B_{P1}\end{array}} & \cdots & \cdots & {\begin{array}{ccc}
A_{11} B_{PQ} & \cdots & A_{1M} B_{PQ}\\
\vdots & \ddots & \vdots \\
A_{N1} B_{PQ} & \cdots & A_{NM} B_{PQ}\end{array}}
\end{array}\right)\nn\\
\end{eqnarray}
with $A$ and $B$ arbitrary matrices (of dimensions $N\times M$ and $P\times Q$, respectively)
\begin{eqnarray}
\begin{array}{c}
A \equiv \left(\begin{array}{ccc}
A_{11} & \cdots & A_{1M}\\
\vdots & \ddots & \vdots \\
A_{N1} & \cdots & A_{NM} \end{array}\right)\,,~~
B \equiv \left(\begin{array}{ccc}
B_{11} & \cdots & B_{1Q}\\
\vdots & \ddots & \vdots \\
B_{P1} & \cdots & B_{PQ} \end{array}\right)
\end{array} \hspace{4cm}
\end{eqnarray}
In particular, we have
\begin{eqnarray}
C \otimes P_+ \equiv \left(\begin{array}{c}
C_1 \\ C_2 \\ 0 \\ 0 \end{array}\right)
\,,\hV C \otimes P_- \equiv \left(\begin{array}{c}
0 \\ 0 \\ C_1 \\ C_2 \end{array}\right)\,,\hI ~ C = \left(\begin{array}{c}
\vspace{0.2cm}C_1 \\ C_2\end{array}\right)\,,
\end{eqnarray}
with $C$ an arbitrary 2-component 1-column matrix.

Let us recall that the mode functions $u_{\bk,\pm}$, $v_{\bk,\pm}$ are not independent from each other, since they are related by the charge conjugation, from where it follows the relation $v_{\mathbf{k},\pm}(t)= \pm\,u_{\mathbf{k},\mp}^{*}(t)$. Thus, making use of this relation and assuming from now on a summation over repeated indices, we can write eqs.~(\ref{eq:u,P+,P-_Appendix}), (\ref{eq:v,P+,P-_Appendix}) in a compact manner like
\begin{eqnarray}\label{eq:u,P+,P-_v2_Appendix}
{\tt u}_{\mathbf{k},r}(t) &=& u_{\mathbf{k},a}(t)S_r \otimes P_a,\\
\label{eq:v,P+,P-_v2_Appendix} 
{\tt v}_{\mathbf{k},r}(t) &=& a\,u_{\mathbf{k},-a}^*(t) S_{-r} \otimes P_a,
\end{eqnarray}
The advantage of this notation will become clear when we perform an explicit spinorial trace calculation, since it will allow us to factorize any trace over 4-component spinors into a product of traces over the projectors $P_\pm$ on one hand, and the 2-component spinors $S_{\pm r}$ on the other hand. This will simplify enormously any calculation. This factorization is possible thanks to the following nice property of the $\otimes$ product just defined
\begin{eqnarray}
(B_1 \otimes B_2)\cdot (B_3 \otimes B_4) = (B_1\cdot B_3) \otimes (B_2 \cdot B_4)
\end{eqnarray}
where $\cdot$ stands for standard matrix multiplication and $B_i$ are arbitrary matrices with appropriate dimensions such that $B_1$ and $B_2$  can multiply $B_3$ and $B_4$, respectively.

Before we move into calculating any trace, let us list a bunch of properties which will aid us in the task. Calling $\mathcal{I}$ the $2\times2$ identity matrix and $\sigma_i$ the Pauli matrices, we find that 
\begin{eqnarray}
S_rS_r^{\rm T} &=& S_{-r}S_{-r}^{\rm T} = \mathcal{I}\,,\\
S_rS_{-r}^{\rm T} &=& - S_{-r}S_{r}^{\rm T} = i\sigma_2\,,\\
S_r^{\rm T}S_r &=& S_{-r}^{\rm T}S_{-r} = 2\,,\\
S_{r}^{\rm T}S_{-r} &=& S_{-r}^{\rm T}S_r  = 0\,,
\end{eqnarray}
as well as
\begin{eqnarray}
P_a^{\rm T}P_b &=& \delta_{ab}\,\\
\sigma_3P_\pm &=& \pm P_\pm\,,\\
\sigma_1P_\pm &=& P_\mp
\end{eqnarray}
Finally, let us note that the flat gamma-matrices $\gamma_\mu$ can be written, in the Dirac basis, as
\begin{eqnarray}
\gamma_0 = \mathcal{I} \otimes \sigma_3\,,\hI \gamma_k = \sigma_k \otimes (-i\sigma_2)\,,  
\end{eqnarray}
and let us recall the Pauli matrix product
\begin{eqnarray}
\sigma_i\sigma_j = i\epsilon_{ijk}\sigma_k + \mathcal{I}\delta_{ij}\,,
\end{eqnarray}
where $\epsilon_{ijk}$ is the totally antisymmetric tensor. In particular it follows that $\sigma_3(i\sigma_2) = \sigma_1$, which we will use repeatedly in the trace calculations below.

\subsection{2-point Functions: Vacuum Expectation Values (VEVs)}
\label{subsec:2point_corr}

The three 2-point functions that we have introduced in the text are $\big\langle 0 \big|\,\Pi_{ij}^{\rm TT}(\bk,t)\,\big|0 \big\rangle_{\rm{reg}}$, $\big\langle 0 \big|\,\mathcal{O}_{2}(\bx,t)\,\big|0 \big\rangle_{\rm{reg}}$, $\big\langle 0 \big|\,\Pi_{ij}^{\rm TT}(\bk,t)\,\big|0 \big\rangle_{\rm{reg}}$ and $\big\langle 0 \big|\,\mathcal{O}_{2}(\bx,t)\,\big|0 \big\rangle$, where $\mathcal{O}_2(\bx,t) \equiv \bar\Psi(\bx,t)\Psi(\bx,t)$. We will begin by the simplest, $\big\langle 0 \big|\,\mathcal{O}_{2}(\bx,t)\,\big|0 \big\rangle$, followed by $\big\langle 0 \big|\,\mathcal{O}_{2}(\bx,t)\,\big|0 \big\rangle_{\rm{reg}}$. Then, as we will show, $\big\langle 0 \big|\,\Pi_{ij}^{\rm TT}(\bk,t)\,\big|0 \big\rangle_{\rm{reg}}$ = $\big\langle 0 \big|\,\Pi_{ij}^{\rm TT}(\bk,t)\,\big|0 \big\rangle = 0$.

Let us then first recall eq.~(\ref{eq:Os_NonReg}),
\begin{equation}\label{eq:Os_NonReg_Appendix}
\big\langle 0 \big|\,\mathcal{O}_2(\bx,t)\,\big|0 \big\rangle = \int\hspace*{-1mm} {d\bp\over(2\pi)^3}~{\bar{\tt v}}_{\mathbf{p},r}{\tt v}_{\mathbf{\mathbf{p}},r}\,.
\end{equation}
The spinorial trace in the integrand is
\begin{eqnarray}\label{eq:Trace_vv}
{\bar{\tt v}}_{\mathbf{p},s}{\tt v}_{\mathbf{\mathbf{p}},s} &=& {{\tt v}}^\dag_{\mathbf{p},s}\gamma_o{\tt v}_{\mathbf{\mathbf{p}},s} = (au^*_{\bp,-a}S_{-r}\otimes P_a)^\dag\cdot\mathcal{I}\otimes\sigma_3\cdot(bu^*_{\bp,-b}S_{-r}\otimes P_b)\nn\\
&=& ab\,u_{\bp,-a}u^*_{\bp,-b}\cdot(S_{-r}^\dag\mathcal{I}S_{-r} \otimes P_a^\dag\sigma_3P_b) = ab\,u_{\bp,-a}u^*_{\bp,-b}\cdot(2\otimes a\delta_{ab})\nn\\
&=& 2a\,|u_{\bp,-a}|^2 = 2\left(|u_{\bp,-}|^2-|u_{\bp,+}|^2\right)\,,
\end{eqnarray}
which coincides with (the integrand of) eq.~(\ref{eq:Os_reg3_I}). 

Recalling eq.~(\ref{eq:Os_regII}), we have
\begin{equation}\label{eq:Os_regII_Appendix}
\big\langle 0 \big|\,\mathcal{O}_{2}(\bx,t)\,\big|0 \big\rangle_{\rm{reg}} = \int\hspace*{-1mm} {d\bp\over(2\pi)^3} \,\left({\bar{\tt v}}_{\mathbf{p},s}{\tt v}_{\mathbf{\mathbf{p}},s}-\bar{\mathcal{V}}_{\mathbf{p},s}\mathcal{V}_{\mathbf{\mathbf{p}},s}\right),
\end{equation}
Thus, we just need to compute the trace $\bar{\mathcal{V}}_{\mathbf{p},s}\mathcal{V}_{\mathbf{\mathbf{p}},s}$, analogously as how we computed ${\bar{\tt v}}_{\mathbf{p},s}{\tt v}_{\mathbf{\mathbf{p}},s}$. We obtain
\begin{eqnarray}\label{eq:Trace_VV}
&& \bar{\mathcal{V}}_{\mathbf{p},r}\mathcal{V}_{\mathbf{\mathbf{p}},r} = {\mathcal{V}}^\dag_{\mathbf{p},s}\gamma_o\mathcal{V}_{\mathbf{\mathbf{p}},r} = (\alpha_{\bp}{\tt v}_{\mathbf{p},r}-\beta_{\bp}{\tt u}_{\mathbf{p},r})^\dag\gamma_o(\alpha_{\bp}{\tt v}_{\mathbf{p},r}-\beta_{\bp}{\tt u}_{\mathbf{p},r})\nn\\
&& = \left(a\alpha_{\bp}u^*_{\bp,-a}(S_{-r}\otimes P_a) - \beta_{\bp}u_{\bp,a}(S_{r}\otimes P_a)\right)^\dag\cdot\mathcal{I}\otimes\sigma_3\cdot\left(b\alpha_{\bp}u^*_{\bp,-b}(S_{-r}\otimes P_b) - \beta_{\bp}u_{\bp,b}(S_{r}\otimes P_b)\right)\nn\\
&& = \left(ab\,\alpha_\bp^2 u_{\bp,-a}u^*_{\bp,-b}\cdot(S_{-r}^\dag\mathcal{I}S_{-r}) + |\beta_\bp|^2 u_{\bp,a}u^*_{\bp,b}\cdot(S_{r}^\dag\mathcal{I}S_{r})\right.\nn\\
&& ~~~~-\left.a\,\alpha_\bp\beta_\bp u_{\bp,-a}u_{\bp,b}\cdot(S_{-r}^\dag\mathcal{I}S_{r}) - b\,\alpha_{\bp}\beta_\bp^* u_{\bp,a}u^*_{\bp,-b}\cdot(S_{r}^\dag\mathcal{I}S_{-r})\right) \otimes (P_a^\dag\sigma_3P_b) \nn\\
&& = \left(\alpha_\bp^2 |u_{\bp,-a}|^2\cdot 2 + |\beta_\bp|^2 |u_{\bp,a}|^2\cdot 2 - a\,\alpha_\bp\beta_\bp u_{\bp,-a}u_{\bp,b}\cdot 0 - b\alpha_{\bp}\beta_\bp^* u_{\bp,a}u^*_{\bp,-b}\cdot 0 \right) \otimes (a\delta_{ab}) \nn\\
&& = 2\left(\alpha_\bp^2-|\beta_\bp|^2\right)\left(|u_{\bp,-}|^2 -  |u_{\bp,+}|^2\right)
\end{eqnarray}
From eq.~(\ref{eq:Trace_VV}), together with eq.~(\ref{eq:Trace_vv}), we conclude then that the regularized integrand characterizing eq.~(\ref{eq:Os_regII_Appendix}) is
\begin{eqnarray}
\left({\bar{\tt v}}_{\mathbf{p},r}{\tt v}_{\mathbf{\mathbf{p}},r}-\bar{\mathcal{V}}_{\mathbf{p},r}\mathcal{V}_{\mathbf{\mathbf{p}},r}\right) = 4|\beta_\bp|^2\left(|u_{\bp,-}|^2 -  |u_{\bp,+}|^2\right) = 2|\beta_\bp|^2{\bar{\tt v}}_{\mathbf{p},r}{\tt v}_{\mathbf{\mathbf{p}},r}\,,
\end{eqnarray}
which coincides with eq.~(\ref{eq:Os_reg3_II}).

Finally, let us consider the VEV of the TT-part of the anisotropic stress-tensor, i.e.~the GW source. Following the prescription described in section~\ref{subsec:Regularization}, we have
\begin{equation}
\left\langle\Pi_{ij}^{\rm TT}(\bk,t)\right\rangle_{\rm reg} = \big\langle 0\big|\,\Pi_{ij}^{\rm TT}({\tt u}_{\bk,r},{\tt v}_{\bk,r})
-{\tilde \Pi}_{ij}^{\rm TT}(\mathcal{U}_{\bk,r},\mathcal{V}_{\bk,r})\,\big|0\big\rangle \equiv \left\langle 0 \big|\,\Pi_{ij,\rm{reg}}^{\rm TT}\,\big|0 \right\rangle\label{eq:Pi_reg.-1_Appen}
\end{equation}
where 
\begin{eqnarray}\label{eq:PI_reg_Appen}
{\Pi}_{ij,{\rm reg}}^{\TT}(\mathbf{k},t) &=& \frac{1}{(2\pi)^{3}a^{2}(t)}\int\hspace*{-1mm} d\mathbf{p}\,\,\Lambda_{ij,lm}(\hat{k})\\
&&\hspace{2cm}\times\left[\,\hat b_{-\mathbf{p},s}\left({\bar{\tt v}}_{\mathbf{p},s}p_{(l}\gamma_{m)}{\tt u}_{\mathbf{\mathbf{k+p}},r}-\bar{\mathcal{V}}_{\mathbf{p},s}p_{(l}\gamma_{m)}\mathcal{U}_{\mathbf{\mathbf{k+p}},r}\right)\hat a_{\mathbf{k+p},r}\right.\nn\\ 
&& \hspace{2.5cm}+ ~\hat b_{-\mathbf{p},s}\left({\bar{\tt v}}_{\mathbf{p},s}p_{(l}\gamma_{m)}{\tt v}_{\mathbf{\mathbf{k+p}},r}-\bar{\mathcal{V}}_{\mathbf{p},s}p_{(l}\gamma_{m)}\mathcal{V}_{\mathbf{\mathbf{k+p}},r}\right)\hat b_{-(\mathbf{k+p}),r}^\dag\nn\\
&& \hspace{2.8cm}+ ~ \hat a_{\mathbf{p},s}^\dag\left({\bar{\tt u}}_{\mathbf{p},s}p_{(l}\gamma_{m)}{\tt u}_{\mathbf{\mathbf{k+p}},r}-\bar{\mathcal{U}}_{\mathbf{p},s}p_{(l}\gamma_{m)}\mathcal{U}_{\mathbf{\mathbf{k+p}},r}\right)\hat a_{\mathbf{k+p},r} \nn\\ 
&& \hspace{3.1cm}+ ~ \left.\hat a_{\mathbf{p},s}^\dag\left({\bar{\tt u}}_{\mathbf{p},s}p_{(l}\gamma_{m)}{\tt v}_{\mathbf{\mathbf{k+p}},r}-\bar{\mathcal{U}}_{\mathbf{p},s}p_{(l}\gamma_{m)}\mathcal{V}_{\mathbf{\mathbf{k+p}},r}\right)\hat b_{-(\mathbf{k+p}),r}^\dag\,\right]\nn
\end{eqnarray}
The non-regularized expression is indeed identical to the above one but with every bilinear product of spinors $\sim (\bar {{\tt v}}\gamma{\tt u}-\bar{\mathcal{V}}\gamma\mathcal{U})$, $(\bar {{\tt v}}\gamma{\tt v}-\bar{\mathcal{V}}\gamma\mathcal{V})$, $(\bar {{\tt u}}\gamma{\tt u}-\bar{\mathcal{U}}\gamma\mathcal{U})$ and $(\bar {{\tt u}}\gamma{\tt v}-\bar{\mathcal{U}}\gamma\mathcal{V})$, replaced by  $\sim\bar{{\tt v}}\gamma{\tt u}, \bar{{\tt v}}\gamma{\tt v}, \bar{{\tt u}}\gamma{\tt u}$ and $\bar{{\tt u}}\gamma{\tt v}$, respectively. Thus, we obtain
\begin{equation}\label{eq:Pi_regII}
\left\langle 0 \left|\,\Pi_{ij,\rm{reg}}^{\rm TT}(\bk,t)\,\right|0 \right\rangle = (2\pi)^3 \left[{1\over a^2(t)}\int\hspace*{-1mm} {d\bp\over(2\pi)^3} \,\Lambda_{ij,lm}(\hat \bk)\,p_{(l}\left({\bar{\tt v}}_{\mathbf{p},s}\gamma_{m)}{\tt v}_{\mathbf{\mathbf{p}},s}-\bar{\mathcal{V}}_{\mathbf{p},s}\gamma_{m)}\mathcal{V}_{\mathbf{\mathbf{p}},s}\right)\right]\delta^{(3)}(\bk)\,,
\end{equation}
and analogously, the non-regularized VEV as
\begin{equation}\label{eq:Pi_NonReg_Appen}
\left\langle 0 \left|\,\Pi_{ij}^{\rm TT}(\bk,t)\,\right|0 \right\rangle = (2\pi)^3 \left[{1\over a^2(t)}\int\hspace*{-1mm} {d\bp\over(2\pi)^3} \,\Lambda_{ij,lm}(\hat \bk)\,p_{(l}\,{\bar{\tt v}}_{\mathbf{p},s}\gamma_{m)}{\tt v}_{\mathbf{\mathbf{p}},s}\right]\delta^{(3)}(\bk)\,,
\end{equation}
Notice that both $\langle 0|\Pi_{ij}^\TT(\bk,t)|0\rangle$ and $\langle 0|\Pi_{ij,{\rm reg}}^\TT(\bk,t)|0\rangle$ are proportional to $\delta^{(3)}(\bk)$, as expected from statistical homogeneity and isotropy. The corresponding VEVs of the GW source in real space reads\footnote{Note that for the homogeneous mode $\bk = {\bf 0}$, the TT-projection given by $\Lambda_{ij,lm}(\hat\bk)$ is ill-defined. Since the transversality condition is automatically verified by a homogeneous mode, it is then sufficient to redefine the projector for this case as $\Lambda_{ij,lm}({\bf 0}) \equiv \delta_{il}\delta_{jm}-{1\over3}\delta_{ij}\delta_{lm}$, which guarantees the tracelessness. For a discussion on how to define the analogous TT-projection on a lattice, suitable for numerical simulations, see~\cite{GWpreheatingTT}.} 
\begin{eqnarray}\label{eq:Pi_regConfigSpace_Appendix}
\left\langle 0 \left|\,\Pi_{ij}^{\rm TT}(\bx,t)\,\right|0 \right\rangle_{\rm reg} &\equiv& {1\over(2\pi)^3}\int d\bk \,e^{-i\bk\bx}\left\langle 0 \left|\,\Pi_{ij,{\rm reg}}^{\rm TT}(\bk,t)\,\right|0 \right\rangle \\
&=& \Lambda_{ij,lm}({\bf 0})\,{1\over a^2(t)}\int\hspace*{-1mm} {d\bp\over(2\pi)^3} \,p_{(l}\left({\bar{\tt v}}_{\mathbf{p},s}\gamma_{m)}{\tt v}_{\mathbf{\mathbf{p}},s}-\bar{\mathcal{V}}_{\mathbf{p},s}\gamma_{m)}\mathcal{V}_{\mathbf{\mathbf{p}},s}\right),\nn
\end{eqnarray}
again with analogous expression for the non-regularized version, but substituting the bilinear $\sim (\bar {{\tt v}}\gamma{\tt v}-\bar{\mathcal{V}}\gamma\mathcal{V})$ by $\sim \bar{{\tt v}}\gamma{\tt v}$. 
As expected, due to statistical homogeneity and isotropy, $\langle 0 |\,\Pi_{ij}^{\rm TT}(\bx,t)\,|0 \rangle_{\rm reg}$ does not depend on $\bx$. Actually $\langle 0 |\,\Pi_{ij}^{\rm TT}(\bx,t)\,|0 \rangle_{\rm reg} = 0,~\forall\,\bx$, since if we split the integration in eq.~(\ref{eq:Pi_regConfigSpace_Appendix}) into angular and radial parts, we obtain
\begin{eqnarray}\label{eq:Pi_regConfigSpaceII_Appendix}
\left\langle 0 \left|\,\Pi_{ij}^{\rm TT}(\bx,t)\,\right|0 \right\rangle_{\rm reg} &\propto&  \int d\Omega_{\hat\bp}\cos\psi_{(l} \times\int\hspace*{-1mm} {dp p^3} \left({\bar{\tt v}}_{\mathbf{p},s}\gamma_{m)}{\tt v}_{\mathbf{\mathbf{p}},s}-\bar{\mathcal{V}}_{\mathbf{p},s}\gamma_{m)}\mathcal{V}_{\mathbf{\mathbf{p}},s}\right)\\
&=& ~~~~~~~~~0 ~~~~~~~~\times~~~~~~~~~~~~~~~~ {\rm (finite)} ~~~~~~~~~~~~~~~~~~~~ =~ 0, \nn
\end{eqnarray}
with $d\Omega_{\hat\bp}$ the differential solid angle in momentum space, and $\psi_l$ the Euler angle of the momentum component $p_l$, i.e.~$\cos\psi_l \equiv p_l/|\bp|$. Notice that only thanks to the regularization, the radial integration over $p_{(l}\left({\bar{\tt v}}_{\mathbf{p},s}\gamma_{m)}{\tt v}_{\mathbf{\mathbf{p}},s}-\bar{\mathcal{V}}_{\mathbf{p},s}\gamma_{m)}\mathcal{V}_{\mathbf{\mathbf{p}},s}\right)$ in eq.~(\ref{eq:Pi_regConfigSpaceII_Appendix}), is finite. The integral over the angles is zero, and therefore $\langle 0 |\,\Pi_{ij}^{\rm TT}(\bx,t)\,|0 \rangle_{\rm reg}$ vanishes $\forall \,\bx$. In the non-regularized version, the analogous radial integration is performed over $p_{(l}{\bar{\tt v}}_{\mathbf{p},s}\gamma_{m)}{\tt v}_{\mathbf{\mathbf{p}},s}$, and it diverges. So even if the angular part integration is zero, still one needs to regularize the VEV in order to render finite and well defined the radial integration.

\subsection{4-point Functions: Unequal Time Correlators (UTCs)}
\label{subsec:4point_corr}

In this section we will obtain eqs.~(\ref{eq:Pi^2(u_r)}) and eq.~(\ref{eq:Pi2_regularized}), representing respectively the non-regularized and regularized VEV of the unequal-time-correlator (UTC), in terms of the mode functions obeying the Dirac equation. It will suffice to obtain the non-regularized version, since the regularized one will follow from there just substituting the mode functions $u_{\bp,\pm}$ by $\tilde u_{\bp,\pm} \equiv \sqrt{2n_{\bp}}\,u_{\bp,\pm}$, $n_{\bp}$ the occupation number, as explained in the main text.

For the sake of clarity, let us reproduce again here the UTC given by eq.~(\ref{eq:Pi^2(u_r)}), which looks like
\begin{equation}
\Pi^{2}(k,t,t')=\frac{1}{a^{2}(t)a^{2}(t')}\int\hspace*{-1mm}\frac{d\mathbf{p}}{\left(2\pi\right)^{3}}\,\left({\bar{\tt v}}_{\mathbf{p},s}(t)p_{(i}\gamma_{j)}{\tt u}_{\mathbf{k-p},r}(t)\Lambda_{ij,lm}(\hat{k}){\bar{\tt u}}_{\mathbf{k-p},r}(t')p_{(l}\gamma_{m)}{\tt v}_{\mathbf{p},s}(t')\right),\label{eq:Pi^2(u_r)_Append}
\end{equation}
Notice that the integrand corresponds to a trace over spinorial-indices. Thanks to $k_{i}\Lambda_{ij,lm}(\hat k) = \Lambda_{ij,lm}(\hat k)k_l = 0$ and to $\Lambda_{ij,lm}(\hat k) = \Lambda_{ij,lm}(-\hat k)$, we can write the integrand in any of the following equivalent manners
\begin{eqnarray}\label{eq:Trace4spinors_v1_Appen}
&& \left({\bar{\tt v}}_{\mathbf{p},s}(t)p_{(i}\gamma_{j)}{\tt u}_{\mathbf{k-p},r}(t)\Lambda_{ij,lm}(\hat{k}){\bar{\tt u}}_{\mathbf{k-p},r}(t')p_{(l}\gamma_{m)}{\tt v}_{\mathbf{p},s}(t')\right) \\
&\sim& \left({\bar{\tt v}}_{\mathbf{p-k},s}(t)p_{(i}\gamma_{j)}{\tt u}_{\mathbf{p},r}(t)\Lambda_{ij,lm}(\hat{k}){\bar{\tt u}}_{\mathbf{p},r}(t')p_{(l}\gamma_{m)}{\tt v}_{\mathbf{p-k},s}(t')\right)
 \\
&\sim& \left({\bar{\tt v}}_{\mathbf{p},s}(t)p_{(i}\gamma_{j)}{\tt u}_{\mathbf{k+p},r}(t)\Lambda_{ij,lm}(\hat{k}){\bar{\tt u}}_{\mathbf{k+p},r}(t')p_{(l}\gamma_{m)}{\tt v}_{\mathbf{p},s}(t')\right)
 \\
&\sim& \left({\bar{\tt v}}_{\mathbf{p+k},s}(t)p_{(i}\gamma_{j)}{\tt u}_{\mathbf{p},r}(t)\Lambda_{ij,lm}(\hat{k}){\bar{\tt u}}_{\mathbf{p},r}(t')p_{(l}\gamma_{m)}{\tt v}_{\mathbf{p+k},s}(t')\right)
\end{eqnarray}
Considering the first of these expressions, eq.~(\ref{eq:Trace4spinors_v1_Appen}), we find
\begin{eqnarray}\label{eq:TraceExplicitCalculation}
&& \left({\bar{\tt v}}_{\mathbf{p},s}(t)p_{(i}\gamma_{j)}{\tt u}_{\mathbf{k-p},r}(t)\Lambda_{ij,lm}(\hat{k}){\bar{\tt u}}_{\mathbf{k-p},r}(t')p_{(l}\gamma_{m)}{\tt v}_{\mathbf{p},s}(t')\right) \nn\\
&=& p_{(i}p_{(l}\Lambda_{ij,lm}(\hat{k})\left({{\tt v}}^\dag_{\mathbf{p},s}(t)\gamma_o\gamma_{j)}{\tt u}_{\mathbf{k-p},r}(t){{\tt u}}^\dag_{\mathbf{k-p},r}(t')\gamma_o\gamma_{m)}{\tt v}_{\mathbf{p},s}(t')\right) \nn\\
&=& p_{(i}p_{(l}\Lambda_{ij,lm}(\hat{k})(au^*_{\bp,-a}(t)S_{-s}\otimes P_a)^\dag\cdot(\mathcal{I}\otimes\sigma_3)\cdot(-i\sigma_{j)}\otimes\sigma_2)\cdot(u_{\bk-\bp,b}(t)S_{r}\otimes P_b)\nn\\
&& ~~~~~~~~~~~~~~~~~\cdot (u_{\bk-\bp,c}(t')S_{r}\otimes P_c)^\dag\cdot(\mathcal{I}\otimes\sigma_3)\cdot(-i\sigma_{m)}\otimes\sigma_2)\cdot(du^*_{\bp,-d}(t')S_{-s}\otimes P_d)\nn\\
&=& ad\,u_{\bp,-a}(t)u_{\bk-\bp,b}(t)u_{\bk-\bp,c}^*(t')u_{\bp,-d}^*(t')(S_{-s}^\dag \mathcal{I} p_{(i}\sigma_{j)}S_{r} \mathcal{I}\Lambda_{ij,lm}(\hat{k}) S_{r}^\dag \mathcal{I} p_{(l}\sigma_{m)}S_{-s})\nn\\
&& \hspace{6cm}~\otimes (P_a^\dag\sigma_3(-i\sigma_2)P_bP_c^\dag\sigma_3(-i\sigma_2)P_d) \nn\\
&=& ad\,u_{\bp,-a}(t)u_{\bk-\bp,b}(t)u_{\bk-\bp,c}^*(t')u_{\bp,-d}^*(t'){\rm Tr}(\Lambda_{ij,lm}(\hat{k})S_{-s}S_{-s}^\dag p_{(i}\sigma_{j)} S_{r}S_{r}^\dag \mathcal{I} p_{(l}\sigma_{m)})\nn\\
&& \hspace{6cm}~\otimes (P_a^\dag\sigma_1P_bP_c^\dag\sigma_1P_d) \nn\\
&=& ad\,u_{\bp,-a}(t)u_{\bk-\bp,b}(t)u_{\bk-\bp,c}^*(t')u_{\bp,-d}^*(t'){\rm Tr}(\Lambda_{ij,lm}(\hat{k}) \mathcal{I} p_{(i}\sigma_{j)} \mathcal{I} p_{(l}\sigma_{m)})\otimes (P_a^\dag P_{-b}P_c^\dag P_{-d}) \nn\\
&=& ad\,u_{\bp,-a}(t)u_{\bk-\bp,b}(t)u_{\bk-\bp,c}^*(t')u_{\bp,-d}^*(t'){\rm Tr}(p_{(i}\sigma_{j)} \Lambda_{ij,lm}(\hat{k})p_{(l}\sigma_{m)})\otimes (\delta_{a,-b}\delta_{c,-d}) \nn\\
&=& ac\,u_{\bp,-a}(t)u_{\bk-\bp,-a}(t)u_{\bk-\bp,-c}^*(t')u_{\bp,-c}^*(t'){\rm Tr}(p_{(i}\sigma_{j)} \Lambda_{ij,lm}(\hat{k})p_{(l}\sigma_{m)}) \nn\\
&=& W_{\bk,\bp}(t)W^*_{\bk,\bp}(t') 2\bp^2\sin^2\theta,
\end{eqnarray}
where in the last line we used, on one hand,
\begin{eqnarray}
{\rm Tr}(p_{(i}\sigma_{j)} \Lambda_{ij,lm}(\hat{k})p_{(l}\sigma_{m)}) = 2|\bp|^2\sin^2\theta,
\end{eqnarray}
with $\theta$ the angle between $\mathbf{k}$ and $\mathbf{p}$, and on the other hand,
\begin{eqnarray}
&& ac\,u_{\bp,-a}(t)u_{\bk-\bp,-a}(t)(u_{\bk-\bp,-c}(t')u_{\bp,-c}(t'))^* \\
&&= \left[u_{\mathbf{k-p},+}(t)u_{\mathbf{p},+}(t)u_{\mathbf{k-p},+}^*(t')u_{\mathbf{p},+}^*(t') + u_{\mathbf{k-p},-}(t)u_{\mathbf{p},-}(t)u_{\mathbf{k-p},-}^*(t')u_{\mathbf{p},-}^*(t') \right.\hV\nonumber\\
&& ~~~~~~~\left. -\,u_{\mathbf{k-p},+}(t)u_{\mathbf{p},+}(t)u_{\mathbf{k-p},-}^*(t')u_{\mathbf{p},-}^*(t') - u_{\mathbf{k-p},-}(t)u_{\mathbf{p},-}(t)u_{\mathbf{k-p},+}^*(t')u_{\mathbf{p},+}^*(t')\right]\nonumber\\
&& = W_{\bk,\bp}(t)W^*_{\bk,\bp}(t')
\end{eqnarray}
with 
\begin{eqnarray}
W_{\bk,\bp}(t) \equiv u_{\mathbf{k-p},+}(t)u_{\mathbf{p},+}(t) - u_{\mathbf{k-p},-}(t)u_{\mathbf{p},-}(t)
\end{eqnarray}

We have derived this way eq.~(\ref{eq:Trace}), which characterizes the non-regularized UTC eq.~(\ref{eq:Pi2_spectra(u_+,-)}), and ultimately determines the spectrum of GW eq.~(\ref{eq:GW_spectra(u_+,-)}). 

Having shown explicitly the calculation of the spinorial trace in eq.~(\ref{eq:Trace4spinors_v1_Appen}), there is no need to repeat the exercise for the regularized case. It is enough to substitute in the final expression of eq.~(\ref{eq:TraceExplicitCalculation}), every mode function $u_{\pm}$ by the regularized ones $\tilde u_{\pm}$ defined in eq.~(\ref{eq:RegModeFunc}).

\section{Time-dependent normal-ordering of bilinears $\mathcal{O}(t)$}
\label{app:AuxFermField}

 
As written in eq. (\ref{eq:tNO}) the tNO regularized vacuum expectation value of a bilinear $\mathcal{O}(t)$  is given by
\begin{equation}
\langle\mathcal{O}(t)\rangle_{{\rm reg}}\equiv\langle0|\mathcal{O}(t)|0\rangle-\langle0_{t}|\mathcal{O}(t)|0_{t}\rangle,
\label{tNO1}
\end{equation}
where $|0\rangle$ is the initial vacuum and $|0_t\rangle$ the vacuum at time $t$. We want to write eq. (\ref{tNO1}) in the form
\begin{equation}
\langle\mathcal{O}(t)\rangle_{{\rm reg}}\equiv\langle0|\mathcal{O}_{{\rm reg}}(t)|0\rangle,\quad
\mathcal{O}_{{\rm reg}}(t)\equiv\mathcal{O}(t)-\tilde{\mathcal{O}}(t)
\label{tNO2}
\end{equation}
where $\mathcal{O}_{{\rm reg}}(t)$ is an "effective regularized" bilinear, and $\tilde{\mathcal{O}}(t)$ satisfies $\langle0|\tilde{\mathcal{O}}(t)|0\rangle=\langle0_{t}|\mathcal{O}(t)|0_{t}\rangle$.
We can find $\tilde{\mathcal{O}}(t)$ by replacing the fields $\Psi(\mathbf{x},t)$, $\bar{\Psi}(\mathbf{x},t)$ in $\mathcal{O}(t)$ by some auxiliary fields $\Phi(\mathbf{x},t)$, $\bar{\Phi}(\mathbf{x},t)$. 
To do this we need to find some operators $\hat{c}_{\mathbf{k},r}(t)$ and $\hat{d}_{\mathbf{k},r}^{\dagger}(t)$, related to $\hat{a}_{\mathbf{k},r}(t)$ and $\hat{b}_{-\mathbf{k},r}^{\dagger}(t)$, which should satisfy for all combinations $\left\{j,k\right\}$ the constraint
\begin{equation}
\langle0|\hat{g}_{j}(t)\hat{g}_{k}(t)|0\rangle = \langle0_{t}|\hat{f}_{j}(t)\hat{f}_{k}(t)|0_{t}\rangle,
\end{equation}
with $\hat{g}_{j}(t)$, $\hat{g}_{k}(t)$ chosen each from the set of operators $\{ \hat{c}_{\mathbf{k},r}(t),\,\hat{c}_{\mathbf{k},r}^{\dagger}(t),\,\hat{d}_{\mathbf{k},r}(t),\,\hat{d}_{\mathbf{k},r}^{\dagger}(t)\}$, and $\hat{f}_{j}(t)$ and $\hat{f}_{k}(t)$ the analogous ones chosen from $\{ \hat{a}_{\mathbf{k},r}(t),\,\hat{a}_{\mathbf{k},r}^{\dagger}(t),\,\hat{b}_{-\mathbf{k},r}(t),\,\hat{b}_{-\mathbf{k},r}^{\dagger}(t)\}$. Considering a linear relation, we have found that the solution to the problem is given univocally by the following operators, which satisfy the required condition,
\begin{equation}
\hat{c}_{\mathbf{k},r}(t)=\alpha_{\mathbf{k}}(t)\hat{a}_{\mathbf{k},r}-\beta_{\mathbf{k}}(t)\hat{b}_{-\mathbf{k},r}^{\dagger}
\end{equation} 
\begin{equation}
\hat{d}_{\mathbf{k},r}^{\dagger}(t)=\alpha_{\mathbf{k}}(t)\hat{b}_{-\mathbf{k},r}^{\dagger}+\beta_{\mathbf{k}}^{*}(t)\hat{a}_{\mathbf{k},r}
\end{equation} 
For example, $\langle0|\hat{c}_{\mathbf{k},r}(t)\hat{d}_{\mathbf{p},s}(t)|0\rangle=\alpha_{\mathbf{k}}(t)\beta_{\mathbf{k}}(t)\delta^{(3)}\left(\mathbf{k}-\mathbf{p}\right)\delta_{r,s}=\langle0_{t}|\hat{a}_{\mathbf{k},r}(t)\hat{b}_{-\mathbf{p},s}^\dag(t)|0_{t}\rangle$ is verified. For the other combinations we find analogous identities.

We can now write the auxiliary fields $\Phi(\mathbf{x},t)$ and $\bar{\Phi}(\mathbf{x},t)$ as the original ones $\Psi(\mathbf{x},t)$ and $\bar{\Psi}(\mathbf{x},t)$, but replacing the operators $\hat{a}_{\mathbf{k},r}$ and $\hat{b}_{-\mathbf{k},r}^{\dagger}$ by $\hat{c}_{\mathbf{k},r}(t)$ and $\hat{d}_{\mathbf{k},r}^{\dagger}(t)$. Thus, we have
\begin{eqnarray}
a^{3/2}\,\Phi(\mathbf{x},t) &\equiv& \int\frac{d\mathbf{k}}{\left(2\pi\right)^{3}}\, e^{-i\mathbf{k\cdot x}}\left[\hat{c}_{\mathbf{k},r}(t){\tt u}_{\mathbf{k},r}(t)+\hat{d}_{\mathbf{k},r}^{\dagger}(t){\tt v}_{\mathbf{k},r}(t)\right]\nonumber\\\label{Phi}
&=& \int\frac{d\mathbf{k}}{\left(2\pi\right)^{3}}\, e^{-i\mathbf{k\cdot x}}\left[\hat{a}_{\mathbf{k},r} \, \mathcal{U}_{\mathbf{k},r}(t) + \hat{b}_{-\mathbf{k},r}^{\dagger} \, \mathcal{V}_{\mathbf{k},r}(t)\right],
\end{eqnarray}
and, correspondingly,
\begin{equation}
\label{barPhi}
a^{3/2}\,\bar{\Phi}(\mathbf{x},t)=a^{3/2}\,\Phi^{\dagger}(\mathbf{x},t)\gamma_{0}=\int\frac{d\mathbf{k}}{\left(2\pi\right)^{3}}\, e^{+i\mathbf{k}\mathbf{x}}\left[\hat{a}_{\mathbf{k},r}^{\dagger}\,\bar{\mathcal{U}}_{\mathbf{k},r}(t)+\hat{b}_{-\mathbf{k},r}\,\bar{\mathcal{V}}_{\mathbf{k},r}(t)\right].
\end{equation}
where we have defined a new set of four spinors as
\begin{equation}
\mathcal{U}_{\mathbf{k},r}(t)=\alpha_{\mathbf{k}}(t){\tt u}_{\mathbf{k},r}(t)+\beta_{\mathbf{k}}^{*}(t){\tt v}_{\mathbf{k},r}(t),
\end{equation}
\begin{equation}
\mathcal{V}_{\mathbf{k},r}(t)=\alpha_{\mathbf{k}}(t){\tt v}_{\mathbf{k},r}(t)-\beta_{\mathbf{k}}(t){\tt u}_{\mathbf{k},r}(t).
\end{equation}
Now the bilinear $\tilde{\mathcal{O}}(t)$ in eq.~(\ref{tNO2}) can be constructed by replacing the fields $\Psi(\mathbf{x},t)$ and $\bar{\Psi}(\mathbf{x},t)$ in $\mathcal{O}(t)$ by the fields $\Phi(\mathbf{x},t)$ and $\bar{\Phi}(\mathbf{x},t)$. Comparing eqs.~(\ref{Phi}), (\ref{barPhi}) with eqs.~(\ref{eq:Phi(u,v)}), (\ref{eq:barPhi(u,v)}), we observe that the only difference between $\Psi(\mathbf{x},t)$, $\bar{\Psi}(\mathbf{x},t)$ and $\Phi(\mathbf{x},t)$, $\bar{\Phi}(\mathbf{x},t)$, is that the 4-spinors ${\tt u}_{\mathbf{k},r}(t)$ and ${\tt v}_{\mathbf{k},r}(t)$ are replaced by $\mathcal{U}_{\mathbf{k},r}(t)$ and $\mathcal{V}_{\mathbf{k},r}(t)$, respectively. Hence, we finally arrive at the functional form of $\tilde{\mathcal{O}}(t)$ by simply replacing ${\tt u}_{\mathbf{k},r}(t)$ and ${\tt v}_{\mathbf{k},r}(t)$ in $\mathcal{O}(t)$, respectively with $\mathcal{U}_{\mathbf{k},r}(t)$ and $\mathcal{V}_{\mathbf{k},r}(t)$.

\section{What if the fermions have a bare mass?}
\label{app:FermionsMass}

In sections~\ref{subsec:YukawaAndScalarDynamics}-\ref{subsec:ThermalEraScenarios} we assumed that the fermions creating the GW had an effective mass $m_{\psi} = h\varphi$, due to the interactions with the scalar field $\varphi$. But what if the fermions had also a bare mass $m_{b}$?, i.e.~a term in the Lagrangian $\mathcal{L} = -m_{b}\bar\Psi\Psi$, with $m_b$ a constant mass parameter. The time-dependent effective mass of the fermionic field would then be 
\begin{equation}
\label{FermMassWithBareMass}
m_{\psi}(t)=m_{b}+h\varphi(t).
\end{equation}
The fermion modes are excited when their effective mass approaches a vanishing mass $m_{\psi} \rightarrow 0$, and hence when $h\varphi(t) \simeq -m_{b}$. All along section~\ref{sec:Applications} we assumed that $m_b = 0$, so that the fermions were excited when $\varphi(t)$ was crossing around zero. Let us recall that we are dealing with scalar fields which behave as $\varphi(t) \simeq \Phi(t)F(t)$, with $\Phi(t)$ the decreasing amplitude and $F(t + 2\pi/\omega) = F(t)$ a periodic function of frequency $\omega$ with an amplitude equal to unity. Fermions will not be excited if the bare mass is so large that the effective mass (\ref{FermMassWithBareMass}) is always nonvanishing. Hence, the fermions are excited only if the following constraint is satisfied
\begin{equation}
\label{FermMassConstraint}
m_b<h\Phi(t)
\end{equation}
In the case of small bare mass  $m_b \ll h\Phi(t)$ we would expect that the produced GW spectra does not differ much from the (bare) massless case $m_b=0$. On the contrary, if the bare mass is large $m_b \geq h\Phi(t)$ we expect that the produced GW spectra is significantly smaller. We have checked this fact in our numerical calculations, by computing the amplitude of GW peak amplitude as a function of $m_b$. We have chosen to demonstrate this effect within the preheating scenario of section~\ref{subsec:FermionicMassivePreheating}, but similar results are obtained for the scenarios in sections~\ref{subsec:FermionicMasslessPreheating} and~\ref{subsec:ThermalEraScenarios}.

We have found numerically that, as expected, the amplitude of the spectrum of GW produced by fermions is almost independent of a very small bare mass $m_{b}\ll h\Phi_{I}$, where $\Phi_{I}$ is the initial amplitude of the scalar field. On the contrary, the GW amplitude is strongly suppressed when the bare mass approaches the scale $h\Phi_I$, since the fermionic decay of the scalar field is then blocked due to the effective mass being forced to be always different than zero. 

\begin{figure}
\centering\includegraphics[width=10.5cm]{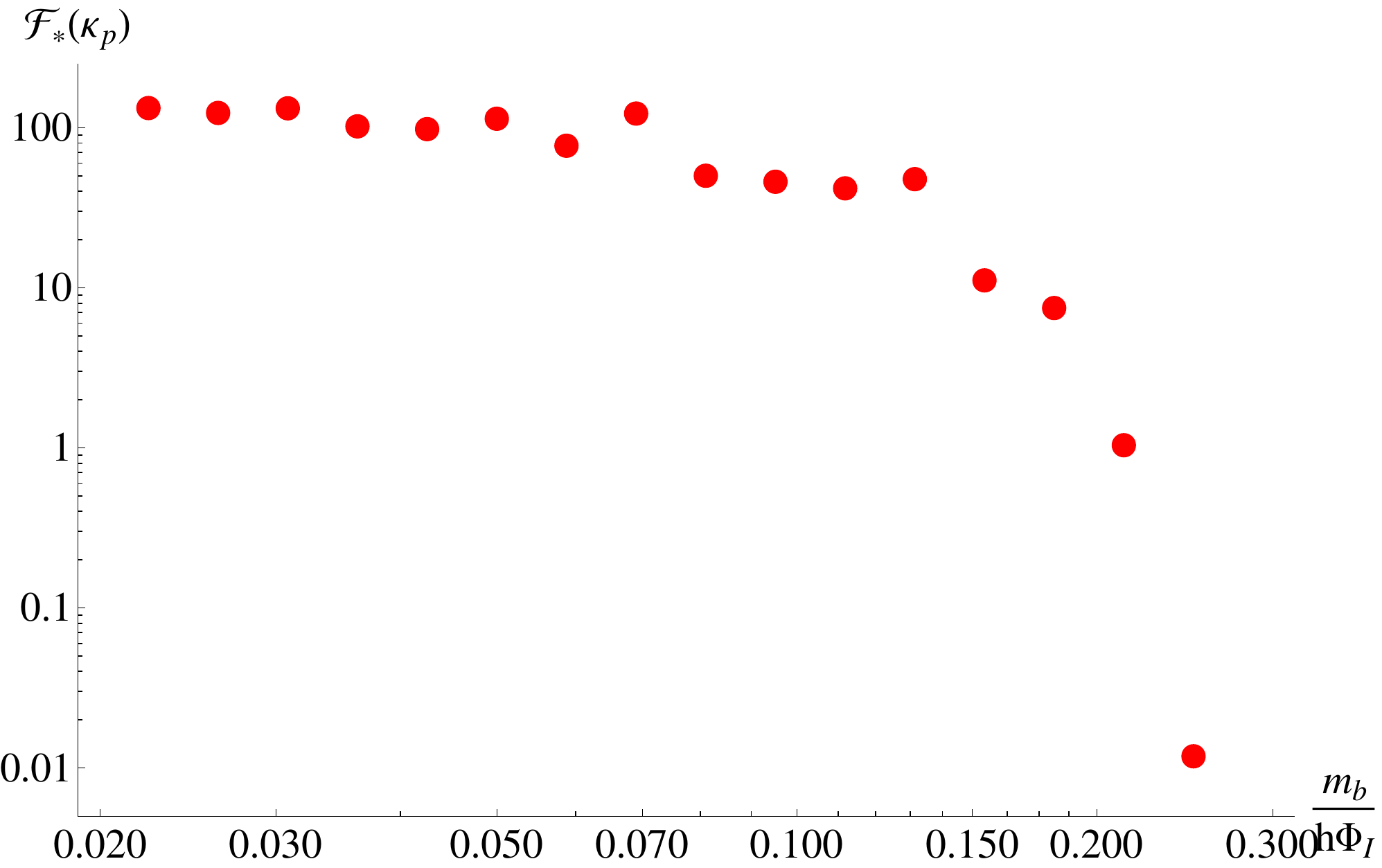}
\caption{The bare mass dependence of $\mathcal{F}(\kappa_p)$ for model considered in section (\ref{subsec:FermionicMassivePreheating}) with parameters $q=10^6$ and $\Phi_{I}={M_{p}\over2\sqrt{\pi}}
$.}
\label{fig:GW_vs_bareMass}
\end{figure}

As can be seen in figure~\ref{fig:GW_vs_bareMass}, the function $\mathcal{F}(\kappa_p)$, which characterizes the peak amplitude of the GW, drops to zero precisely when the bare mass $m_b$ approaches (from below) the scale $h\Phi_I$. Actually, it can be clearly appreciated that the amplitude of $\mathcal{F}(\kappa_p)$ starts falling already at $m_b \gtrsim 0.1\,h\Phi_I$. This occurs, because then the constraint (\ref{FermMassConstraint}) is satisfied only during a short period of time and thus the effective mass (\ref{FermMassWithBareMass}) crosses around zero only few times. In fact when $m_b \gtrsim 0.24\,h\Phi_I$, the fermion's effective mass simply never crosses around zero, and thus there is no fermion excitation nor production of GW. When $m_b$ is sufficiently smaller than $h\Phi_I$, the amplitude $\mathcal{F}(\kappa_p)$, and hence the GW background, is insensitive for the bare mass as indicated by the plateau for small bare mass in figure~(\ref{fig:GW_vs_bareMass}). Actually, the amplitude $\mathcal{F}(\kappa_p)$ is not exactly independent from the bare mass but there is a mild negative slope. The reason for this is similar as above for the relatively high bare mass. The amplitude $\Phi(t)$ decays in time and eventually $h\Phi(t)$ becomes smaller than $m_b$. Therefore this sets the end of GW generation earlier than than what it would be if there was no bare mass. The smaller $m_b$ the later this happens, thus the bigger the GW amplitude, and hence the negativeness of the slope (with increasing $m_b$). 

In summary, it is clear that the effect of a fermionic bare mass on the amplitude of the GW is simply, either negligible if $m_b \ll h\Phi_I$ or, on the contrary, a terminator if $m_b \gtrsim h\Phi_I$. Therefore, this justifies completely our assumption of taking $m_b = 0$ in the calculations in sections~\ref{subsec:YukawaAndScalarDynamics}-\ref{subsec:ThermalEraScenarios}.

\section{Comparison - Gravitational waves produced by bosons}
\label{app:GWbosons}

Instead of the Dirac equation (\ref{eq:Dirac}), a scalar field $\mathcal{X}$ in a FRW background evolves according to the Klein-Gordon equation, as
\begin{equation}
\mathcal{X}''+2\mathcal{H}\mathcal{X}'-\nabla^{2}\mathcal{X}+a^{2}m_{\chi}^{2}\mathcal{X} = 0,
\end{equation}
where $m_{\mathcal{X}}^{2} \equiv {d^2V(\mathcal{X})\over d\mathcal{X}^2}$ is its effective mass, and $V(\mathcal{X})$ its potential and over primes denoting derivatives with respect conformal time. If the scalar field $\mathcal{X}$ is coupled to another scalar field $\varphi$, $e.g.$ the inflaton at preheating, or an spectator field in the thermal era, the dynamics of the latter will provide an effective mass to the former. For example, if the two fields interact as $\mathcal{L}_{\rm int} = -\frac{1}{2} g^2\varphi^2\mathcal{X}^2$, then we have $m_{\mathcal{X}}^{2}=g^{2}\varphi^{2}$. Rescaling the scalar field as $\chi=a\mathcal{X}$, we then obtain the $eom$ at sub-horizon scales ($k \gg \mathcal{H}, {a''\over a}$) for $\mathcal{X}$, as
\begin{equation}
\chi''-\nabla^{2}\chi+a^{2}m_{\chi}^{2}\chi = 0
\label{eomChi}
\end{equation}
The quantized scalar field can be written like
\begin{equation}
\label{ChiQuant}
\chi(\mathbf{x},t)=\int\frac{d \mathbf{k}}{\left(2\pi\right)^{3}}e^{-i\mathbf{k\cdot x}}\left[\hat{a}_{\mathbf{k}}\chi_{\mathbf{k}}(t)+\hat{a}_{-\mathbf{k}}^{\dagger}\chi_{\mathbf{k}}^{*}(t)\right],
\end{equation}
where the creation/annihilation operator satisfy the canonical commutation relations 
\begin{equation}\label{eq:commutation}
[\hat{a}_{\mathbf{k}},\hat{a}_{\mathbf{k'}}^{\dagger}]=\left(2\pi\right)^{3}\delta^{(3)}(\mathbf{k-k'}),
\end{equation}
with other commutators vanishing. The definition of the (initial) vacuum state is given by $\hat{a}_{\mathbf{k}}|0\rangle = 0$, as usual. From eq. (\ref{eomChi}) we obtain $eom$ for $\chi_{\mathbf{k}}$ as
\begin{equation}
\chi_{\mathbf{k}}''+\left(k^{2}+a^{2}m_{\chi}^{2}\right)\chi_{\mathbf{k}}=0.
\end{equation}
The source of gravitational waves is the TT-part of the anisotropic stress tensor, given by
\begin{equation}
\label{StressTensorChi}
\Pi_{ij}^{\TT} = \left\lbrace \partial_{i}\mathcal{X}\,\partial_{j}\mathcal{X}\right\rbrace^{\TT}.
\end{equation}
By substituining the quantized field eq.~(\ref{ChiQuant}) in eq.~(\ref{StressTensorChi}), we obtain in Fourier space
\begin{equation}
\Pi_{ij}^{\TT}(\mathbf{k},t) = {\Lambda_{ij,lm}(\hat{k})\over a^2(t)}\int\hspace*{-0.1cm}\frac{d\mathbf{p}}{\left(2\pi\right)^{3}}~p_{l}p_{m}\left(\hat{a}_{\mathbf{p}}\chi_{\mathbf{p}}(t)+\hat{a}_{-\mathbf{p}}^{\dagger}\chi_{\mathbf{p}}^{*}(t)\right)\left(\hat{a}_{\mathbf{k-p}}\chi_{\mathbf{\mathbf{k-p}}}(t)+\hat{a}_{-\mathbf{\left(k-p\right)}}^{\dagger}\chi_{\mathbf{k-p}}^{*}(t)\right),
\end{equation}
where $\Lambda_{ij,lm}(\hat{k})$ is the TT-projection operator, defined in eq.~(\ref{projector}). To obtain the GW spectrum we need to calculate the Unequal-Time-Correlator (UTC) 
\begin{equation}
\langle \text{0}| \Pi_{ij}^{\TT}(\mathbf{k},t)\Pi_{ij}^{\TT^{\hspace*{0.2mm}*}}\hspace*{-0.7mm}(\mathbf{k'},t') |0\rangle \equiv (2\pi)^3 \Pi^2(k,t,t') \delta^{(3)}(\bk-\bk)
\end{equation}
The only combination of creation/annihilation operators which contributes to such VEV, turns out to be
\begin{equation}\label{eq:survivingCombinationBosons}
	\langle \text{0} |\hat{a}_{\mathbf{p}}\hat{a}_{\mathbf{k-p}}\hat{a}_{\mathbf{q}}^{\dagger}\hat{a}_{\mathbf{k'-q}}^{\dagger} |0\rangle =\left(2\pi\right)^{6}\big[\delta^{(3)}(\mathbf{k-p-q}) +\delta^{(3)}(\mathbf{p-q})\big]\delta^{(3)}(\mathbf{k-k'}),
\end{equation}
\begin{equation}
\langle \text{0}|\hat{a}_{\mathbf{p}}\hat{a}_{\mathbf{-(k-p)}}^{\dagger}\hat{a}_{\mathbf{q}}\hat{a}_{-(\mathbf{k'-q})}^{\dagger}|0\rangle = (2\pi)^{6}\delta^{(3)}(\mathbf{k})\delta^{(3)}(\mathbf{k'}-\bk),
\end{equation}
where we have used the commutation rule eq.~(\ref{eq:commutation}). The last term, proportional to $\delta^{(3)}(\mathbf{k})\delta^{(3)}(\mathbf{k'})$, does not contribute to $\Pi^{2}(k,t,t')$ at finite momenta $k = k' \neq {0}$. Thus only the term eq.~(\ref{eq:survivingCombinationBosons}) survives, and we then find the UTC for bosons as
\begin{equation}
\Pi^{2}(k,t,t') = \frac{1}{4\pi^{2}a^{2}(t)a^{2}(t')}\int dp\, d\theta\, p^{6}\sin^{5}\theta\,\chi_{\mathbf{p}}(t)\chi_{\mathbf{\mathbf{k-p}}}(t)\chi_{\mathbf{k-p}}^{*}(t')\chi_{\mathbf{\mathbf{p}}}^{*}(t'),
\end{equation}
where we have used the result
\begin{equation}
\Lambda_{ij,lm}(\hat{k})\big(p_{i}(k-p)_{j}(k-p)_{l}p_{m}+p_{i}(k-p)_{j}p_{l}(k-p)_{m}\big) = p^{4}\sin^{4}\theta
\end{equation}
with $\theta$ the angle between $\bp$ and $\bk$.

The spectrum of GW eq.~(\ref{eq:GW_spectra(Pi)}) is then given by
\begin{equation}
\frac{d\rho_{\mathrm{\GW}}}{d\log k}\left(k,t\right)=\frac{Gk^{3}}{2\pi^{3}a^{4}(t)}\int dp\, d\theta\, p^{6}\sin^{5}\theta\,\left(\left|I_{(c)}(k,p,\theta,t)\right|^{2}+\,\left|I_{(s)}(k,p,\theta,t)\right|^{2}\right),
\end{equation}
with
\begin{equation}
I_{(c)}(k,p,\theta,t)\equiv\int_{t_{I}}^{t}\frac{dt'}{a(t')}\cos(kt')\chi_{\mathbf{\mathbf{k-p}}}(t')\chi_{\mathbf{p}}(t')
\end{equation} 
\begin{equation}
I_{(s)}(k,p,\theta,t)\equiv\int_{t_{I}}^{t}\frac{dt'}{a(t')}\sin(kt')\chi_{\mathbf{\mathbf{k-p}}}(t')\chi_{\mathbf{p}}(t')
\end{equation}
The outcome is very similar, in the form, with the spectrum of gravitational waves produced by fermions given in eqs.~(\ref{eq:Pi^2(u_+,-)})-(\ref{eq:F_and_I_functions(u_+,-)}). We see that in the bosonic case, apart from different multiplicative factors, there is a power $p^{6}$ instead of $p^{4}$ in the momentum integrand. The 'extra' power $p^{2}$ arises in the presence of bosons because the bosonic energy-momentum tensor has two spatial derivatives, and not only one like in the fermionic case. Moreover, in the case of the bosons there is a $\sin^5\theta$ angular dependence, versus a $\sin^3\theta$ dependence in the case of Fermions. This is again due to the structure of derivatives in the energy-momentum tensors, which determine how many momenta variables are contracted with the TT-projector $\Lambda_{ij,lm}$, and thus ultimately how many powers of $\sin\theta$ appear. Besides, in the bosonic case there are no spinoral $dof$, and consequently there is no polarization indices $\pm$, as in the fermions. 
The GW by (scalar) bosons are sourced by scalar modes $\chi_{\mathbf{k}}(t)$ satisfying the Klein-Gordon eq.~(\ref{eomChi}), instead of by fermionic mode functions $u_{\mathbf{k},\pm}(t)$ satisfying the Dirac equation. The integrand in the $I_{(x)}$ functions, $\left(u_{\mathbf{k-p},+}(t)u_{\mathbf{p},+}(t) - u_{\mathbf{k-p},-}(t)u_{\mathbf{p},-}(t)\right)$, is then simply replaced by $\chi_{\mathbf{k-p}}(t)\chi_{\mathbf{k}}(t)$. 

Note that both bosonic vacuum expectation values, like the UTC $\Pi^{2}(k,t,t')$, also require regularization. This has not been an issue for bosonic sources in the literature, since the bosonic UTC's are typically introduced as theoretical Ansatzs, which are already regularized. Or in the case of lattice simulations of bosonic fields, the ultraviolet (UV) modes causing the divergence are not captured in the simulations, simply because the lattice spacing is finite. Thus, despite the fact that one would also require a similar regularization procedure for bosons as the one introduced for fermions in Section~\ref{subsec:Regularization}, in practice there has been no real need. The great difference between bosons and fermions sourcing GW, is that the excited modes in the bosons -- due to their bosonic nature --, develop a huge hierarchical ratio of relative amplitudes between IR and UV modes, the IR having a much greater amplitude than the UV ones. This is the case, for example, both in preheating and in phase 
transitions. In the former the UV tail of the distributions are exponentially suppressed, and in the latter the UV tail are power-law suppressed. If we were to include all the infinite tower of bosonic UV modes (so, well beyond the numerically obtained UV tails), we would encounter that they give rise to a divergence. However, due to the hierarchy of amplitudes, it has been as common practice to extend the UV tails (obtained only for a limited range of momenta) all the way up to infinite. This way, the otherwise divergent contribution form the UV bosonic vacuum fluctuations (which are not excited in the GW production process), is automatically removed. In the fermionic case, due to Pauli blocking, there is no such a hierarchy of amplitude between the IR and the UV modes, since every fermionic mode is 'Pauli blocked'. Therefore, in the case of fermions, one must necessarily deal with regularization, as indeed we showed in detail in Section~\ref{subsec:Regularization}.


\end{document}